%% file: paper.tex
\documentclass[twocolumn,showpacs,aps,prd]{revtex4}

\usepackage{graphicx}
\usepackage{dcolumn}
\usepackage{amsmath}
\usepackage{epsfig}
\usepackage{hhline}

\def\EPJ{{Eur. Phys. Jour.} C}

\def\NIMA{{Nucl. Instr. Meth.} A}
\def\NPB{{Nucl. Phys. B\,}}
\def\PLB{{Phys. Lett.} B}

\def\PRD{{Phys. Rev.} D}
\def\PRL{{Phys. Rev. Lett.}}

\def\ZPC{{Z. Phys.} C}
\def\etal{{\it et al.}}

\usepackage{relsize}
\def\babar{\mbox{\slshape B\kern-0.1em{\smaller A}\kern-0.1em
    B\kern-0.1em{\smaller A\kern-0.2em R}}}
\def\pep2{PEP-II}

\mathchardef\Upsilon="7107
\def\Y#1S{\ensuremath{\Upsilon{(#1S)}}\xspace}
\def\FourS {\Y4S}
\RequirePackage{xspace}
\def\ra {\ensuremath{\rightarrow}\xspace}
\def\epem {\ensuremath{e^+e^-}\xspace}
\def\pt {\mbox{$p_T$}\xspace}
\def\BB {\ensuremath{B\Bbar}\xspace}
\def\qqbar {\ensuremath{q\overline q}\xspace}
\def\taup {\ensuremath{\tau^+}\xspace}
\def\nub {\ensuremath{\overline{\nu}}\xspace}
\def\nutb {\ensuremath{\nub_\tau}\xspace}
\newcommand{\mc}{{\rm Monte Carlo}}

\def\Kpi{\mbox{$\Kmi\pipl$}}
\def\Ktwopi{\mbox{$\Kmi\pipl\pizero$}}
\def\Kthreepi{\mbox{$\Kmi\pipl\pipl\pimi$}}
\def\ppeta{\mbox{$\pipl\pimi\eta$}}

\def\etaM{\mbox{$\eta$}}
\def\etaPM{\mbox{$\eta^{\prime}$}}
\def\omegaM{\mbox{$\omega$}}
\def\De{\mbox{$\Delta E$}}
\def\mes{\mbox{$m_{\rm ES}$}}

\def\Acor{\mbox{${\cal A}_{\rm{corr}}$}}
\def\Bsec{\mbox{${\cal B}_{\rm sec}$}}
\def\Ncand{\mbox{${\cal N}_{\rm cand}$}}
\def\Pback{\mbox{${\cal N}_{\rm pb}$}}
\def\NPback{\mbox{${\cal N}_{\rm npb}$}}
\def\Ncf{\mbox{${\cal N}_{\rm CF}$}}

\providecommand{\thetathr}{\mbox{$\theta_{\rm thr}$}}
\providecommand{\thetasph}{\mbox{$\theta_{\rm sph}$}}
\providecommand{\thetahel}{\mbox{$\theta_{\rm hel}$}}
\providecommand{\thetadali}{\mbox{$\theta_{\rm D}$}}
\providecommand{\thetan}{\mbox{$\theta_{\rm N}$}}
\providecommand{\thetabst}{\mbox{$\theta_{\rm B^{\ast}}$}}

\def\stat{{\rm stat.}}
\def\syst{{\rm syst.}}

\providecommand{\rad}{\mbox{\rm rad}}

\providecommand{\mm}{\mbox{\rm mm}}

\providecommand{\invfb}{\mbox{${\rm fb^{-1}}$}}

\providecommand{\MeV}{\mbox{\rm MeV}}
\providecommand{\MeVc}{\mbox{${\rm MeV}/c$}}
\providecommand{\MeVcsq}{\mbox{${\rm MeV}/c^2$}}
\providecommand{\GeV}{\mbox{\rm GeV}}
\providecommand{\GeVc}{\mbox{${\rm GeV}/c$}}
\providecommand{\GeVcsq}{\mbox{${\rm GeV}/c^2$}}
\providecommand{\dedx}{\mbox{$dE/dx$}}
\providecommand{\Br}{\mbox{${\cal B}$}}
\providecommand{\chisq}{\mbox{$\chi^2$}}

\providecommand{\epl}{\mbox{$e^+$}}
\providecommand{\emi}{\mbox{$e^-$}}
\providecommand{\eplemi}{\mbox{\epl \emi}}

\providecommand{\hzero}{\mbox{$h^0$}}
\providecommand{\hmi}{\mbox{$h^-$}}

\providecommand{\Wmi}{\mbox{$W^-$}}
\providecommand{\gamgam}{\gamma \gamma}
\providecommand{\pipl}{\mbox{$\pi^+$}}
\providecommand{\pimi}{\mbox{$\pi^-$}}
\providecommand{\pizero}{\mbox{$\pi^0$}}

\providecommand{\twopi}{\mbox{\pipl \pimi}}
\providecommand{\threepi}{\mbox{\pipl \pimi \pizero}}

\providecommand{\etap}{\mbox{$\eta^\prime$}}
\providecommand{\rhopl}{\mbox{$\rho^+$}}
\providecommand{\rhomi}{\mbox{$\rho^-$}}

\providecommand{\qbar}{\mbox{$\overline{q}$}}

\providecommand{\cbar}{\mbox{$\overline{c}$}}

\providecommand{\bbar}{\mbox{$\overline{b}$}}

\providecommand{\Kbar}{\kern 0.18em\overline{\kern -0.18em
K}{}\xspace} 
\providecommand{\Kzerobar}{\mbox{$\Kbar^0$}}

\providecommand{\Kmi}{\mbox{$K^-$}}

\providecommand{\Kbar}{\mbox{$\overline{K}$}}

\providecommand{\Kstar}{\mbox{$K^*$}}

\providecommand{\pt}{\mbox{p_T}}
\providecommand{\bbar}{\mbox{\bbar}}
\providecommand{\cbar}{\mbox{\cbar}}
\providecommand{\qbar}{\mbox{\qbar}}

\providecommand{\Dzero}{\mbox{$D^0$}}

\providecommand{\Dstarpl}{\mbox{$D^{*+}$}}

\providecommand{\Dstarzero}{\mbox{$D^{*0}$}}

\providecommand{\Dstpl}{\mbox{$D^{(*)+}$}}
\providecommand{\Dstze}{\mbox{$D^{(*)0}$}}
\providecommand{\B}{\mbox{$B$}}
\providecommand{\Bbar}{\kern 0.18em\overline{\kern -0.18em
B}{}\xspace}
\providecommand{\BBbar}{\mbox{$B \Bbar$}}
\providecommand{\Bzero}{\mbox{$B^0$}}
\providecommand{\Bzerobar}{\mbox{$\Bbar^0$}}
\providecommand{\Bpl}{\mbox{$B^+$}}
\providecommand{\Bmi}{\mbox{$B^-$}}

\def\figurebox#1#2#3{%
    \def\arg{#3}%
    \ifx\arg\empty
    {\hfill\vbox{\hsize#2\hrule\hbox to #2{\vrule\hfill\vbox to #1{\hsize#2\vfill}\vrule}\hrule}\hfill}%
    \else
    {\hfill\epsfbox{#3}\hfill}%
    \fi}

\long\def\inst#1{\par\nobreak\kern 4pt\nobreak
    {\it #1}\par\vskip 10pt plus 3pt minus 3pt}

\newcommand{\BaBarYear}       {03}
\newcommand{\BaBarNumber}     {20}
\newcommand{\SLACPubNumber} {10193}

\begin{document}

{\pagestyle{empty}
\begin{flushleft}
\babar-PUB-\BaBarYear/\BaBarNumber  \\
SLAC-PUB-\SLACPubNumber \\
\end{flushleft}
}

\title
{\large \bf
Measurement of Branching Fractions of Color-Suppressed Decays \\
of the \boldmath{${\Bzerobar}$} Meson to $\mathbf{\Dstze
\pizero}$, $\mathbf{\Dstze \eta}$, $\mathbf{\Dstze \omega}$,  and
$\mathbf{\Dzero \eta^{\prime}}$
}

%
\author{B.~Aubert}
\author{R.~Barate}
\author{D.~Boutigny}
\author{J.-M.~Gaillard}
\author{A.~Hicheur}
\author{Y.~Karyotakis}
\author{J.~P.~Lees}
\author{P.~Robbe}
\author{V.~Tisserand}
\author{A.~Zghiche}
\affiliation{Laboratoire de Physique des Particules, F-74941
Annecy-le-Vieux, France }
% ...
\author{(\babar\ Collaboration)}
\noaffiliation

\date{\today}

\begin{abstract}
\input{abstract}
\end{abstract}



\maketitle

\section{Introduction}
\label{sec:Introduction}

\begin{figure}
\begin{center}
\epsfig{file=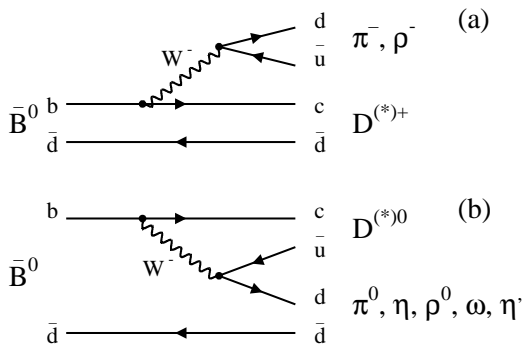,width=0.4\textwidth}
\end{center}
\caption[decaydiagram]{The (a) color-allowed and (b)
color-suppressed spectator tree diagrams for $\Bzerobar \ra D \ h$
decays. \label{fig:Feynman}}
\end{figure}

Weak decays like $\Bzerobar \ra \Dstpl \hmi$ can proceed through
the emission of a virtual $\Wmi$, which then can materialize as a
charged hadron~\cite{ref:footnote1}. Because the $\Wmi$ carries no
color, no exchange of gluons with the rest of the final state is
required. Such decays are called color-allowed, though
color-favored might be more apt. By contrast, decays like
$\Bzerobar \ra \Dstze \hzero$ cannot occur in this fashion. The
quark from the decay of the virtual $\Wmi$ must be combined with
some anti-quark other than its partner from the $\Wmi$. However,
other anti-quarks will have the right color to make a color
singlet only one-third of the time. As a result, these decays are
``color-suppressed''. The tree level diagrams for the
color-allowed and color-suppressed decays are shown in
Fig.~\ref{fig:Feynman}.

\begin{table}
\caption{\label{tab:existingBFs} Prior measurements of branching
fractions for $\Bzerobar$ color-allowed and color-suppressed
decays. When two uncertainties are given, the first uncertainty is
statistical and the second systematic. We also quote the 90\%
confidence upper limits (UL) when the statistical significance of
the measurement is less than four standard deviations.}
\begin{center}
\begin{tabular}{lcccc}
\hline \hline \\
$\Bzerobar$ mode      & { \ } & {$\Br \ (\times 10^{-4})$} & { \ } & {UL ($\times 10^{-4}$)} \ \\ \hline \\
$D^+ \pi^-$          && $26.8 \pm 1.2 \pm 2.7$~\cite{ref:cleoDpi} && - \ \\
$\Dzero \pizero$     && $2.9 \pm 0.5$~\cite{ref:PDG} && - \ \\
$D^{\ast +} \pi^-$   && $27.6 \pm 2.1$~\cite{ref:PDG} && - \ \\
$\Dstarzero \pizero$ && $2.5 \pm 0.7$~\cite{ref:PDG} && - \ \\
$D^+ \rho^-$         && $78 \pm 14$~\cite{ref:PDG} && -  \ \\
$\Dzero \rho^0$      && $2.9 \pm 1.0 \pm 0.4$~\cite{ref:Belle2} && - \ \\
$D^{\ast +} \rho^-$  && $73 \pm 15$~\cite{ref:PDG} && - \ \\
$\Dstarzero \rho^0$  && - && $<5.1$~\cite{ref:Belle2} \ \\
$\Dzero \eta$        && $1.4^{+0.5}_{-0.4} \pm 0.3$~\cite{ref:Belle} \ \\
$\Dstarzero \eta$    && $2.0^{+0.9}_{-0.8} \pm 0.4$~\cite{ref:Belle}  && $<2.6$~\cite{ref:PDG} \ \\
$\Dzero \omega$      && $1.8 \pm 0.5^{+0.4}_{-0.3}$~\cite{ref:Belle} && - \ \\
$\Dstarzero \omega$  && $3.1^{+1.3}_{-1.1} \pm 0.8$~\cite{ref:Belle} && $<7.4$~\cite{ref:PDG} \ \\
$\Dzero \eta^{\prime}$     && - && $<9.4$~\cite{ref:PDG} \ \\
$\Dstarzero \eta^{\prime}$ && - && $<14$~\cite{ref:PDG} \ \\
\hline \hline
\end{tabular}
\end{center}
\end{table}

The decays of $\Bzerobar$ into $\Dstze \pizero$ have been observed
by the CLEO collaboration~\cite{ref:CLEO}, while the $\Bzerobar$
decays into $\Dstze \pizero$, $\Dzero \eta$, $\Dzero \omega$, and
$\Dzero \rho^0$ have also been measured by the Belle
collaboration~\cite{ref:Belle,ref:Belle2}. We present in
Table~\ref{tab:existingBFs} the prior measurements of branching
fractions of the $\Bzerobar$ color-allowed and color-suppressed
decays. The level of color suppression can be estimated from the
branching fractions for the $D^{(\ast)} \pi$ and $D^{(\ast)} \rho$
decay modes.

Since QCD calculations of decay rates from first principles are at
present not possible, we must rely on models to describe the above
processes. In an early model~\cite{ref:BSW,ref:Deandrea1}, the
``naive'' (or ``generalized'') factorization model, which is very
successful in describing charmed meson decays, the decay
amplitudes of exclusive two-body non-leptonic weak decays of heavy
flavor mesons are estimated by replacing hadronic matrix elements
of four-quark operators in the effective weak Hamiltonian by
products of current matrix elements. These current matrix elements
are determined in terms of form factors describing the transition
of the $B$ meson into the meson containing the spectator quark,
and a factor proportional to a decay constant describing the
creation of a single meson from the remaining quark--anti-quark
pair. In this approach, the decay amplitudes corresponding to
Figs.~\ref{fig:Feynman}(a) and \ref{fig:Feynman}(b) are
proportional to $a_1$ and $a_2$~\cite{ref:NeuSte}, respectively,
where the $a_i$ are effective QCD Wilson coefficients. As an
example, using the naive factorization model, the decay amplitude
for the $\Bzerobar \ra D^+ \pi^-$ mode corresponding to
Fig.~\ref{fig:Feynman}(a) can be written as~\cite{ref:NeuSte}
\begin{multline}
{\cal A}^f{(\Bzerobar \ra D^+ \pi^-)}= \\ i{{G_F} \over
\sqrt{2}}V_{cb} V^{\ast}_{ud} (m^2_B - m^2_D) a_1  f_{\pi} F_0^{B
\ra D}(m^2_{\pi}),
\end{multline}
while the decay amplitude for the $\Bzerobar \ra D^0 \pizero$ mode
corresponding to Fig.~\ref{fig:Feynman}(b) can be expressed
as~\cite{ref:NeuPet,ref:ChengYang}
\begin{multline}
\sqrt{2}{\cal A}^f{(\Bzerobar \ra D^0 \pizero)}= \\  i{{G_F} \over
\sqrt{2}}V_{cb} V^{\ast}_{ud} (m^2_B - m^2_{\pi}) a_2 f_{D} F_0^{B
\ra \pi}(m^2_{D}),
\end{multline}
where $G_F$ is the Fermi coupling constant, $V_{cb}$ and $V_{ud}$
are CKM matrix elements, $f_\pi$ and $f_D$ are the decay constants
of the $\pi$ and $D$ mesons, and $F_0^{B \ra M}(q^2)$ are the
longitudinal form factors of the $B$-meson decays to  $M$ mesons
at momentum transfer $q^2$. The coefficients $a_1$ and $a_2$ are
real in the absence of final-state interactions (FSI) and ideally
would be process
independent~\cite{ref:BSW,ref:Deandrea1,ref:NeuSte,ref:NeuPet}.

The color-allowed $\Bzerobar \ra D^{(*)+} (\pi^-,\rho^-,a_1^-)$
and $B \ra D^{(*)} D_s^{(*)}$ decays, the color-suppressed $B \ra
(c\cbar) (K^{(*)},\pi)$ decays, and the mixed $B^- \ra \Dstze
(\pi^-,\rho^-,a_1^-)$ decays can all be accommodated by universal
constants $a_1 = 1.1\pm0.2$ and $a_2 \simeq$~0.2--0.3
\cite{ref:Beneke,ref:NeuSte,ref:KKW,ref:PDG}. This no longer holds
for color-suppressed $B$ decays with one $c$-quark only, like
$D^{(*)}\pi$, where measurements listed in
Table~\ref{tab:existingBFs} are inconsistent with a universal
value of $a_2$ in the absence of FSI~\cite{ref:NeuPet}. The naive
factorization
model~\cite{ref:Beneke,ref:NeuSte,ref:NeuPet,ref:Chua,ref:Rosner,
ref:Deandrea,ref:ChRos} predicts too small values for the
branching fractions of the color-suppressed modes, in the range
$(0.3$--$1.7) \times 10^{-4}$  and corresponding to a factor
$(a_2/a_1)^2 \simeq$~0.03--0.09.

Final state interactions, however, may change this picture
significantly and, thus, may increase substantially these rates,
as rescattering effects can connect the final states shown in
Fig.~\ref{fig:Feynman}(a) and Fig.~\ref{fig:Feynman}(b) (see, for
example, Ref.~\cite{ref:Chua}). Similar effects have already in
the past completely changed the conclusions of the models that
describe non-leptonic $\Dzero$ decays, especially for decay modes
such as $\Dzero \ra \Kzerobar \pizero$~\cite{ref:Lipkin}.
Therefore, in the case of large FSI, a description in terms of
isospin amplitudes is more appropriate and will be used in
Sec.~\ref{sec:IsospinRel} to discuss our results.

This situation is an impetus for higher precision results and the
investigation of additional channels that might provide clues to
the underlying mechanisms. In this paper we report on the
branching fraction measurements of the seven color-suppressed
$\Bzerobar$-meson decays to $\Dstze \pizero$, $\Dstze \eta$,
$\Dstze \omega$, and $\Dzero \eta^{\prime}$. We also report on a
search for the $\Bzerobar \ra \Dstarzero \eta^{\prime}$ decay.
These results are based upon an integrated luminosity equivalent
to $88.8 \times 10^6$ $\BBbar$ events. This corresponds to about
nine times that used for the earlier measurement by
CLEO~\cite{ref:CLEO} ($9.7 \times 10^6$ $\BBbar$ events) and about
four times that used for the earlier measurements by
Belle~\cite{ref:Belle} ($23.1 \times 10^6$ $\BBbar$ events).
Recently, with $31.3 \times 10^6$ $\BBbar$ events, the Belle
collaboration has reported branching fraction measurements for
$\Bzerobar \ra \Dstze \pi^+ \pi^-$ decays, including the $\Dzero
\rho^0$ mode, as already discussed, and the investigation of the
$\Dstarzero \rho^0$ channel~\cite{ref:Belle2}. We present the
first measurement of the $\Bzerobar \ra \Dstarzero \eta$,
$\Dstarzero \omega$, and $\Dzero \eta^{\prime}$ modes with more
than five-sigma statistical significance.

\section{The \babar\ detector and data sample}
\label{sec:babar}

The \babar\ detector is located at the \pep2\ \epem\ storage rings
operating at the Stanford Linear Accelerator Center. At PEP-II
9.0-$\GeV$ electrons collide with 3.1-$\GeV$ positrons to produce
a center-of-mass energy of $10.58\, \GeV$, the mass of the \FourS.
The data used in this analysis were collected with the \babar\
detector and correspond to an integrated luminosity of $81.9\,
\invfb$ recorded at the $\FourS$ resonance.

The \babar\ detector is described in detail in
Ref.~\cite{ref:babar}. Surrounding the interaction point is a
5-layer double-sided silicon vertex tracker (SVT), which gives
precision spatial information in three dimensions for charged
particles and measures their energy loss $(\dedx)$. The SVT is the
primary detection device for low-momentum charged particles.
Outside the SVT, a 40-layer drift chamber (DCH) provides
measurements of the polar angles and of the transverse momentum
(\pt) of charged particles with respect to the beam direction,
together with the SVT. The resolution of the \pt\ measurement for
tracks with momenta above $1\, \GeVc$ is $\sigma_{\pt} / \pt =
0.13\% \times \pt + 0.45\%$, where $\pt$ is measured in $\GeVc$.
The drift chamber measures \dedx\ with a precision of 7.5\%.
Beyond the outer radius of the DCH  is a detector of internally
reflected Cherenkov radiation (DIRC), which is used primarily for
charged-hadron identification. The detector consists of quartz
bars in which Cherenkov light is produced when relativistic
charged particles traverse the material. The light is internally
reflected along the length of the bar into a water-filled volume
mounted on one end of the detector. The Cherenkov rings expand in
the water volume and are measured with an array of photomultiplier
tubes mounted on its outer surface. A CsI(Tl) crystal
electromagnetic calorimeter (EMC) is used to detect photons and
neutral hadrons, as well as to identify electrons. The resolution
of the calorimeter can be expressed as $\sigma_E / E = 2.3\% /
(E)^{\frac{1}{4}} \oplus 1.9\%$, where $E$ is measured in $\GeV$.
The EMC detects photons with energies down to $20\, \MeV$. The EMC
is surrounded by a superconducting solenoid, which produces at
1.5-T magnetic field. The instrumented flux-return (IFR) consists
of multiple layers of resistive plate chambers (RPC) interleaved
with the flux-return iron. The IFR is used in the identification
of muons and long-lived neutral hadrons.

Signal and generic background \mc\ events are generated using the
\babar\ particle decay simulation package~\cite{ref:Lange}, the
``EvtGen'' package. The interactions of the generated particles
traversing the detector are simulated using the
GEANT4~\cite{ref:GEANT} program. Beam-induced backgrounds, which
varied from one data-taking period to the next, are taken into
account in the simulation of the detector response. This is done
by adding the signals generated  by these beam-induced backgrounds
to the simulation of the various physics events.

\section{Particle reconstruction and Counting of \boldmath $\BBbar$ events}
\label{sec:Bcounting}

Charged-particle tracks are reconstructed from measurements in the
SVT and/or the DCH. The tracks must have at least 12 hits in the
DCH and $\pt > 100$~$\MeVc$~\cite{ref:footnote2}. In the case of
the tracks used to reconstruct $\rho^\pm$ mesons, we also use
tracks reconstructed with the SVT alone (see
Sec.~\ref{subsec:pi0select}). The tracks must extrapolate to
within $20\,\mm$ of the $\eplemi$ interaction point in the plane
transverse to the beam axis and to within $50\,\mm$ along the beam
axis. Charged-kaon candidates are identified using a likelihood
function that combines ${\rm d}E/{\rm d}x$ and DIRC information.
The likelihood function is used to define tight and loose kaon
criteria as pion vetos. To satisfy the tight kaon criterion, the
track must also have $p > 250$~$\MeVc$ and make an angle with
respect to the electron beam direction, which is used as the
reference axis for all the polar angles, between 0.45 and $2.50\,
\rad$ so that the candidate is within the fiducial region of the
DIRC. Photons are identified by energy deposits in contiguous
crystals in the EMC. Each photon must have an energy greater than
30~$\MeV$ and a lateral shower shape consistent with that of an
electromagnetic shower.

The measurement of branching fractions depends upon an accurate
measurement of the number of $\BB$ meson pairs in the data sample.
We find the number of \BB\ pairs by comparing the rate of
spherical multi-hadron events in data recorded on the \FourS\
resonance to that in data taken off-resonance. This latter data
sample is collected 40~$\MeV$ below the $\FourS$ resonance and
corresponds to an integrated luminosity of about $10\, \invfb$.

The purity of the multi-hadrons events is enhanced by requiring
the events to pass selection criteria based on all tracks
(including those reconstructed in the SVT only), detected in the
fiducial region $0.41 <\theta < 2.54 \ \rad$ and on neutral
clusters with an energy greater than 30~$\MeV$, in the fiducial
region $0.410 < \theta < 2.409 \  \rad$:
\begin{itemize}
\item There must be at least three tracks in the fiducial region.
The total energy of the charged and neutral particles in the
fiducial region must be greater than 4.5~$\GeV$.

\item The ratio of the second to the zeroth Fox-Wolfram
moment~\cite{ref:FoxWolf} must be less than 0.5. All tracks and
neutral clusters defined above are used.

\item The event vertex must be within $5\, \mm$ of the nominal
beam-spot position in the plane transverse to the beam and within
$60\, \mm$ along the beam direction.

\end{itemize}
These requirements are about 95.4\% efficient for \BB\ events as
estimated from \mc\ simulation. The systematic uncertainty on the
number of $\BBbar$ events is 1.1\%.

\section{Meson candidate selection}

\subsection {General considerations}

The color-suppressed $\Bzerobar$ meson decay modes are
reconstructed from $\Dzero$ or $\Dstarzero$ meson candidates that
are combined with light neutral-meson candidates $\hzero$
($\pizero$, $\eta$, $\omega$, and $\etap$). Events are required to
pass the selection criteria used for $\BB$ counting listed in
Sec.~\ref{sec:Bcounting}. Additional requirements discussed below
are applied to the signal sample.

We combine tracks and/or neutral clusters to form candidates for
the mesons produced in the $B$ decays. Vertex constraints are
applied to charged daughters before computing their invariant
masses. At each step in the decay chain we require that mesons
have masses consistent with their assumed particle type. If
daughter particles are produced in the decay of a parent meson
with a natural width that is small relative to the reconstructed
width, we constrain the meson's mass to its nominal value. This
fitting technique improves the resolution of the energy and the
momentum of the $\Bzerobar$ candidates as they are calculated from
improved energies and momenta of the $\Dstze$ and $\hzero$.

We select $\Dstarzero$, $\Dzero$, $\hzero$, and $\Bzerobar$
candidates using only well-understood discriminating variables in
order to reduce the systematic uncertainties for the branching
fraction measurements. We choose selection criteria that maximize
the quality factor $Q = S / \sqrt{S+B}$, where $S$ and $B$ are the
expected number of signal and background events. The values of $S$
and $B$ are estimated from signal and background \mc\ simulation
and data in the signal sidebands, but not from data in the signal
regions. When optimizing the cuts, the values of $S$ have been
estimated using the previous branching fraction measurements
obtained by the CLEO~\cite{ref:CLEO} and Belle~\cite{ref:Belle}
collaborations. For the $\Dstze \eta^{\prime}$ analyses, a
conservative value for the branching fractions equal to $10^{-4}$
has been assumed. In most cases we find that $Q$ does not change
significantly when selection criteria are varied near their
optimal values. This allows us to choose selection criteria that
are common to most final states.

\subsection {Selection of \boldmath{$\hzero$} and $\rho^\pm$ candidates}
\label{sec:lightHad}

The momentum of the $\hzero$ candidate must satisfy the condition
$1.3 < p^{\ast} < 3.0$~\GeV/c. This requirement is loose enough
that various sources of background populate the sidebands of the
signal region. These sidebands are used in the background estimate
for the signal.

\subsubsection{\bf \boldmath{$\pizero$} and $\rho^\pm$ selection}
\label{subsec:pi0select}

The $\pizero$ meson is reconstructed from photon pairs. We
consider three sources of $\pizero$ with decreasing momenta:
$\pizero$ originating from $\Bzerobar$ decays, from $\Dzero$,
$\eta$, and $\omega$ decays, and directly from $\Dstarzero$
decays. The latter two sources are discussed below. The mass
resolution of $\pizero$ candidates from $\Bzerobar$ decays with
momenta $p^{\ast}$ near 2~$\GeVc$ is dominated by the uncertainty
in the opening angle between the two photons and is approximately
8~$\MeVcsq$.

These $\pizero$s are also combined with charged pions to attempt
the reconstruction of $\rhomi$ mesons. The charged pions are not
required to satisfy our regular selection criteria for tracks.
Thus we retain also low momentum charged pions that are
reconstructed with the SVT alone. A  $\pizero \pi^-$ pair is
selected if its mass is reconstructed within 250~$\MeVcsq$ of the
nominal $\rhomi$ meson mass. The $\rhomi$ candidates are used to
reconstruct the color-allowed $\Bmi \ra \Dstze \rho^-$ decays that
form a significant background for $\Bzerobar \ra \Dstze \pizero$.
The color-allowed decays have branching fractions about fifty
times that for $\Bzerobar \ra \Dstze \pizero$ and they mimic the
latter through an asymmetric $\rhomi$ decay in which the $\pizero$
carries most of the available energy. We veto events with a
reconstructed $\Bmi \ra \Dstze \rhomi$. A discussion of the veto
is deferred until Secs.~\ref{subsec:d0pi0}~and~B.

\subsubsection{\bf \boldmath{$\eta$} selection}
\label{subsec:etaselect}

The $\eta$ candidate is reconstructed in the $\gamgam$ and
$\threepi$ decay modes. The branching fraction in the $\gamgam$
mode is almost twice as large as that of the $\threepi$ decay
channel and the efficiency for the $\gamgam$ mode is greater since
there are fewer particles to detect.

In the $\gamgam$ decay mode we require that the photons have
energies greater than 200~$\MeV$. A photon is not used if it can
be paired with another photon with energy greater than 150~$\MeV$
to form a $\pizero$ candidate with an invariant mass in the range
120--150~$\MeVcsq$. The mass resolution for $\eta \ra \gamgam$ is
approximately 15~$\MeVcsq$.

In the $\threepi$ decay mode, the $\eta$ meson is reconstructed
employing a vertex constraint that requires a $\chi^2$ probability
greater than 0.1\%. To reduce combinatorial background the
charged-pion candidates must have momentum greater than
250~$\MeVc$ and they must fail the tight kaon criterion, while the
$\pizero$ must have an energy greater than 300~$\MeV$ and a mass
in the range 115--150~$\MeVcsq$. The mass resolution for $\eta \ra
\threepi$ is approximately 4~$\MeVcsq$.

\subsubsection{\bf \boldmath{$\omega$} selection}

The $\omega$ meson is reconstructed in its $\threepi$ decay mode,
employing a vertex constraint that requires a $\chi^2$ probability
greater than 0.1\%. To reduce combinatorial background, the
charged pion candidates must have momentum greater than
200~$\MeVc$ and they must fail the tight kaon criterion, while the
$\pizero$ must have an energy greater than 250~$\MeV$ and a mass
in the range 120--150~$\MeVcsq$. The mass resolution of the
$\omega$ is dominated by its natural width of approximately $10\,
\MeVcsq$. The use of additional angular properties in the $\omega$
meson decays will be described in Sec.~\ref{sec:evtshapAndangl}.

\subsubsection{\bf \boldmath{$\eta^{\prime}$} selection}

We reconstruct the $\eta^{\prime}$ meson in its $\ppeta (\ra
\gamgam)$ decay mode. The product of the branching fractions of
secondary decays in this channel is 17.5\%~\cite{ref:PDG}. This
limits the signal efficiency, so a separate event selection for
$\Dstze \etap$ is used. We use the $\ppeta$ decay mode rather than
the dominant $\rho^0 \gamma$ mode as it provides a much cleaner
signal.

The two photons used to reconstruct the $\eta$ candidate are
required to have energies greater than 100~$\MeV$. A photon is not
used to reconstruct the $\eta$ meson if it can be paired with
another photon with energy greater than 100~$\MeV$ to form a
$\pizero$ candidate with mass in the range 120--150~$\MeVcsq$. We
select $\eta$ candidates with a mass in the range
495--600~$\MeVcsq$. To obtain the highest possible signal
efficiency we rely on the high purity of the signal and impose
neither a momentum nor any particle-identification requirement on
the charged pions. For the same reason, a vertex constraint is
applied to the $\twopi$ pair when computing the energy and the
momentum of an $\eta^{\prime}$ meson candidate, but there is no
requirement on the $\chi^2$ probability of the vertex. The mass
resolution for $\eta^{\prime} \ra \ppeta (\ra \gamgam)$ is
approximately 4~$\MeVcsq$.

\subsection{Selection of \boldmath{$\Dzero$ and $\Dstarzero$} candidates}
\label{sec:CharmMesons}

The momentum of the $\Dstze$ mesons must satisfy the condition
$p^{\ast} > 1.5$~\GeV/c. As for the light neutral-hadron
selection, this requirement retains sidebands, which can be used
to evaluate backgrounds.

\subsubsection{\bf \boldmath{$\Dzero \to \Kpi$}, $\Ktwopi$, and $\Kthreepi$
selection}

The $\Dzero$ mesons are reconstructed in three decay modes:
$\Kpi$, $\Ktwopi$, and $\Kthreepi$. The $\chi^2$ probability for
the vertex fit of the charged pions is required to be greater than
0.1\%. In the $\Kpi$ final state the kaon candidate must satisfy
the pion veto requirement, while in the $\Ktwopi$ and $\Kthreepi$
final states the kaon candidate must satisfy the tight kaon
criterion because of the increased background present in these
combinations. All pion candidates must fail the tight kaon
criterion.

To reduce combinatorial background in the $\Ktwopi$ final state we
use the results of the Fermilab E691
experiment~\cite{ref:DaliKpipi0}, which determined the
distribution of events in the Dalitz plot. This distribution is
dominated by the two possible $\Kstar$ resonances ($K^{*0} \ra
\Kmi \pi^+$ or $K^{*-} \ra \Kmi \pizero$) and by the $\rhopl(\ra
\pipl \pizero)$ resonance. We select only those events that fall
in the enhanced regions of the Dalitz plot as determined by
experiment E691. Reconstructed $\pizero$ mesons are required to
have masses in the range 115--150~$\MeVcsq$. The mass resolution
is approximately 6.5~$\MeVcsq$. To increase the signal purity only
$\pizero$ mesons with energy greater than 300$~\MeV$, as defined
in the laboratory frame, are retained.

The $\Dzero$ mass resolutions are approximately 6.7, 10.7, and
5.0~$\MeVcsq$ for the $\Kpi$, $\Ktwopi$, and $\Kthreepi$ decay
modes, respectively.

\subsubsection{\bf \boldmath{$\Dstarzero \to \Dzero \pizero$} selection}

The $\Dstarzero$ mesons are reconstructed in the $\Dzero \pizero$
decay mode. The $\Dzero$ candidates are selected as described
above. The $\pizero$ candidates are required to have momenta that
satisfy the condition $70< p^{\ast} < 300$~\MeV/c and a mass in
the range 115--150~$\MeVcsq$. The mass resolution for the soft
$\pizero$ daughter is approximately 6.5~$\MeVcsq$. The resolution
of the $\Dstarzero$--$\Dzero$ mass difference is approximately
1~$\MeVcsq$.


\subsection{Selection of \boldmath{$B$} candidates} \label{sec:ExclB}

\subsubsection{\bf Event shape and angular distributions}
\label{sec:evtshapAndangl}

Both \BB\ events and $u$, $d$, $s$, and $c$ quark-antiquark events
contribute to the combinatorial background that does not peak near
the nominal $B$ mass. To reject $u$, $d$, $s$, and $c$ components
we use shape variables and angular distributions that distinguish
these from the signal \BB\ events.

Because the $u$, $d$, $s$, and $c$ continuum events are jet-like,
while $B$ meson decays produce spherical events, we can suppress
them by requiring that the ratio of the second to the zeroth
Fox-Wolfram moment~\cite{ref:FoxWolf} must be less than 0.5 as
described in Sec.~\ref{sec:Bcounting}. For each reconstructed
$\Bzerobar$ candidate we compute the thrust and sphericity axes of
both the candidate and the rest of the event, using only the
tracks and neutral clusters as defined in
Sec.~\ref{sec:Bcounting}. We define the angles $\thetathr$ and
$\thetasph$ between the axes of the $\Bzerobar$ candidate and the
rest of the event. The distributions of $|\cos{\thetathr}|$ and
$|\cos{\thetasph}|$ peak near 1.0 for $u$, $d$, $s$, and $c$
background while they are nearly flat for $B$ decays. Thus we
require at least one of the conditions $|\cos{\thetasph}| < 0.85$
or $|\cos{\thetathr}| < 0.85$ to be true for the $\Dstze \pizero$,
$\Dstze \eta$, and $\Dstze \omega$ modes. Since the two angles
$\thetathr$ and $\thetasph$ are strongly but not completely
correlated for signal events, the relative signal efficiency for
this requirement is close to 92\%. This is larger than the
relative signal efficiency of about 85\% if only the requirement
$|\cos{\thetathr}| < 0.85$ is applied, while the background
rejection is about the same.

For the $\Dstze \pizero$, $\Dstze \eta$, and $\Dstze \omega$ final
states we also take advantage of the $\sin^2{\thetabst}$
distribution of the polar angle $\thetabst$. This quantity is the
angle between the $B$ momentum vector and the beam axis in the
\FourS rest frame. Therefore we only keep the candidates that
satisfy $|\cos{\thetabst}| < 0.8$ as the distribution is almost
flat in $|\cos{\thetabst}|$ for combinatorial background.

For the $\Dstze \eta^{\prime}$ channels, we have seen that the
event yield is expected to be small. In order to keep the signal
acceptance as high as possible, we use a more complex scheme. We
require $|\cos{\thetathr}| < 0.9$ and then calculate a Fisher
discriminant ($\cal F$) that combines eleven
variables~\cite{ref:CLEOCharmless}. Two of these are the two polar
angles $\thetabst$ and $\theta_{\rm T}$, where $\theta_{\rm T}$ is
the angle between the $B$ candidate thrust axis and the beam axis
in the \FourS\ rest frame. The other nine are the scalar sums of
the energies of all charged tracks and neutral showers (except
those used  in the $B$ candidate reconstruction) binned in nine
$10^\circ$ polar angle intervals relative to the $B$ candidate
thrust axis. The separation between the means of the signal and
$q\bar{q}$ background distributions of the ${\cal F}$ variable is
1.2--1.3 times the width of either distribution.

For the $\Dzero \omega$ channel where the $\omega$ is necessarily
longitudinally polarized, we use the properties of the
distributions of two additional angles. The angle $\theta_{\rm N}$
is the angle between the normal to the plane of the three daughter
pions in the $\omega$ center-of-mass frame and the line-of-flight
of the $B$-meson in the $\omega$  rest  frame. The angle
$\thetadali$ is the angle, in the rest frame of one dipion,
between the third pion and either of the other two. The signal
events are distributed as $\cos^2{\thetan}$ and
$\sin^2{\thetadali}$, while the corresponding $\cos{\thetan}$ and
$\cos{\thetadali}$ distributions are nearly flat for combinatorial
background. We select only events in a region of the
three-dimensional parameter space of the angles $\thetabst$,
$\theta_{\rm N}$, and $\thetadali$ that has high signal
efficiency. This region is defined by
\begin{equation}
\label{equ:dalitehaB}
 \cos{\thetadali}^2 + \cos{\thetabst}^2 < 0.64,
\end{equation}
\begin{equation}
{ \left( {\cos{\thetadali} \over 0.8} \right) }^2 + {\left( {\vert
\cos{\theta_{\rm N}} \vert - 1.0 \over 0.5} \right) }^2 < 1,
\end{equation}
and
\begin{equation}
{\left( {\cos{\thetabst} \over 0.8} \right) }^2 + {\left( {\vert
\cos{\theta_{\rm N}} \vert - 1.0 \over 0.5} \right) }^2 < 1.
\end{equation}
In the $\Dstarzero \omega$ channel, the $\omega$ polarization is
not known a priori and we apply only the requirement given by
Eq.~(\ref{equ:dalitehaB}).

For the $\Dstarzero \hzero$, $\hzero=\pizero$, $\eta$, and
$\eta^{\prime}$ modes where the $\Dstarzero$ is longitudinally
polarized, we use the angular decay distribution to reject
combinatorial background. The angle $\thetahel$ is defined as the
angle between the line-of-flight of the $\Dzero$ and the one of
the $\Dstarzero$, both evaluated in the $\Dstarzero$ rest frame.
The distribution is almost flat in $\cos{\thetahel}$ for
combinatorial background, while signal events are distributed as
$\cos^2{\thetahel}$. For the $\Dstarzero \pizero$ and $\Dstarzero
\eta$ channels we require
\begin{equation}
{\left( {\cos{\thetabst} \over 0.8} \right)}^2 + {\left( {\vert
\cos{\thetahel} \vert - 1.0 \over 0.6}\right)} ^2 < 1.
\end{equation}
For the $\Dstarzero \eta^{\prime}$ final state we only require
$|\cos{\thetahel}| > 0.4$ since the angle $\thetabst$ is already
included in the definition of $\cal F$.

\subsubsection{\bf Multiple \boldmath{$B$} candidates}

After applying the above selection criteria, a small fraction of
events have more than one $B$ candidate. The average multiplicity
of $B$ candidates for the data events is between 1.01 and 1.18,
depending on the $\Dzero$ decay mode. The average multiplicity is
slightly higher for the $\Dstarzero \hzero$ modes than for the
$\Dzero \hzero$ modes. With the exception of the $\Dstze
\eta^{\prime}$ final states we select the $B$ candidate with the
lowest value of
\begin{equation*}
\chisq_B =  \left( {\frac{m_D - m^{\rm nom}_D}
             {\sigma_{m_D}}} \right)^2 +
\left( {\frac{ m_h - m^{\rm nom}_h} {\sigma_{m_h}}} \right)^2 +
\nonumber
\end{equation*}
\begin{equation}
\label{equ:chi2}
 \left(  {\frac {\Delta m_{D^{\ast}D} - \Delta m^{\rm
 nom}_{D^{\ast}D}}
 {\sigma_{\Delta m_{D^{\ast}D}}}}
\right)^2,
\end{equation}
where $\sigma_{m_D}$ and $\sigma_{m_h}$ are the resolutions of the
measured $\Dzero$ and $\hzero$ masses. The last term in the
equation is only present for $\Dstarzero$ decays and
$\sigma_{\Delta m_{D^{\ast}D}}$ is the average resolution of the
measured $\Dstarzero$--$\Dzero$ mass difference. The mass
resolutions depend on the decay modes and are slightly different
for data and \mc\ simulation. Each of the three terms is found to
be approximately Gaussian with mean value near zero and standard
deviation near one.

In order to reduce combinatorial backgrounds, we require that each
of the terms in Eq.~(\ref{equ:chi2}) is less than $2.5^2$ (a $\pm
\ 2.5\, \sigma$ requirement). In the case of the $\omega$ mesons,
the candidates must have a reconstructed invariant mass within
$25$~\MeVcsq\ ($\pm \ 2.5$ times the $\omega$ natural width) of
the nominal value.

For the  $\Dstze \eta^{\prime}$ channels, the signal acceptance is
relatively lower than for other modes, but the background level is
also much smaller. Therefore we keep all the candidates in the
events and weight them by $1/N$ where $N$ is the number of $B$
candidates in the event. In order to reduce the combinatorial
background for these two channels the invariant mass of the
$\eta^{\prime}$ candidate is required to be within $2.5\, \sigma$
of its nominal value. The $\Dzero$ candidates are required to have
a reconstructed mass within 2--3~$\sigma$ (depending on the decay
mode) of their nominal value. We reject $\Dstarzero$ candidates
whose $\Dstarzero$--$\Dzero$ mass difference is not within $3\,
\sigma$ of its nominal value.

\begin{table*}
\caption{\label{tab:yield} The number of candidates (${\cal
N}_{\rm cand}$), the number of non-peaking (${\cal N}_{\rm npb}$)
and peaking (${\cal N}_{\rm pb}$) background events, the number of
cross-feed ($\Ncf$) background events from other color-suppressed
modes, the number of signal events ($S$) after peaking and
cross-feed backgrounds are subtracted, and the statistical
significance of the signals ($S / \sqrt{S+{\cal N}_{\rm bckgd}}$).
We obtain ${\cal N}_{\rm npb}$ from a fit to the data \mes\
distribution, while ${\cal N}_{\rm pb}$ is estimated from the \mc\
simulation. The statistical uncertainty on $S$ includes the
uncertainty on ${\cal N}_{cand}$ as obtained from the ML \mes\
fit. The statistical uncertainty on ${\cal N}_{\rm pb}$ and the
estimated uncertainties for $\Ncf$ are accounted for in the
systematic uncertainties of the branching fractions. For the
$\Dstze \eta^{\prime}$ modes, the number of candidates is small;
therefore Poisson statistics rather than Gaussian statistics are
used. The statistical significance is defined as $\sqrt{2 \ln
({\cal L}_{max}/ {\cal L}(0))}$, where ${\cal L}_{max}$ is the
likelihood at the nominal signal yield and ${\cal L}(0)$ is the
likelihood with the signal yield set to 0. In the table, the
symbol ``-'' means that the corresponding number can be
neglected.}
\begin{center}
\begin{tabular}{lcrclcrclcrclcrclcrclcc}
\hline \hline \\ $\Bzerobar$ mode  &{ \ }&
\multicolumn{3}{c}{\Ncand} &{ \ }& \multicolumn{3}{c}{$\NPback$}
&{ \ }& \multicolumn{3}{c}{$\Pback$} &{ \ }&
\multicolumn{3}{c}{$\Ncf$} & { \ } & \multicolumn{3}{c}{$S$} &{ \
} & statistical \\
(decay channel) && && && && && && && && && && && significance \\  \hline \\
$\Dzero \pizero$              && 556&$\pm$&34 && 603&$\pm$&22 && 51&$\pm$&9 && 18&$\pm$&4 && 487&$\pm$&34 && $14.3$      \ \\
$\Dstarzero \pizero$          && 102&$\pm$&12 &&  32&$\pm$&6  && 11&$\pm$&5 &&  2&$\pm$&1 &&  88&$\pm$&12 && $7.6$       \ \\
$\Dzero \eta(\ra\gamgam)$     && 200&$\pm$&20 && 181&$\pm$&12 && 17&$\pm$&3 && 10&$\pm$&2 && 173&$\pm$&20 && $8.9$       \ \\
$\Dzero \eta(\ra\threepi)$    &&  76&$\pm$&12 &&  69&$\pm$&7  &&    &-&     &&  2&$\pm$&1 &&  74&$\pm$&12 && 6.2       \ \\
$\Dstarzero \eta(\ra\gamgam)$ &&  43&$\pm$&7  &&   8&$\pm$&2  &&    &-&     &&  4&$\pm$&1 &&  40&$\pm$&7  && 5.5       \ \\
$\Dzero \omega$               && 207&$\pm$&18 && 136&$\pm$&10 &&  4&$\pm$&3 &&  5&$\pm$&1 && 198&$\pm$&18 && $10.7$      \ \\
$\Dstarzero \omega$           &&  75&$\pm$&12 &&  58&$\pm$&7  &&    &-&     &&  5&$\pm$&1 &&  70&$\pm$&12 && 6.1       \ \\
$\Dzero \eta^{\prime}$        &&  27&$\pm$&6  &&  10&$\pm$&1  &&    &-&     &&     &-&    &&  27&$\pm$&6  && 6.3 \ \\
$\Dstarzero \eta^{\prime}$    &&   4&$\pm$&2  &&  &-& && &-& &&
&-& &&  4&$\pm$&2 && 3.0 \
\\  \hline\hline
\end{tabular}
\end{center}
\end{table*}

\subsubsection{\bf \boldmath{$B$} candidates and background yields}
\label{sec:yieldextract}

Two kinematic variables are used to isolate the $B$-meson signal
for all modes. One is $\mes$, the beam-energy-substituted mass.
The other is $\De$, the difference between the reconstructed
energy of the $B$ candidate and the beam energy in the $\eplemi$
center-of-mass frame. Both quantities use the strong constraint
given by the precisely known beam energy (the beam energy is known
to within a fraction of an~\MeV). The beam-energy-substituted mass
is defined as
\begin{equation}
\label{equ:mES} \mes = \sqrt{{\left( {\frac {{s / 2}
+\vec{p}_0.\vec{p}_B} {E_0}}  \right)^2}-\vert \vec{p}_B \vert
^2},
\end{equation}
and the energy difference is
\begin{equation}
\label{equ:deltaE} \De = E_D^{\ast} + E_h^{\ast} - \sqrt{s}/2.
\end{equation}
Where $\sqrt{s}$ is the $\eplemi$ center-of-mass energy. The small
variations of the beam energy over the duration of the run are
taken into account when calculating \mes. For the momentum
$\vec{p}_i$ ($i=0,B$) and the energy $E_0$, the subscripts $0$ and
$B$ refer to the $\eplemi$ system and the reconstructed $B$ meson,
respectively. The energies $E_D^{\ast}$ and $E_h^{\ast}$ are
calculated from the measured $\Dstze$ and $\hzero$ momenta. Signal
events have $\mes \simeq m_{B^0}$ and $\De \simeq 0$, within their
respective resolutions.

We limit the selection of the $\Bzerobar$ candidates to the
``signal neighborhood'', defined by $\mid $\De$ \mid <350\,\MeV$
and $5.2<\mes<5.3$ \GeVcsq. The $\mes$ resolution is dominated by
the beam energy spread and is approximately $3\,\MeVcsq$,
depending slightly on the $B$ decay mode. The $\De$ resolution for
the $\Dstze \pizero$ and $\Dstze \eta(\ra \gamgam)$ modes is
dominated by the angular and energy resolution of the EMC. The
$\De$ resolution is approximately 37--44~\MeV\ for the $\Dstze
\pizero$ modes and 28--35~\MeV\ for the $\Dstze \eta(\ra \gamgam)$
modes, depending on the $\Dstarzero$ and $\Dzero$ decay mode. The
$\De$ resolution is better for the $\Dzero \eta(\ra \threepi$),
$\Dstze \omega$, and $\Dstze \eta^{\prime}$ modes because the
angular and the momentum resolution for charged tracks is better
than for photons. For these modes it is approximately 15--20~\MeV.

We define the signal region using the resolutions in $\mes$ and
$\De$ obtained from the \mc. The limits of the signal region are
$5.270 < \mes < 5.290\, \GeVcsq$ (about $\pm \ 3 \ \sigma$ around
the $B$ mass) and $\vert \De \vert <  3 \ \sigma$. In the case of
$\Bzerobar \ra \Dstze \pizero$ decay modes, we reduce the
contribution from the color-allowed $B^- \ra \Dstze \rho^-$
background by requiring $\De$ to be in the region from $-90$ to
$100\, \MeV$. We change these requirements slightly for the
$\Dstze \eta^{\prime}$ channels where we want to optimize the
statistical significance. Here the signal region is defined by
$\vert \De \vert<$~2--3~$\sigma$ depending on the $\Dzero$ decay
mode and $5.273 < \mes < 5.286\, \GeVcsq$. The number of signal
candidates is computed in the signal region for each $\Bzerobar$
decay mode and the signal \mc\ simulation is used to determine the
acceptance.

We perform an unbinned maximum likelihood (ML) fit to the $\mes$
distribution to extract the number of signal candidates (\Ncand).
A fit to the $\mes$ distribution allows us to model the signal and
background shapes with a well known, simple, and universal
function, independent of the $B$ decay mode analysed.

In the fit the signal component is modeled by a Gaussian
distribution whose $\sigma$ is constrained to the value obtained
from the signal \mc\ separately for each $\Bzerobar$ decay mode.
The value of \Ncand\ is computed from the fit within the $\mes$
signal region defined earlier. The background component is modeled
by an empirical phase-space distribution~\cite{ref:Argus}
(henceforth referred to as the ARGUS distribution):
\begin{eqnarray}
\label{eq:ARGUS}
A(\mes;m_0,\xi,\alpha)& = &\alpha \ \mes\sqrt{1-(\mes/m_0)^2} \times \nonumber \\
             &         & \exp(\xi (1-(\mes/m_0)^2)),
\end{eqnarray}
where $m_0$ is set to a typical beam energy (5.29\, \GeV),
$\alpha$ is the fitted normalization parameter, and $\xi$ is the
fitted parameter describing the shape of the function.

The ML fit is performed within the limits of the signal region in
$\De$, as defined above, and for $\mes$ between $5.2$ and $5.3\,
\GeVcsq$. For the $\Dstze \eta^{\prime}$ modes, in addition to
using the $\mes$ resolution obtained from the \mc\ simulation, the
mean of the Gaussian distribution is also constrained in the ML
fit to the nominal $B$ mass. The value of the $\xi$ parameter in
the ARGUS function is fixed to the value obtained from a ML fit to
the $\mes$ data in the $\De$ sideband: $200 < \vert \De \vert <
350\, \MeV$ and $5.2 < \mes < 5.3\, \GeVcsq$.

The ARGUS function accounts for random combinatorial background
originating from $u$, $d$, $s$, and $c$ continuum events,
${\tau}^+ {\tau}^-$ events, two-photon processes, and $B \Bbar$
events but not for ``peaking background'' from $\Bzero\Bzerobar$
and $\Bpl\Bmi$ decays, which  have distributions that peak in the
same location as signal events do. The number of
non-peaking-background events (${\cal N}_{\rm npb}$) is determined
from the fit to the data in the full $5.2 < \mes < 5.3\, \GeVcsq$
interval and the $\De$ signal region by integrating the ARGUS
function over the much smaller signal region.

The number of peaking-background events (${\cal N}_{\rm pb}$) is
small relative to the non-peaking background but it is dangerous
because the peaking-background events lie in the signal region.
Peaking background comes also from color-suppressed decays in
$\Bzero\Bzerobar$ events that are incorrectly reconstructed. This
small contribution (${\cal N}_ {CF}$) is evaluated separately and
thus does not contribute to the value of ${\cal N}_{\rm pb}$, as
discussed in Sec.~\ref{sec:Backgrounds}. Altogether we write the
total number of background events (${\cal N}_{\rm bkgd}$) in the
signal region as
\begin{equation}
{\cal N}_{\rm bkgd} ={\cal N}_{\rm npb}+{\cal N}_{\rm pb} + {\cal
N}_ {CF}.
\end{equation}
Finally, the number of signal events is calculated as
\begin{equation}
S={\cal N}_{\rm cand} - {\cal N}_{\rm pb} - {\cal N}_ {CF}.
\end{equation}
The values of \Ncand, ${\cal N}_{\rm npb}$, ${\cal N}_{\rm pb}$,
${\cal N}_ {CF}$, $S$, and the statistical significance of the
signals for the $\Bzerobar$ decay channels studied in this paper
are listed in Table~\ref{tab:yield}.


\section{Background Estimation}
\label{sec:Backgrounds}

\subsection{Peaking backgrounds from \boldmath{$B \Bbar$} decays
other than color-suppressed modes} \label{sec:pkBckgd}

To investigate backgrounds that peak at the $B$ mass in the $\mes$
distribution, we use two types of \mc\ samples: a sample that
contains only $\Bmi \ra \Dstze \rhomi$ and a sample that contains
all other charged and neutral $B$-meson decays, except the
color-suppressed $\Bzerobar$ decay modes reported in this paper.
In the next section we describe how we estimate the cross-feed
from the color-suppressed $\Bzerobar$ decay modes.

The peaking background is estimated with a ML fit to the \mc\
samples, using a Gaussian distribution for signal and an ARGUS
background distribution, just as for the data (see
Sec.~\ref{sec:yieldextract}). We constrain the ARGUS shape
parameter $\xi$ to be the same as the one obtained for the
corresponding data \mes\ distribution. The normalization of the
ARGUS function is a free parameter as are all parameters of the
Gaussian. The values of the parameters of the Gaussian
distribution for the peaking-background events are expected to be
different than that for signal events. The mean value of the
Gaussian distribution is possibly different from the $B$ mass and
the resolution is expected to be larger than the nominal value for
signal events, which is about $3\,\MeVcsq$.

The peaking background is taken to be the area under the Gaussian
distribution in the signal region $5.270 < \mes < 5.290\, \GeVcsq$
($5.273 < \mes < 5.286\, \GeVcsq$ for $\Dstze \eta^{\prime}$
channels), normalized to the luminosity of the data.
Table~\ref{tab:yield} gives the estimate of the number of
peaking-background events to be subtracted from the fitted
candidate event yields in the data for each of the various
channels. For each channel, the number is the sum of the various
contributions estimated from the $B \Bbar$ background \mc\
samples. As this number is extracted from \mc\ simulations, we use
the statistical uncertainty associated with this quantity as a
systematic uncertainty for the branching fraction measurements.

The systematic uncertainty due to the constraint applied to the
ARGUS parameter $\xi$, which is fixed to the data value in the ML
fit to the various \mc\ \mes\ distributions used for the
peaking-background computation, is small or negligible. This
systematic uncertainty is estimated by recalculating the peaking
background when using two other fixed values for $\xi$. These two
values are computed from ML fits to two \mes\ distributions
obtained with the \mc\ simulation. One distribution corresponds to
the sum of all the normalized contributions from the various
background sources (peaking or non-peaking) only. The second one
also includes the expected contribution from the signal events. It
is found that the values of $\xi$ for the two types of \mc\ \mes\
distributions are very close (within the statistical
uncertainties) to the corresponding data value.

\subsection{Peaking backgrounds from other color-suppressed modes}
\label{sec:cfcs}

Signal event yields must be corrected for cross-feed between
color-suppressed modes. Cross-feed occurs when a true decay chain
of type $k$ is erroneously reconstructed as a candidate decay
chain of type $j$.  This will bias the signal yield for events of
type $j$ if such events of type $k$ enter the signal region.
Cross-feed to each signal from $\Bzerobar \ra \Dstze \hzero$
decays is investigated using signal \mc\ samples for these decay
modes. In the end, we find that the contribution of cross-feed is
for the most part less than half the statistical uncertainty in
the signal.

For each light neutral hadron type, $\hzero$, the dominant
contribution to $\Bzerobar \ra \Dzero (\Dstarzero) \hzero$ arises
from the associated $\Bzerobar \ra \Dstarzero (\Dzero) \hzero$
mode. In the case of the $\Dstarzero \hzero$ decay modes, since we
only consider the $\Dstarzero \to \Dzero \pizero$ channel, the
contribution from the final state $\Dstarzero \to \Dzero \gamma$
is non negligible. These cross-feed contributions peak at the same
$\mes$ as the signal, but are shifted in $\De$.

The number $N_{k\ra j}$ of events of type $k$ entering the signal
region for type $j$ is given by
\begin{equation}
    N_{k\ra j} = N(\BBbar) {\cal B}_k {\cal A}_{k\ra j},
\end{equation}
where $N(\BBbar)$ is the number of $\BBbar$ pairs and ${\cal B}_k$
is the branching fraction of the decay chain $k$ including the
$\Bzerobar$ branching fraction. ${\cal A}_{k\ra j}$ denotes the
probability for an event of type $k$ to enter the signal region
for decay mode $j$. The probability ${\cal A}_{k\ra j}$ is
estimated from the \mc\ simulation as
\begin{equation}
    {\cal A}_{k\ra j} = \frac{{S}_{{\rm MC },k\ra j}}
                                 {N_{{\rm gen},k}} \mbox{.}
\end{equation}
Here, ${S}_{{\rm MC},k\ra j}$ is the number of events of type $k$
entering the signal region for decay mode $j$ and $N_{{\rm
gen},k}$ is the number of generated \mc\ events. It is convenient
to introduce the fractional cross-feed quantity
\begin{equation}
\label{ref:ratio_N}
    {\cal R}_{k,j} = \frac{N_{k\ra j}}
                        {N_{j\ra j}} =
           \frac{ {\cal B}_k {\cal A}_{k\ra j} }
                { {\cal B}_j {\cal A}_{j\ra j}}.
\end{equation}
For a given candidate event of type $j$, the probability that it
is generated by one of the possible cross-feed contributions can
be expressed by the fraction ${\cal F}_{\rm CF}(j)$ given by
\begin{equation}
    {\cal F}_{\rm CF}(j) = {{\sum_{k \neq j } N_{k \ra j}} \over
    {N_{j\ra j} + \sum_{l \neq j } N_{l \ra j}}} {\rm ,}
\end{equation}
or, using Eq.~\ref{ref:ratio_N}, by
\begin{equation}
    {\cal F}_{\rm CF}(j) = {{\sum_{k \neq j} {\cal R}_{k,j}}
        \over {1 + \sum_{l \neq j} {\cal R}_{l,j}}} {\rm .}
\end{equation}
In what follows, ${\cal F}_{\rm CF}(j)$ is simply written as
${\cal F}_{\rm CF}$ for each color-suppressed decay mode $j$.

In order to calculate $N_{k \ra j}$, we must know the branching
fractions of the investigated decay modes. We use recently
measured values for the branching fractions of the $\hzero$,
$\Dzero$, and $\Dstarzero$ decays chains~\cite{ref:PDG}. We
consider $63$ color-suppressed $\Bzerobar \ra \Dstze \hzero$ decay
chains. The light neutral hadron $\hzero$ is a $\pizero(\ra
\gamgam)$, an $\eta(\ra \gamgam$ or $\threepi)$, an $\omega(\ra
\threepi)$, a $\rho^0(\ra \pi^+ \pi^-)$, or an $\eta^{\prime}(\ra
\ppeta (\ra \gamgam)$ or $\rho^0 \gamma)$ meson. The $\Dzero$
mesons are reconstructed in the modes $\Kpi$, $\Ktwopi$, and
$\Kthreepi$, and the $\Dstarzero$ mesons in the channels $\Dzero
\pizero$ and $\Dzero \gamma$. For the $\Bzerobar \ra \Dstze
\hzero$ branching fractions we use the values measured in this
analysis (summarized in Table~\ref{tab:BF}). These final branching
fractions are determined after several iterations because the
cross-feed estimate depends upon the branching fractions being
measured. Therefore, we iterate the calculation of the background
from cross-feed until the values of the computed branching
fractions do not change by more than $10^{-6}$. For the
contributions from $\Dzero \rho^0$ and $\Dstarzero \rho^0$
channels we use the results obtained recently by
Belle~\cite{ref:Belle2}: $(2.9 \pm 1.0 (\stat) \pm 0.4 (\syst) )
\times 10^{-4}$ and the upper limit $5.1\times 10^{-4}$,
respectively. In the latter case the assumption of such a large
value for the branching fraction is likely to be an overestimate;
yet the $\Dstze \rho^0$ decays do not generate any significant
cross-feed contributions to any of the modes studied in this
paper.

Table~\ref{tab:CFres} shows the total contributions from
cross-feed to each mode reported in this study. The dominant
sources are also shown in decreasing order of importance. The
number of cross-feed events, $\Ncf$, is calculated as the
difference between the number of candidates in the data and the
number of other peaking-background events estimated from the \mc\
simulation, which includes no signal, multiplied by the fractional
cross-feed:
\begin{equation}
\Ncf = ({\cal N}_{\rm cand} - {\cal N}_{\rm pb}) \times {\cal
F}_{\rm CF} {\rm .}
\end{equation}
The corresponding number of cross-feed events is listed in
Table~\ref{tab:yield} for each mode.

The cross-feed contributions for the $\Bzerobar \ra \Dstze
\eta^{\prime}$ analyses are found to be negligible. This is due to
both the good mass resolution of the mode $\eta^{\prime}\ra \ppeta
(\ra \gamgam)$ and to the complexity of the signature used to
reconstruct these signals.

\begin{figure*}
\begin{tabular}{lr} \hspace{-1cm}
\epsfig{file=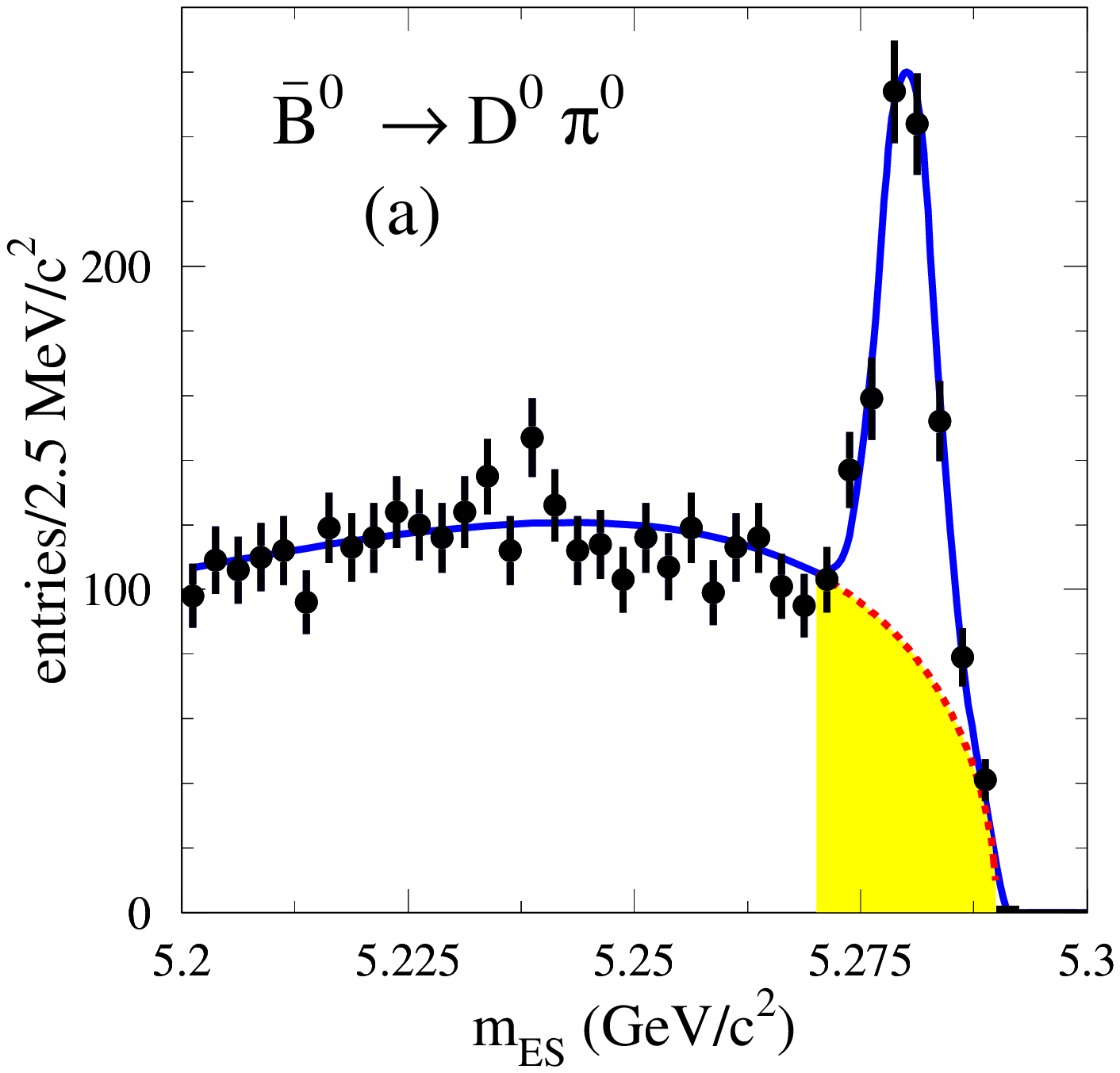,width=0.35\textwidth}
&\hspace{-0.25cm}
\epsfig{file=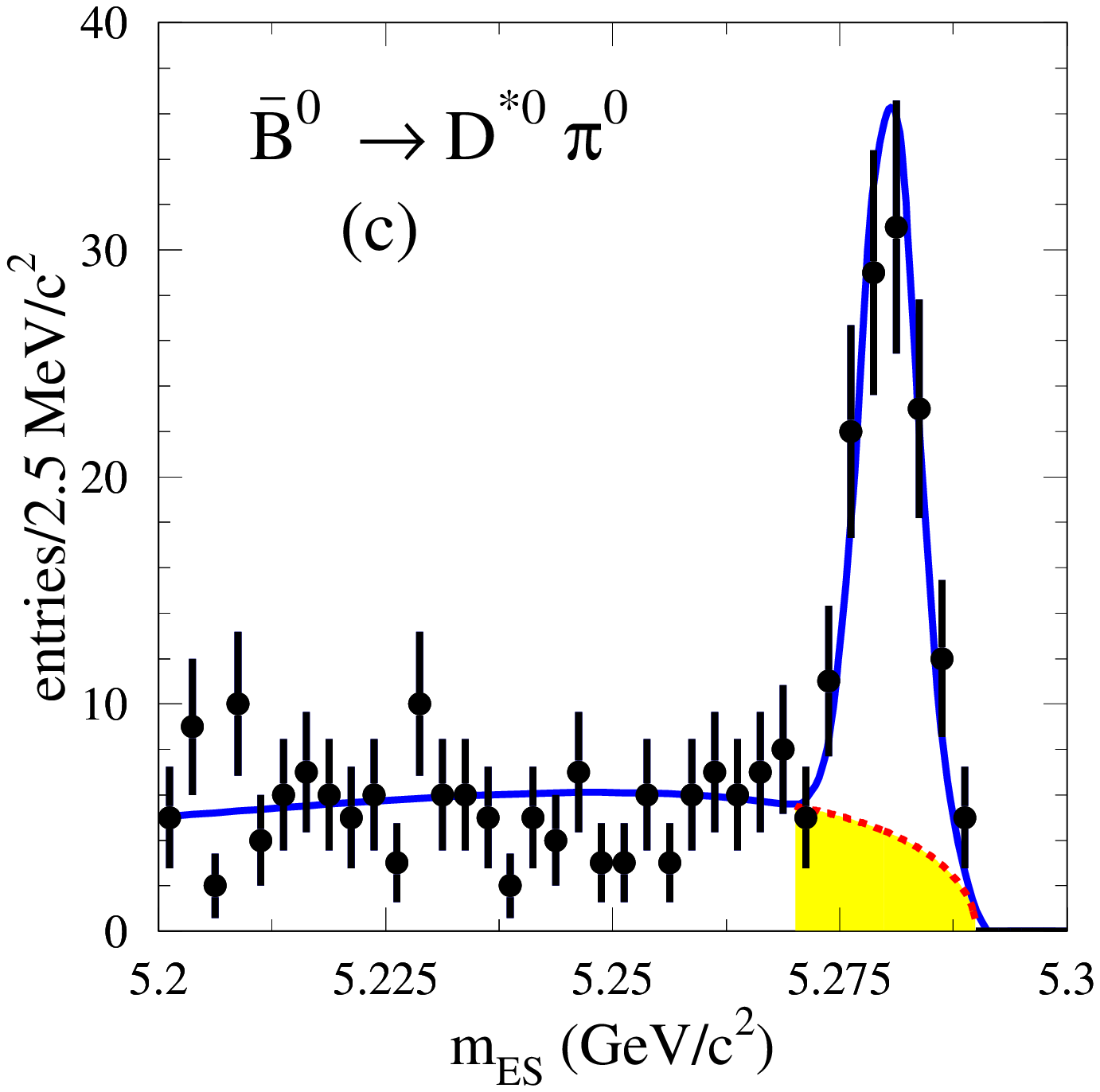,width=0.35\textwidth}
\\ \hspace{-1cm}
\epsfig{file=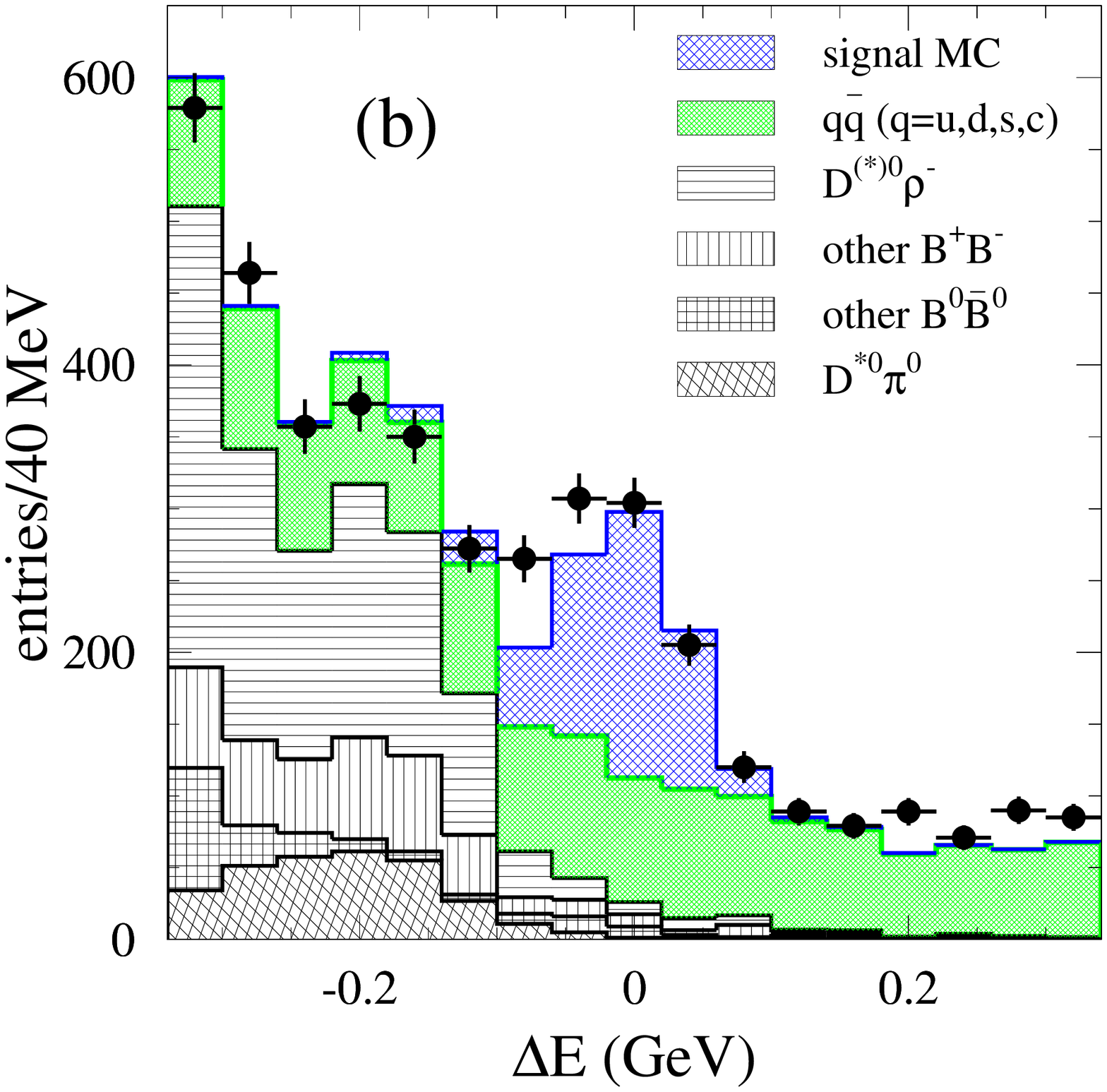,width=0.35\textwidth}
&\hspace{-0.25cm}
\epsfig{file=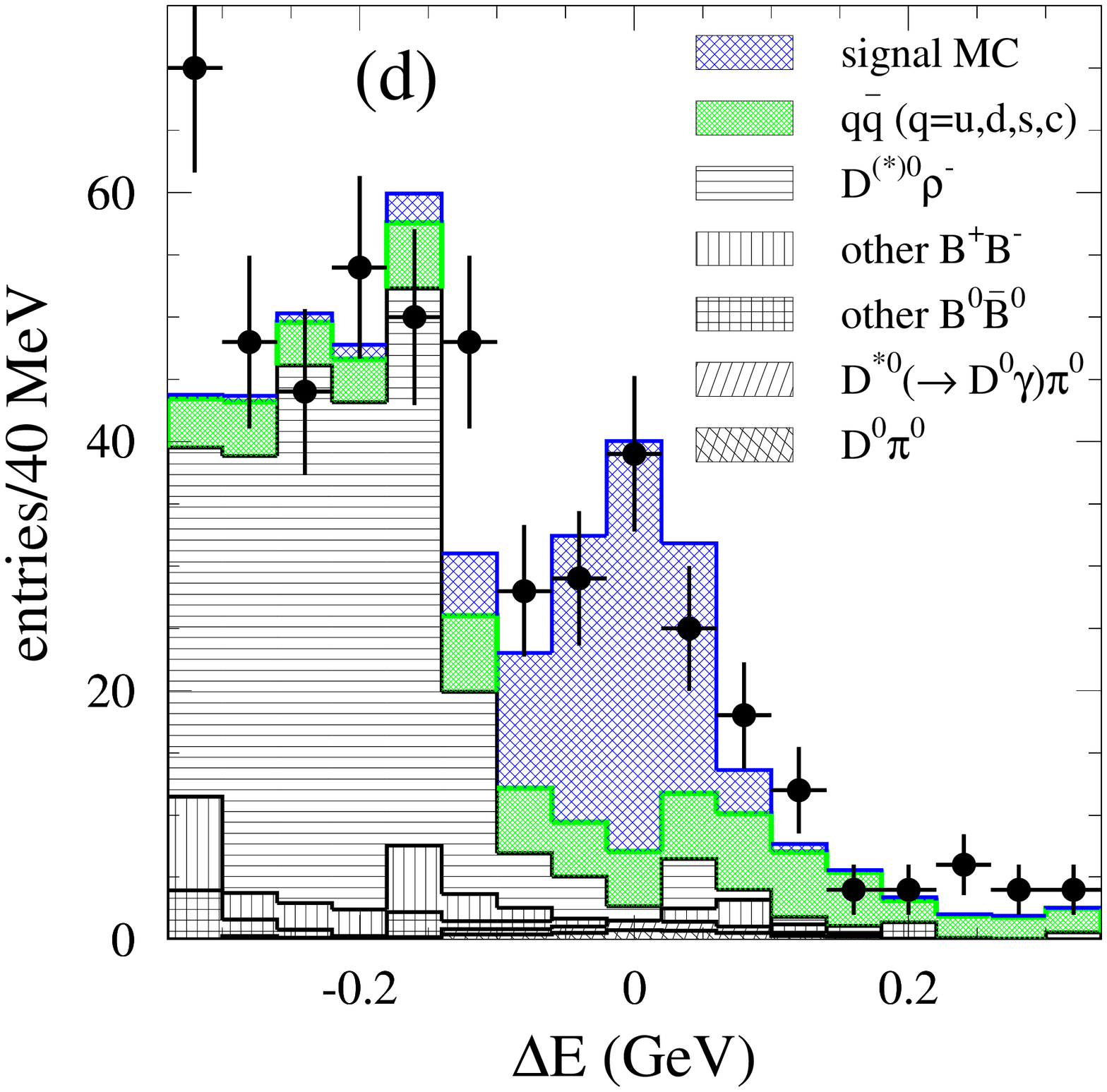,width=0.35\textwidth}
\end{tabular}
\begin{center}
\caption[D0stpi0]{Distributions of \mes\ and of $\De$\ for (a, b)
candidate $\Bzerobar \ra \Dzero \pizero$ events and (c, d)
candidate $\Bzerobar \ra \Dstarzero \pizero$ events. The dots with
error bars correspond to data. In the \mes\ distribution, the
ARGUS and Gaussian ML fits are superimposed. The number of signal
candidates (${\cal N}_{cand}$), which includes peaking-background
and cross-feed contributions, is the area of the Gaussian function
in the signal region $5.270 < \mes < 5.290\, \GeVcsq$. The
non-peaking background ($\NPback$) is represented by the shaded
region. The hatched histograms in the $\De$ distributions
represent the simulated events, and are shown separately for
signal and the various backgrounds from \BB\ and $\qqbar$
$(q=u,d,s,c)$ events. \label{fig:d0_dstar0_pi0}}
\end{center}
\end{figure*}

\begin{table}
\caption{\label{tab:CFres} Total fractional cross-feed $({\cal
F})$ expressed in percent (see text for definition) observed in
the \mc\ simulation. The dominant sources that contribute are
shown in decreasing order of importance.}
\begin{center}
\begin{tabular}{lcccl}
\hline \hline
\\
$\Bzerobar$ mode & & ${\cal F}_{\rm CF}(\%)$ & & Dominant sources \\
\hline \\
$\Dzero \pizero$                           && 3.6 && $\Dstarzero \pizero$                                                 \ \\
$\Dstarzero(\Dzero \pizero) \pizero$       && 2.6 && $\Dstarzero(\Dzero \gamma) \pizero$, $\Dzero \pizero$             \ \\
$\Dzero\eta(\gamgam)$                      && 5.4 && $\Dstarzero \eta$, $\Dzero \pizero$, $\Dstarzero \pizero$        \ \\
$\Dzero \eta(\threepi)$                    && 2.2 && $\Dstarzero \eta$, $\Dzero \pizero$, $\Dzero \omega$             \ \\
$\Dstarzero(\Dzero \pizero) \eta(\gamgam)$ && 8.8 && $\Dstarzero(\Dzero \gamma) \eta$, $\Dstarzero \pizero$, $\Dzero\eta$ \ \\
$\Dzero \omega$                            && 2.5 &&   $\Dstarzero \omega$                                                \ \\
$\Dstarzero(\Dzero \pizero)\omega$  && 6.5 && $\Dstarzero(\Dzero \gamma) \omega$, $\Dzero \omega$,                        \ \\
&&&& and  $\Dzero\eta(\threepi)$ \ \\  \hline \hline
\end{tabular}
\end{center}
\end{table}


\begin{figure*}
\begin{tabular}{lcr} \hspace{-1cm}
\epsfig{file=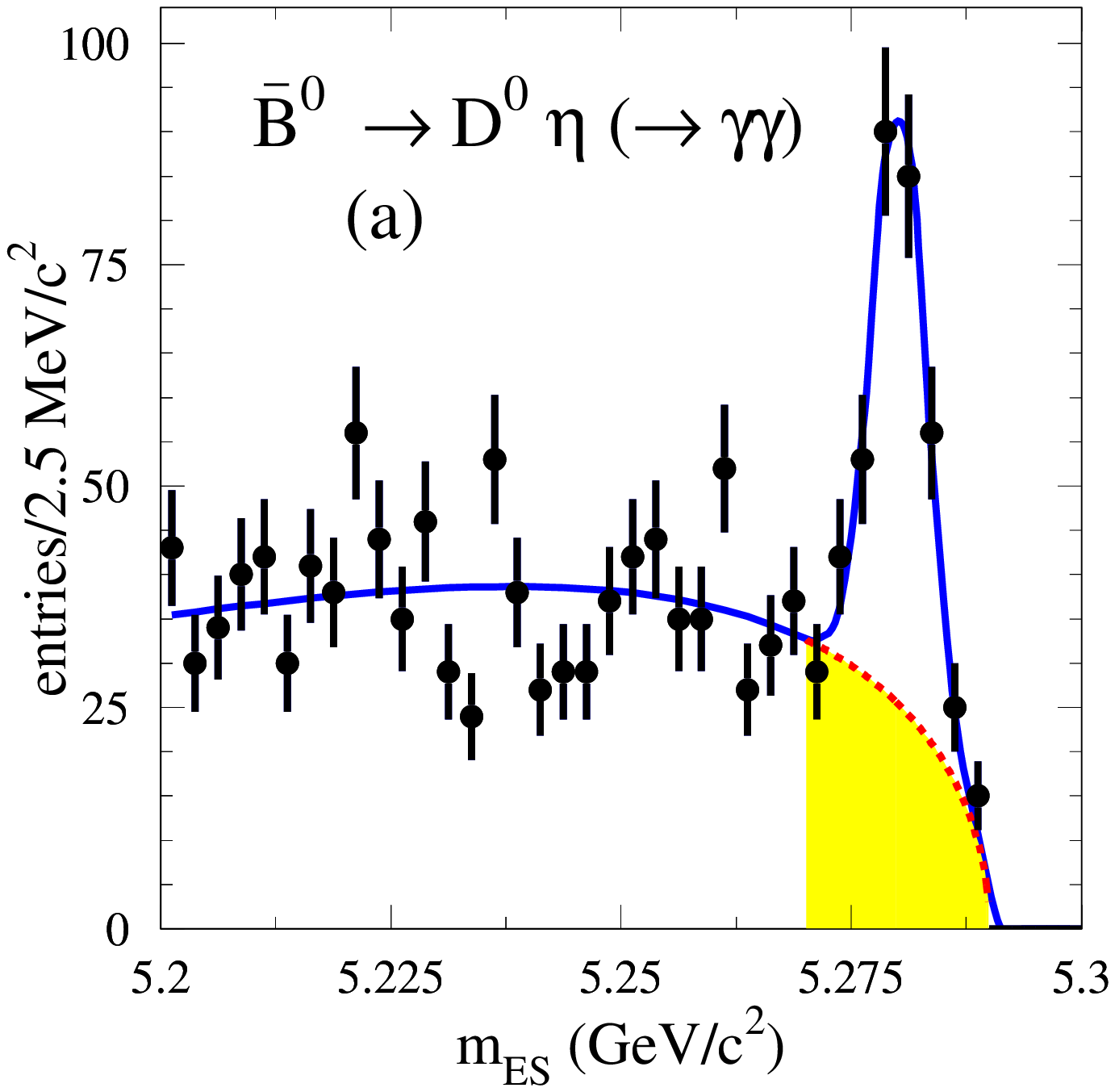,width=0.35\textwidth}
&\hspace{-0.25cm}
\epsfig{file=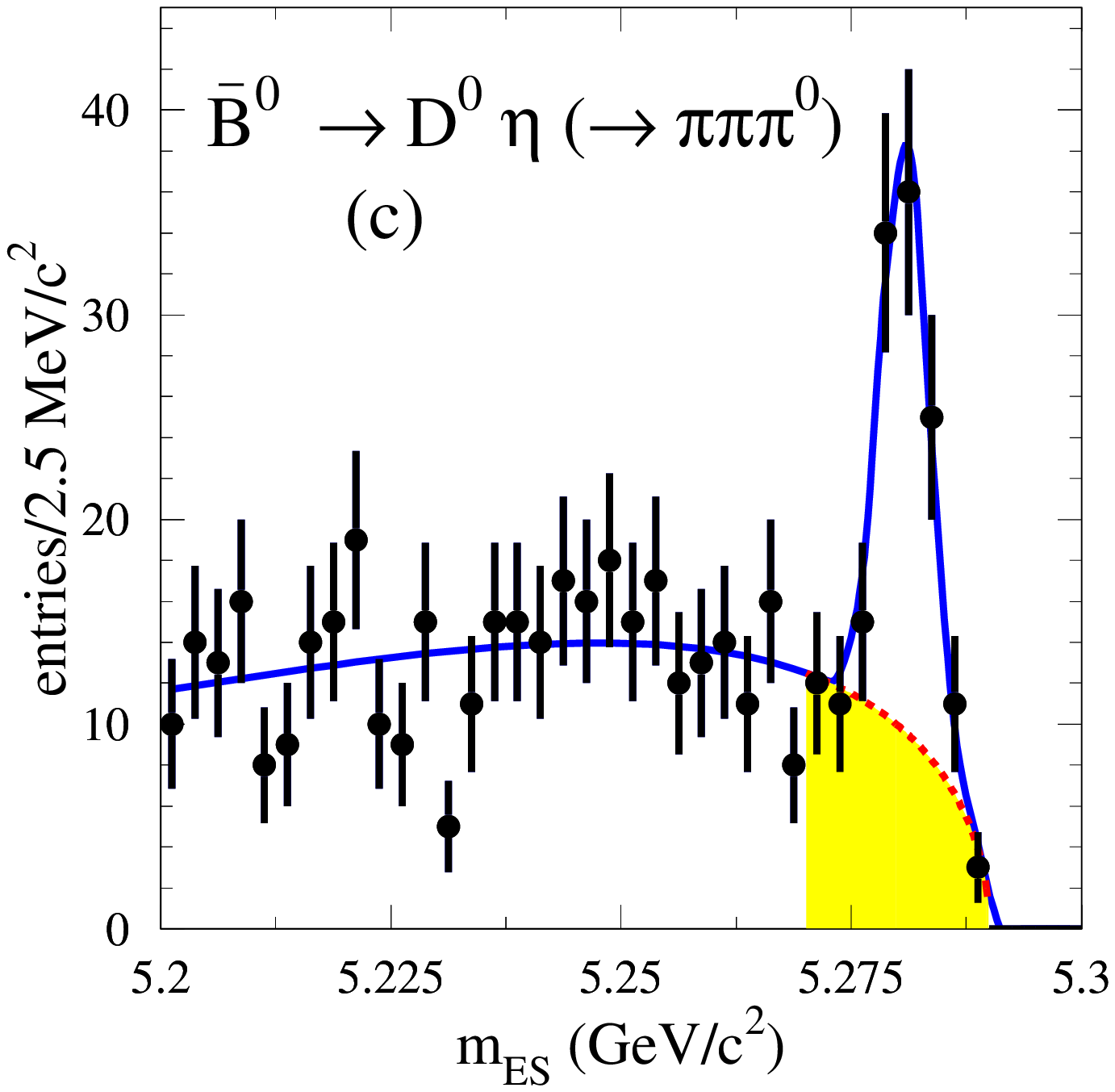,width=0.35\textwidth}&\hspace{-0.25cm}
\epsfig{file=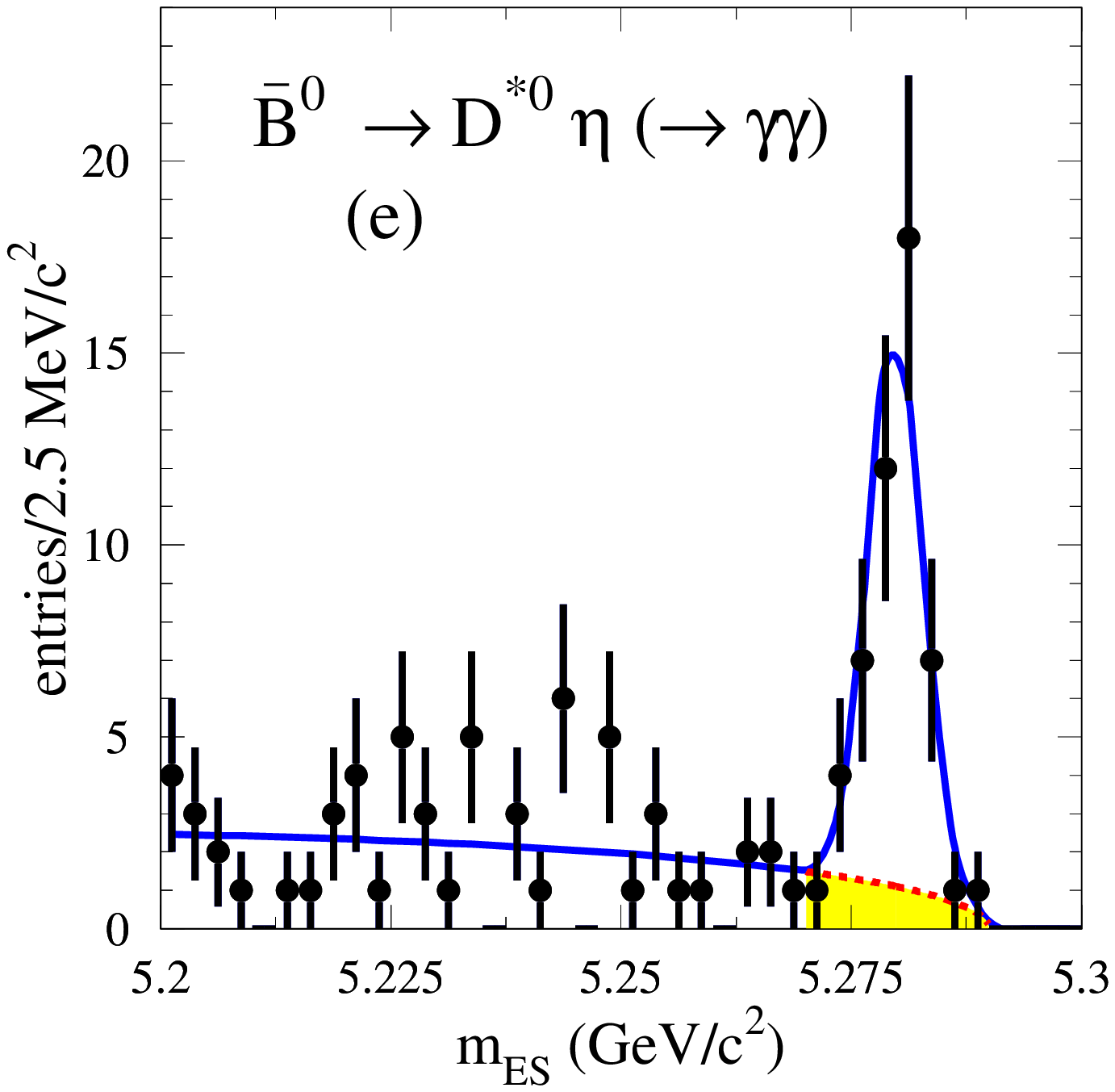,width=0.35\textwidth}
\\ \hspace{-1cm}
\epsfig{file=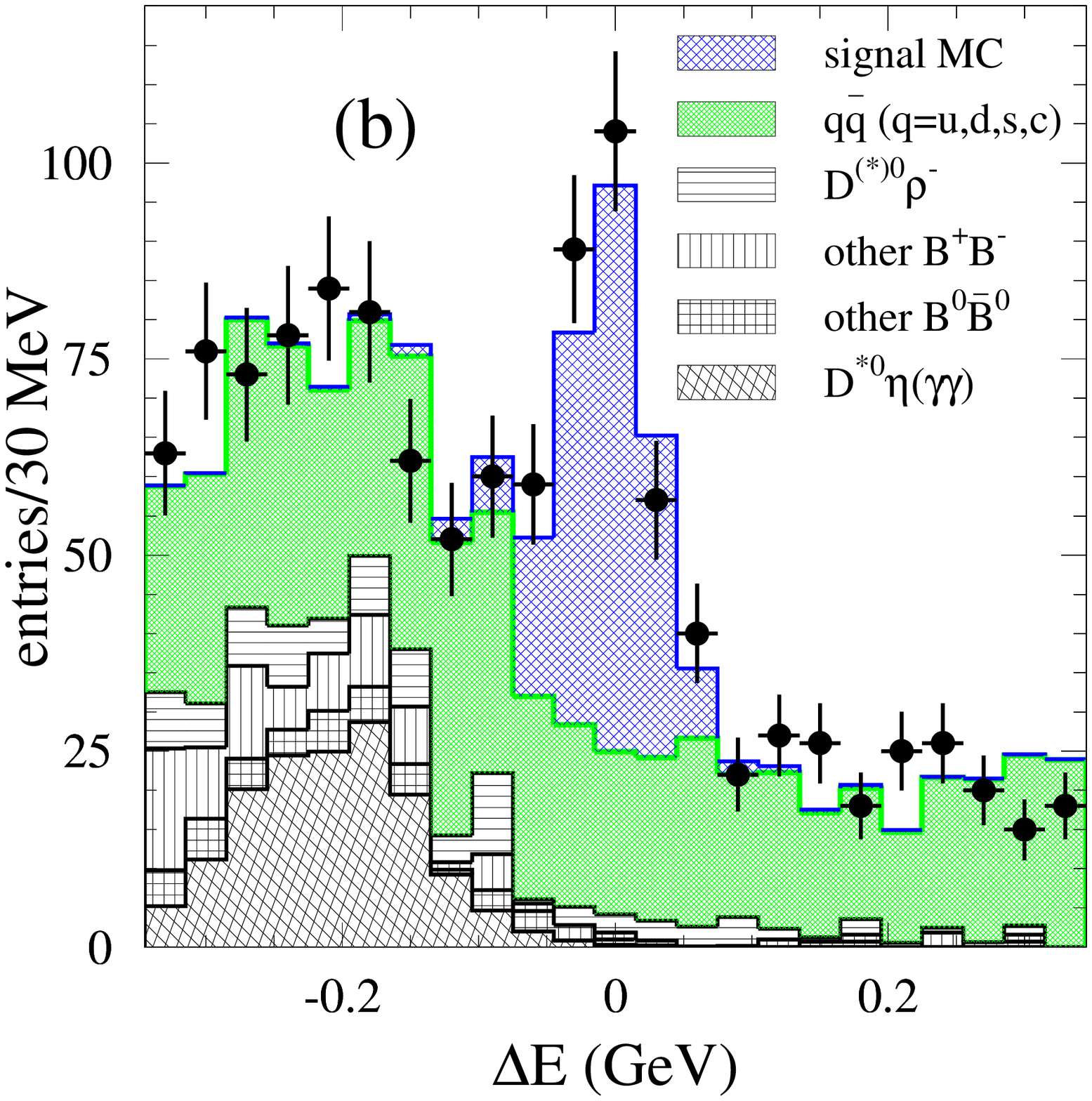,width=0.35\textwidth}
&\hspace{-0.25cm}
\epsfig{file=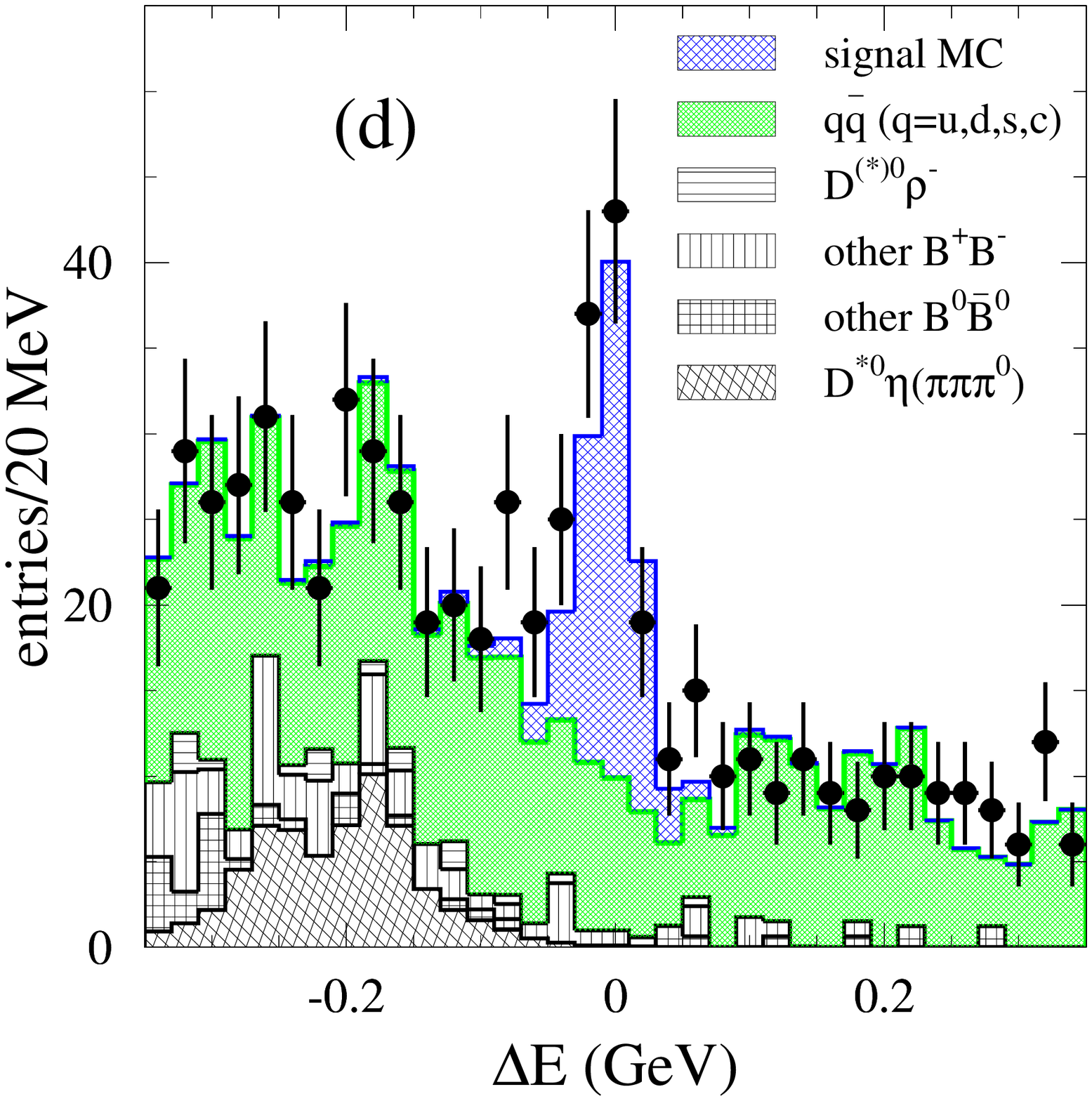,width=0.35\textwidth}&\hspace{-0.25cm}
\epsfig{file=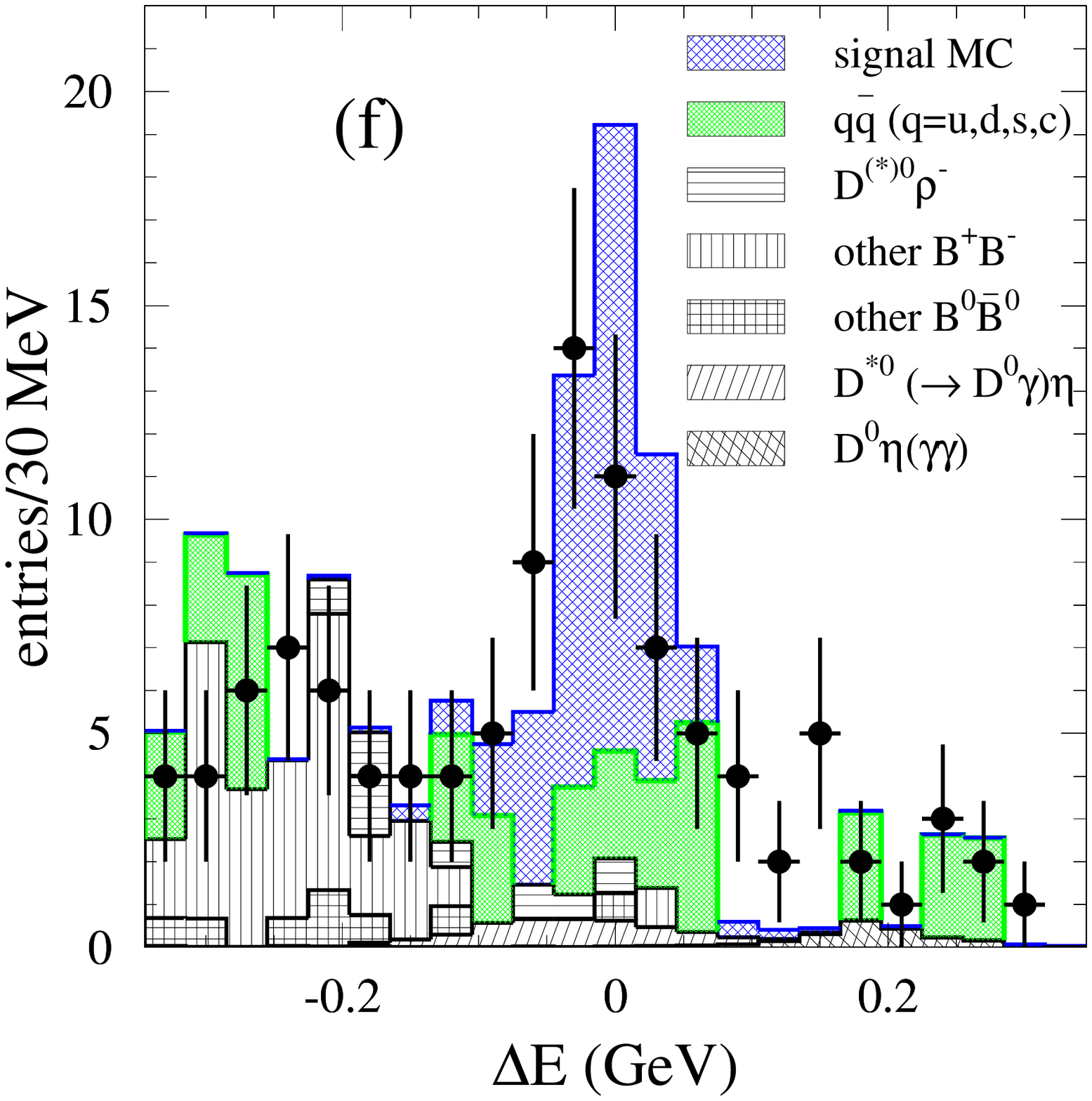,width=0.35\textwidth}
\end{tabular}
\begin{center}
\caption[D0steta]{Distributions of \mes\ and of $\De$\ for (a, b)
candidate $\Bzerobar \ra \Dzero \eta$ ($\eta \ra \gamgam$) events,
(c, d) candidate $\Bzerobar \ra \Dzero \eta$ ($\eta \ra \threepi$)
events, and (e, f) candidate $\Bzerobar \ra \Dstarzero \eta$
($\eta \ra \gamgam$) events. The various contributions are shown
as in Fig.~\ref{fig:d0_dstar0_pi0}.\label{fig:d0_dstar0_eta}}
\end{center}
\end{figure*}

\begin{figure*}
\begin{tabular}{lr} \hspace{-1cm}
\epsfig{file=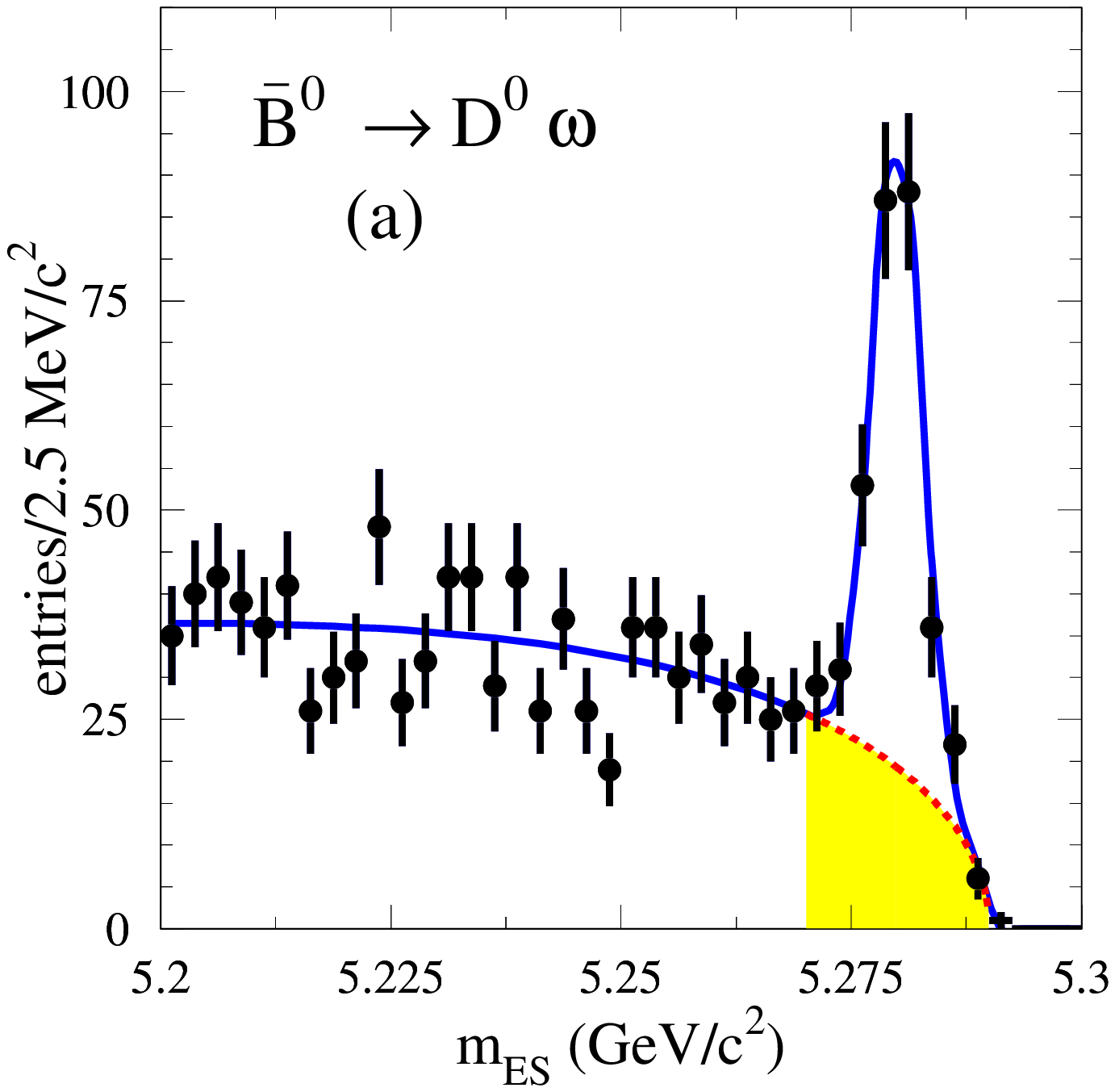,width=0.35\textwidth} &
\hspace{-0.25cm}
\epsfig{file=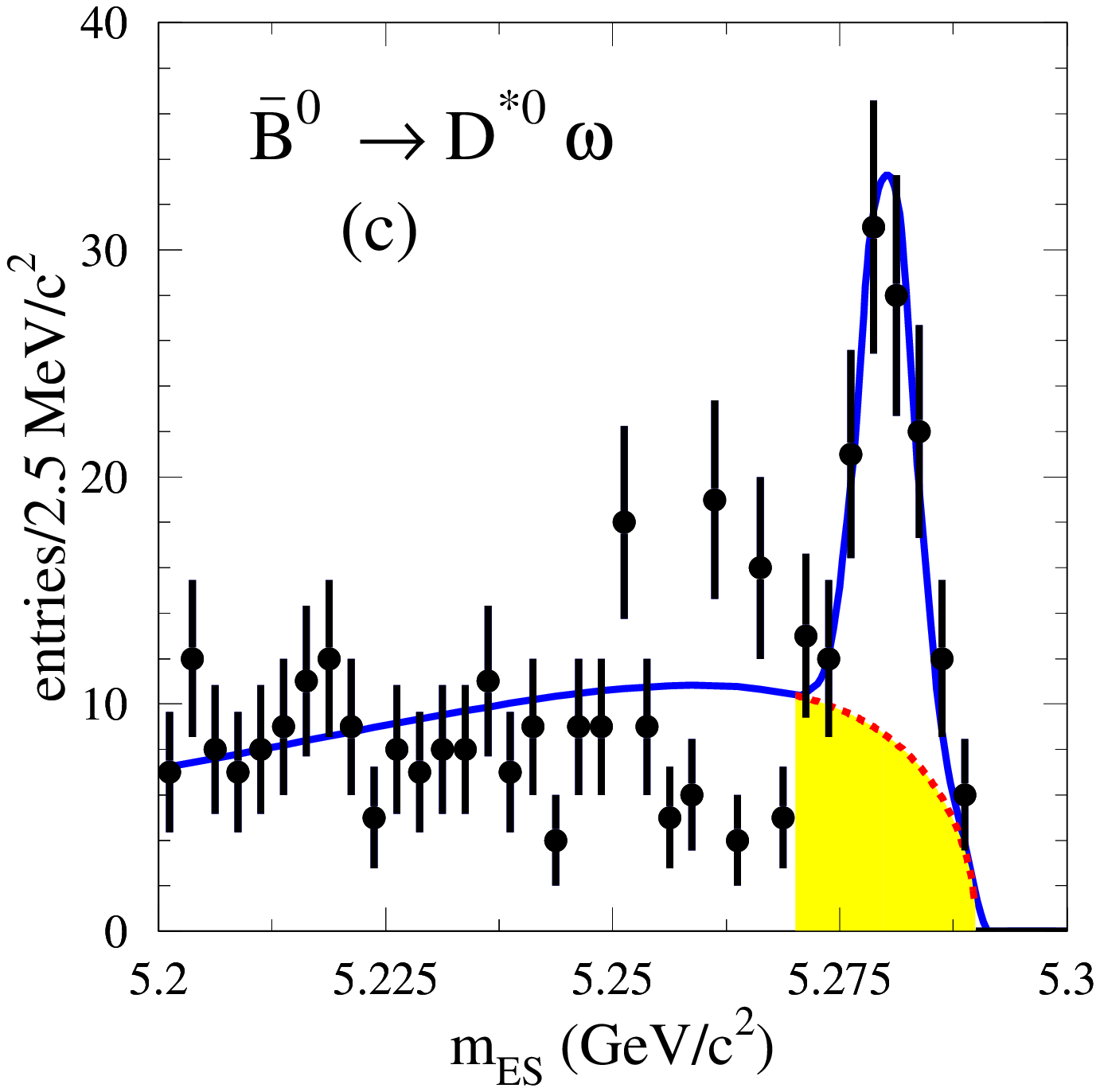,width=0.35\textwidth} \\
\hspace{-1cm}
\epsfig{file=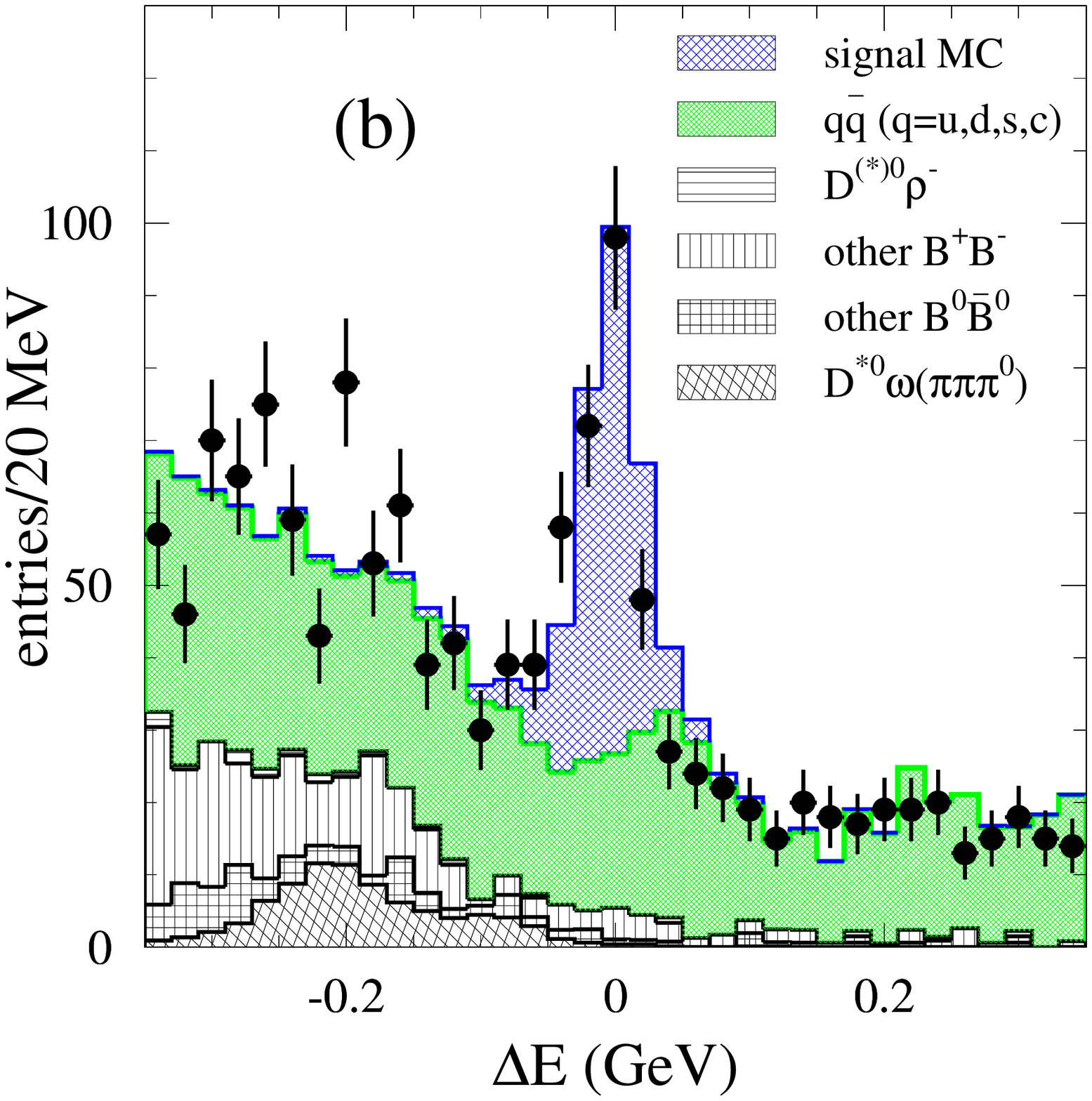,width=0.35\textwidth} &
\hspace{-0.25cm}
\epsfig{file=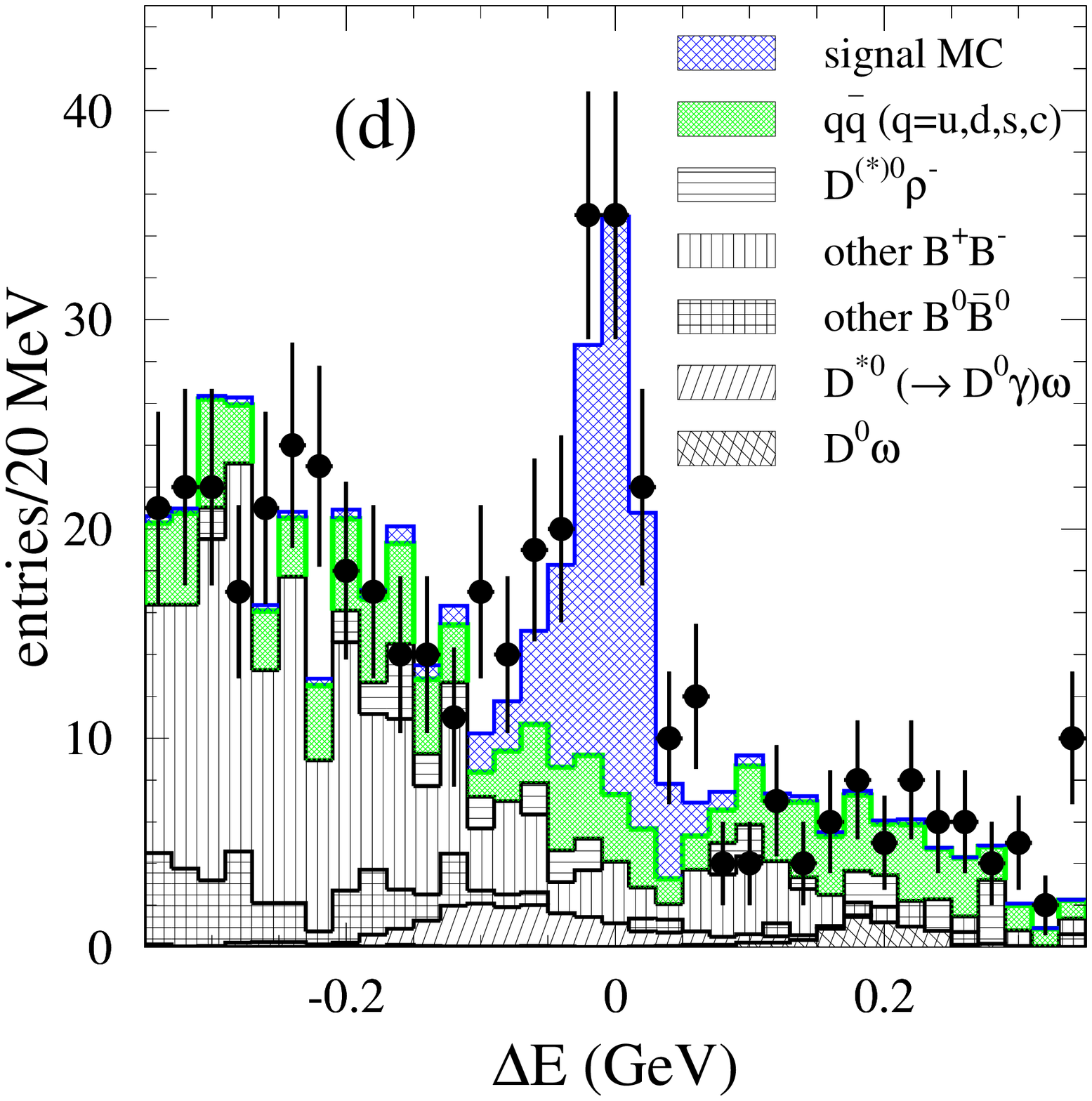,width=0.35\textwidth}
\end{tabular}
\begin{center}
\caption[D0stomega]{Distributions of \mes\ and of $\De$\ for (a,
b) candidate $\Bzerobar \ra \Dzero \omega$ events and (c, d)
candidate $\Bzerobar \ra \Dstarzero \omega$ events. The various
contributions are shown as in Fig.~\ref{fig:d0_dstar0_pi0}.
\label{fig:d0_dstar0_omega}}
\end{center}
\end{figure*}

\begin{figure*}
\begin{tabular}{lr}
\hspace{-1cm}
\epsfig{file=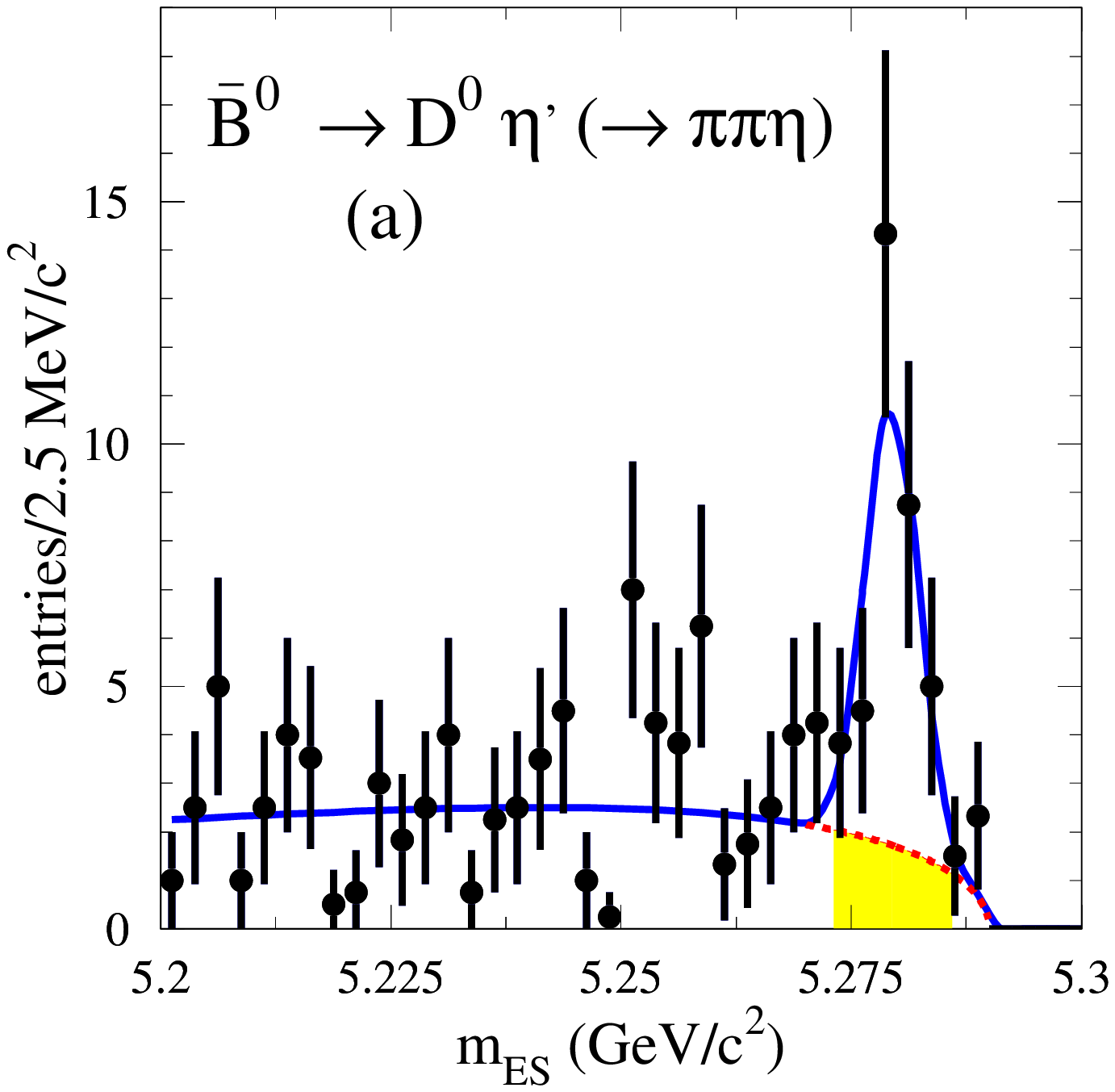,width=0.35\textwidth} &
\hspace{-0.25cm}
\epsfig{file=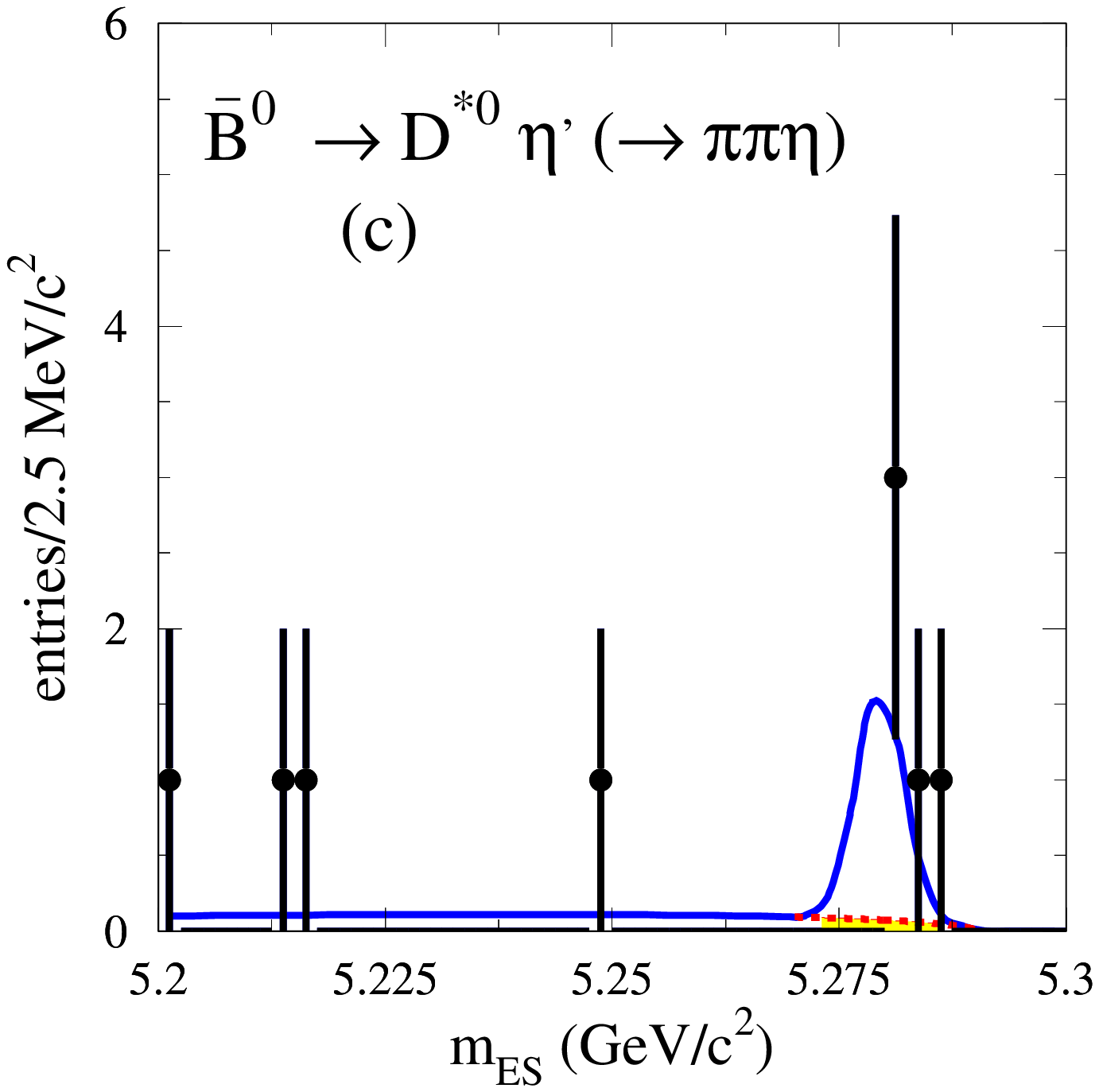,width=0.35\textwidth}
 \\
\hspace{-1cm}
\epsfig{file=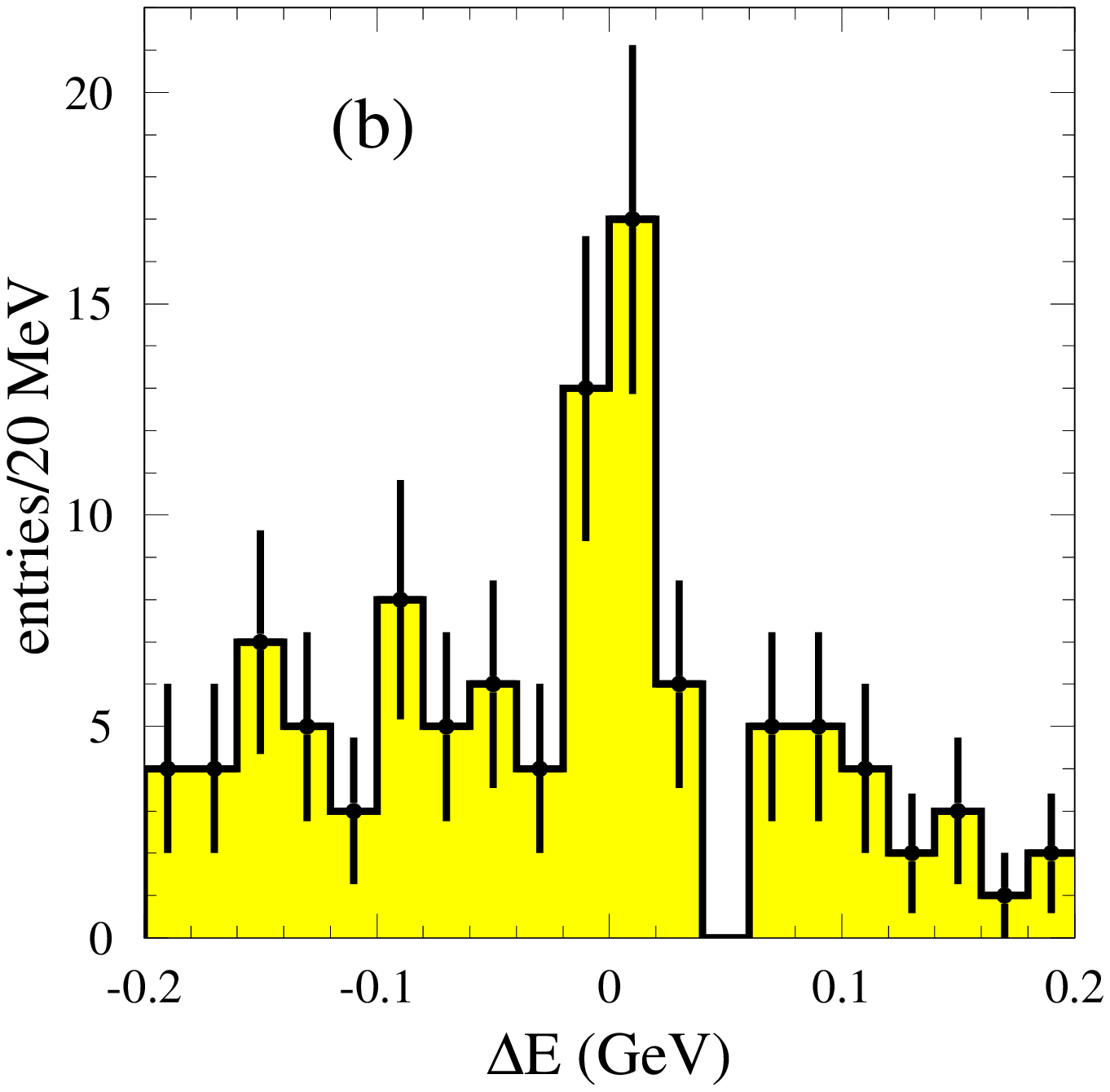,width=0.35\textwidth} &
\hspace{-0.25cm}
\epsfig{file=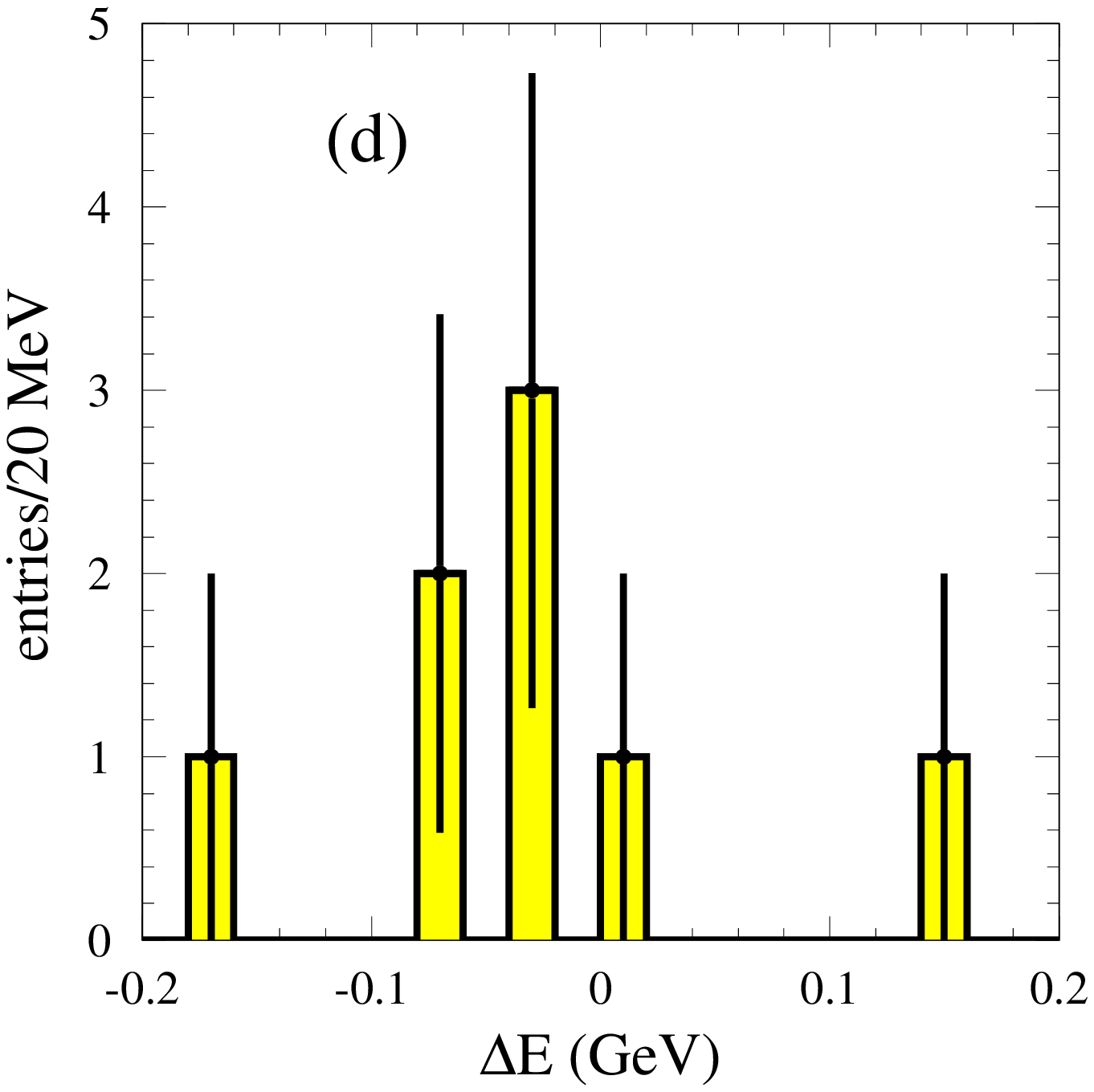,width=0.35\textwidth}
\end{tabular}
\begin{center}
\caption[D0stetap]{Distributions of \mes\ and of $\De$ of (a, b)
candidate $\Bzerobar \ra \Dzero \eta^{\prime}$ events and (c, d)
candidate $\Bzerobar \ra \Dstarzero \eta^{\prime}$ events.
\label{fig:d0_dstart0_etaP}}
\end{center}
\end{figure*}

\section{\boldmath{$\Bzerobar$} candidates in the various
color-suppressed decay modes}

\subsection{\boldmath{$\Bzerobar \ra \Dzero \pizero$}}
\label{subsec:d0pi0}

Figures~\ref{fig:d0_dstar0_pi0}(a) and \ref{fig:d0_dstar0_pi0}(b)
show the distributions in \mes\ with $-90 < \De < 100$~$\MeV$ and
in $\De$ with $5.270 < \mes < 5.290\, \GeVcsq$ for candidate
$\Bzerobar \ra \Dzero \pizero$ events.  The solid line in
Fig.~\ref{fig:d0_dstar0_pi0}(a) represents the ML fit to the sum
of the ARGUS and Gaussian functions. In
Fig.~\ref{fig:d0_dstar0_pi0}(b) the hatched histograms represent
the simulated events for the signal and separately for the various
backgrounds from \BB\ and $\qqbar$ $(q=u,d,s,c)$ events.

Peaking backgrounds originate from color-allowed $\Bmi \ra \Dzero
\rhomi$ decays where the $\pimi$ from the $\rhomi \ra \pimi
\pizero$ decay has very low momentum and is missed in the
reconstruction of the $\Dzero \pizero$ final state. This type of
background populates the $\De$ plot in the region that is at least
one pion mass below the signal region. It produces a peak in the
$\mes$ distribution in and slightly below the signal region.
Resolution effects in $\De$ will cause some events to migrate from
below the signal region into the signal region and thus contribute
to the signal peak in the $\mes$ distribution.

A veto on the color-allowed $\Bmi \ra \Dzero \rhomi$ decays is
applied as part of the selection of the $\Bzerobar$ candidates. A
$\Bzerobar$ candidate is rejected if it can be reconstructed as a
$\Bmi \ra \Dzero \rhomi$ candidate with the following properties:
\begin{itemize}
\item It uses the same $\Dzero$ and $\pizero$ as the $\Bzerobar$
candidate and the $\rho^-$ meson is selected as described in
Sec.~\ref{subsec:pi0select}. \item The \mes\ is within 9~$\MeVcsq$
of the nominal $\Bmi$ mass and $|\De| < 100$~$\MeV$.
\end{itemize}
According to the \mc\ simulation, this veto removes only a few
percent of signal events, while it rejects about 70\% of $\Dzero
\rhomi$ events and 60\% of $\Dstarzero \rhomi$ events. This
background reduction occurs nearly entirely in the $\De$ region
below approximately one pion mass and the veto is less effective
in the signal region, where only a few percents of the background
events are rejected.

The veto is nevertheless very useful because it decreases the
$\De$ distribution in the region just below the signal region,
thereby reducing the likelihood that the finite energy resolution
will shift events from the negative  $\De$ region into the signal
region. The precise determination of the resolution here is
related to the resolution of the EMC for relatively energetic
$\pizero$ mesons. Removing a large fraction of these background
events at and below the lower signal region limit reduces
substantially this uncertainty.  Even after the veto is applied,
as it can be seen in Fig.~\ref{fig:d0_dstar0_pi0}(b), the shape of
the $\De$ distribution for this background changes abruptly at
about minus one pion mass and that below this limit the magnitude
of the $\Bmi \ra \Dstze \rhomi$ background can still not be
neglected.

The yield of the  fitted candidate $\Dzero \pizero$ events and the
numbers for the various background contributions to this decay
mode are listed in Table~\ref{tab:yield}.

\subsection{\boldmath{$\Bzerobar \ra \Dstarzero \pizero$}}
\label{subsec:dst0pi0}

Figures~\ref{fig:d0_dstar0_pi0}(c) and \ref{fig:d0_dstar0_pi0}(d)
show the distributions in \mes\ with $-90 < \De < 100\, \MeV$ and
in $\De$ with $5.270 < \mes < 5.290\,\GeVcsq$ for the candidate
$\Bzerobar \ra \Dstarzero \pizero$ events.

The $\Dstarzero \pizero$ candidates are contaminated by
color-allowed $\Bmi \ra \Dstze \rhomi$ decays. Events from $\Bmi
\ra \Dstze \rhomi$ can enter the signal region when the soft
$\pimi$ from the $\rhomi$ decay is missed. In the case of $\Dzero
\rhomi$ events an unrelated $\pizero$ is used to reconstruct the
$\Dstarzero$ meson. For this mode we veto both $\Bmi \ra \Dzero
\rhomi$ and $\Dstarzero \rhomi$ decays. The criteria used to veto
$\Bzerobar$ candidates are the same as for the veto described in
the $\Dzero \pizero$ subsection except that for $\Dstarzero
\pizero$ the $\Bzerobar$ candidate is rejected if there is a $\Bmi
\ra \Dstze \rhomi$ candidate that uses the same $\Dstze$ and
$\pizero$ mesons as the $\Bzerobar$ candidate.

According to the \mc\ simulation this veto rejects about 65\% of
$\Dzero \rhomi$ events and 70\% of $\Dstarzero \rhomi$ events and
the  signal efficiency is close to 80\%. The veto is relatively
less effective in the signal region of the $\De$ distribution,
where 60\% of $\Dzero \rhomi$ events and 40\% of $\Dstarzero
\rhomi$ events are rejected. As in the $\Dzero \pizero$ case
discussed above, the veto reduces the systematic uncertainty
related to the background estimate.

The yield of the  fitted candidate $\Dstarzero \pizero$ events and
the numbers for the various background contributions to this decay
mode are listed in Table~\ref{tab:yield}.

\subsection{\boldmath{$\Bzerobar \ra \Dzero \eta$}}

Figures~\ref{fig:d0_dstar0_eta}(a) and \ref{fig:d0_dstar0_eta}(b)
show the distributions in \mes\ with $\vert \De \vert < 89\, \MeV$
(3 times the $\De$ resolution measured in the \mc\ simulation) and
in $\De$ with $5.270 < \mes < 5.290\, \GeVcsq$ for candidate
$\Bzerobar \ra \Dzero \eta$ events, where the $\eta$ meson is
reconstructed in the $\gamgam$ decay channel.
Figures~\ref{fig:d0_dstar0_eta}(c) and \ref{fig:d0_dstar0_eta}(d)
show the same distributions when the $\eta$ meson is reconstructed
in the $\threepi$ decay channel. Here the selection $\vert \De
\vert < 54\, \MeV$ is applied (again 3 times the $\De$ resolution)
in the \mes\ distribution.

In the $\eta \ra \gamgam$ case, the contribution to the peaking
background from $\Bmi \ra \Dstze \rhomi$ decays is dominant. It
corresponds to 80\% of the peaking background. In this case a
photon from the fast $\pizero$ in the $\rhomi$ decay is combined
with another photon to form an $\eta$ candidate. This background
is sufficiently suppressed by the $\pizero$ veto described in
Sec.~\ref{subsec:etaselect} so that no additional requirements are
imposed.

According to the \mc\ simulation, the peaking background is
negligible in the $\threepi$ decay channel. The \mc\ simulation
includes processes such as $D^{(\ast)} \pi^- \pi^- \pi^+ \pizero$
that may fake a $\Bzerobar \ra \Dzero \eta (\ra \threepi)$ signal
if one charged $\pi$ is lost in the reconstruction of the
$\Bzerobar$ meson. The branching fractions for these modes have
been measured recently by the CLEO
collaboration~\cite{ref:cleoDrhoprime}. Because these backgrounds
are shifted in $\De$ by more than the mass of the missing $\pi$
and because the $\eta$ mass selection is quite tight, the \mc\
simulation indicates that no events originating from such modes
are selected within the signal region. We checked the effect of
widening the signal region to $\vert \De \vert < 110\, \MeV$. Due
to resolution effects more background events in the $\De$ sideband
region migrate into the wider $\De$ signal region; we observe that
in that latter case about 10\% of the total $B \Bbar$ background
is generated by ${\rm D}^{(\ast)} \eta(\ra \threepi) \pi^-$
decays.

The yields of the  fitted candidate $\Dzero \eta$ events for the
$\eta \ra \gamgam$ and $\threepi$ decay modes and the numbers for
the various background contributions to these decay modes are
listed in Table~\ref{tab:yield}.

\subsection{\boldmath{$\Bzerobar \ra \Dstarzero \eta$}}

Figures~\ref{fig:d0_dstar0_eta}(e) and \ref{fig:d0_dstar0_eta}(f)
show the distributions in \mes\ with $\vert \De \vert < 92\, \MeV$
(3 times the $\De$ resolution measured in the \mc\ simulation) and
in $\De$ with $5.270 < \mes < 5.290\, \GeVcsq$ for candidate
$\Bzerobar \ra \Dstarzero \eta$ events in which the $\eta$ meson
is reconstructed in the $\gamgam$ channel.

According to the \mc\ simulation, the peaking background is
negligible. The yield of the  fitted candidate $\Dstarzero \eta$
events and the numbers for the various background contributions to
this decay mode are listed in Table~\ref{tab:yield}. The
statistical significance of the signal is $5.5$.

\subsection{\boldmath{$\Bzerobar \ra \Dzero \omega$}}

Figures~\ref{fig:d0_dstar0_omega}(a) and
\ref{fig:d0_dstar0_omega}(b) show the distributions in \mes\ with
$\vert \De \vert < 61\, \MeV$ (3 times the $\De$ resolution
measured in the \mc\ simulation), and in $\De$ with $5.270 < \mes
< 5.290\, \GeVcsq$ for the candidate $\Bzerobar \ra \Dzero \omega$
events. Due to the tight $\omega$ mass selection and the angular
selections, the peaking background is small.

For the peaking-background determination, we have included
possible contributions from $D^{(\ast)} \pi^- \pi^- \pi^+ \pizero$
decays. CLEO~\cite{ref:cleoDrhoprime} reports the observation of
these processes, gives branching fractions for $D^{\ast} \pi^-
\pi^- \pi^+ \pizero$ and $D^{(\ast)} \omega \pi^-$, and provides
evidence for $D^{(\ast)} \rho^{\prime -} (\ra \omega \pi^-)$.
These measurements have been performed for both charged and
neutral $B$ decays. If the additional $\pi^-$ from the
$\rho^{\prime -}$ decay is missed, these decays can fake
$\Bzerobar \ra \Dzero \omega$ events. But \mc\ simulation
indicates that the $\De$ distribution for this background is
shifted by more than the mass of the missing pion and rarely falls
in the signal region. We estimate from the \mc\ simulation that
about 11\% of the total $\Bzero \Bzerobar$ background in the
signal region originates from ${\rm D}^{(\ast)+} \omega \pi^-$
modes; similarly, 13\% of the total $\Bpl\Bmi$ background is from
$\Dstze \omega \pi^-$ decays. These fractions remain the same if
the $\De$ signal range is extended to $\vert \De \vert < 100\,
\MeV$, thus indicating that the $D^{(\ast)} \pi^- \pi^- \pi^+
\pizero$ background is randomly distributed in $\De$ over the
signal region. We also find that $D^{(\ast)}\rho$ events
contribute about 5\% of the total $B \Bbar$ background.

The yield of the  fitted candidate $\Dzero \omega$ events and the
numbers for the various background contributions to this decay
mode are listed in Table~\ref{tab:yield}.

\subsection{\boldmath{$\Bzerobar \ra \Dstarzero \omega$}}

Figures~\ref{fig:d0_dstar0_omega}(c) and
\ref{fig:d0_dstar0_omega}(d) show the distributions in \mes\ with
$\vert \De \vert < 61\, \MeV$ (3 times the $\De$ resolution
measured in the \mc\ simulation) and in $\De$ with $5.270 < \mes <
5.290\, \GeVcsq$ for candidate $\Bzerobar \ra \Dstarzero \omega$
events.

As for the $\Dzero \omega$ analysis, when determining the peaking
background, the effect of $D^{(\ast)} \pimi \pimi \pipl \pizero$
decays has been evaluated. In this case the mode $\Bmi \ra \Dzero
\omega \pi^-$ may contaminate the signal when the $\pi^-$ is
replaced by a $\pizero$ to fake a $\Dstarzero$ meson. However, the
kinematics of the soft $\pizero$ in the $\Dstarzero$ decay for the
$\Dstarzero \omega$ signal is very different from those of the
$\pi^-$ where the momentum can be large. In addition the
relatively small branching fraction for the $\Dzero \omega \pi^-$
decays implies that the contribution from this background in the
signal region is not expected to be important. We estimate from
the \mc\ simulation that 16\% of the total $\Bpl\Bmi$ background
in the signal region originates from $\Dstze \omega \pi^-$ modes.
No contribution to the $\Bzero \Bzerobar$ background from the
${\rm D}^{(\ast)+} \omega \pi^-$ decays has been found. The
fractions remain the same when the $\De$ range of the signal
region is extended to $\vert \De \vert < 100\, \MeV$. Again, this
confirms that this type of background is uniformly distributed in
$\De$ over the signal region and rules out any significant
contribution to the peaking background from these decays. We also
find that $D^{(\ast)}\rho$ events contribute about 5\% of the
total $B \Bbar$ background. The peaking background for this decay
mode is found to be negligible.

The yield of the  fitted candidate $\Dstarzero \omega$ events and
the numbers for the various background contributions to this decay
mode are listed in Table~\ref{tab:yield}. The statistical
significance of the signal is $6.1$.

\subsection{\boldmath{$\Bzerobar \ra \Dzero \eta^{\prime}$}}

Figure~\ref{fig:d0_dstart0_etaP}(a) shows the \mes\ distribution
for $\Dzero \ra \Kpi$ with $\vert \De \vert < 60\, \MeV$ and for
$\Dzero \ra \Ktwopi$ and $\Kthreepi$ with $\vert \De \vert < 40\,
\MeV$. Figure~\ref{fig:d0_dstart0_etaP}(b) shows the $\De$
distribution with $5.273 < \mes < 5.286\, \GeVcsq$. According to
the \mc\ simulation, the peaking background in this channel is
negligible. As reported in Table~\ref{tab:yield}, the fit yields
$\Ncand = 26.6 \pm 6.0$ candidate $\Dzero \etap$ events and
$\NPback = 10.4 \pm 1.1$ combinatorial-background events. The
statistical significance of the signal, calculated from Poisson
statistics, is $6.3$.

\subsection{\boldmath{$\Bzerobar \ra \Dstarzero \eta^{\prime}$}}

Figure~\ref{fig:d0_dstart0_etaP}(c) shows the \mes\ distribution
for $\Dzero \ra \Kpi$ with $\vert \De \vert < 60\, \MeV$ and for
$\Dzero \ra \Ktwopi$ and $\Kthreepi$ with $\vert \De \vert < 40\,
\MeV$. Figure~\ref{fig:d0_dstart0_etaP}(d) shows the $\De$
distribution with $5.273 < \mes < 5.286\, \GeVcsq$. According to
the \mc\ simulation the peaking background is negligible. As
reported in Table~\ref{tab:yield}, the fit yields $\Ncand = 4.0
\pm 2.2$ candidate $\Dstarzero \etap$ events and $\NPback = 0.5
\pm 0.3$ combinatorial-background events. The statistical
significance of the signal, calculated from Poisson statistics, is
only $3.0$.


\begin{table}
\caption{\label{tab:eff} Acceptance (${\cal A}$), corrected
acceptance ($\Acor$) obtained after differences between Monte
Carlo simulation of detector response and data are taken into
account, and overall efficiency $({\cal E})$ that includes
branching fractions from secondary decays. The uncertainties
associated with these numbers are discussed in
Sec.~\ref{sec:SystUncert}.}
\begin{center}
\begin{tabular}{lcccccc}
\hline \hline
\\
$\Bzerobar$ mode & { \ }& ${\cal A}$ (\%)&{ \ }& $\Acor$ (\%) &{ \
}& ${\cal E}(\%)$
\\
({decay channel})  & & & & & &
\\
\hline
\\
$\Dzero \pizero$                && 9.1  && 7.9  && 1.87 \ \\
$\Dstarzero \pizero$            && 2.7  && 2.3  && 0.34 \ \\
$\Dzero \eta(\ra \gamgam)$      && 9.7  && 8.6  && 0.82 \ \\
$\Dzero \eta(\ra \threepi)$     && 6.5  && 5.6  && 0.30 \ \\
$\Dstarzero \eta(\ra \gamgam)$  && 3.3  && 2.8  && 0.17 \ \\
$\Dzero \omega$                 && 4.2  && 3.5  && 0.75 \ \\
$\Dstarzero \omega$             && 1.7  && 1.4  && 0.19 \ \\
$\Dzero \eta^{\prime}$          && 5.0  && 4.2  && 0.18 \ \\
$\Dstarzero \eta^{\prime}$      && 1.6  && 1.4  && 0.035 \ \\
\hline \hline
\end{tabular}
\end{center}
\end{table}

\begin{table}
\caption{\label{tab:etaggeff} Values of ${\Acor}$, ${\cal
B}(\Dzero)$ (the branching fraction of the various $\Dzero$ decay
modes~\cite{ref:PDG}), $\Bsec$ (the product of the branching
fractions associated with the secondary decays of the $\eta\ra
\gamgam$ and the $\Dzero$), and ${\cal E}$ for the $\Bzero \ra
\Dzero \eta(\gamgam)$ decay mode. The branching fraction for the
$\eta \ra \gamgam$ is taken to be 39.4\%~\cite{ref:PDG}. The
uncertainties associated with these numbers are discussed in
Sec.~\ref{sec:SystUncert}.}
\begin{center}
\begin{tabular}{lcccccccc}
\hline \hline
\\
$\Dzero$ decay && $\Acor(\%)$ && ${\cal B}(\Dzero)(\%)$ && $\Bsec(\%)$ && ${\cal E}(\%)$ \\
\hline
\\
$\Kpi$         && 19.5   && 3.8   && 1.5    && 0.29         \ \\
$\Ktwopi$      && 6.0    && 13.1  && 5.1    && 0.31         \ \\
$\Kthreepi$    && 7.4    &&  7.5  && 2.9    && 0.22         \ \\
all            && 8.6    &&   -   && 9.5    && 0.82         \ \\
\hline \hline
\end{tabular}
\end{center}
\end{table}

\section{Branching Fractions}
\label{sec:effcy}

The acceptance ${\cal A}$ for signal events is estimated from
signal \mc\ as
\begin{equation}
{\cal A} = \frac{{S}_{\rm MC}} {N_{\rm gen}} \mbox{.}
\label{eq:accep}
\end{equation}
Where ${S}_{\rm MC}$ is the number of events in the signal region
that pass the selection criteria and $N_{\rm gen}$ is the number
of generated signal \mc\ events.

The selection efficiencies for each mode are obtained from
detailed \mc\ studies in which the detector response is simulated
using the GEANT4~\cite{ref:GEANT} program. The efficiencies of
tracking, detection and reconstruction in the EMC, vertex fitting,
and particle identification have been measured in control sets of
data and compared with their \mc\ simulation. We correct the
acceptance for differences between data and \mc\ simulation of
these effects by using precise correction factors that are applied
to each track (for track reconstruction efficiency), to each
photon, $\pizero$, $\eta(\gamgam)$ (for neutral cluster detection
efficiency and energy resolution), to each kaon candidate (for
particle identification efficiency), and to each vertex-fit (for
vertex-fit efficiency). Most of these corrections depend upon the
polar angle and momenta of the tracks and neutral clusters and
some also depend on the running conditions.

Tracking efficiencies are determined by identifying tracks in the
SVT and measuring the fraction of tracks that are reconstructed in
the DCH. The $\gamma$ and $\pizero$ efficiencies are measured by
comparing the ratio of the number of events $N(\taup \to \nutb
h^+\pizero)$ and $N(\taup \to \nutb h^+\pizero\pizero)$ to the
known branching fractions~\cite{ref:cleotau}. The kaon
identification efficiency is estimated from a sample of $\Dstarpl
\ra \Dzero \pipl$, $\Dzero\ra \Kpi$ decays that are identified
kinematically. Based on a similar selection, a sample of
$\Bzerobar \ra \Dstarpl \pimi$, $\Dstarpl \ra \Dzero \pipl$,
$\Dzero \ra \Kpi$, $\Ktwopi$, or $\Kthreepi$ decays is used to
determine the vertex-fit efficiency corrections.

The acceptances ${\cal A}$ obtained with Eq.~(\ref{eq:accep}) and
the corrected acceptances $\Acor$ are listed in
Table~\ref{tab:eff}. The last column in Table~\ref{tab:eff} lists
the values of the overall efficiency ${\cal E}$ defined as
\begin{equation}
{\cal E} = \Acor \times \Bsec {\rm ,}
\end{equation}
where
\begin{equation}
\begin{split}
{\Bsec} = \Br(\Dstarzero \ra \Dzero \pizero) \times
\Br(\pizero \ra \gamgam) \times \ \\
\Br(\hzero \ra Y) \times \sum_X{\Br(\Dzero \ra X)}
\end{split}
\end{equation}
is the product of the branching fractions associated with the
secondary decays of the $\Dstarzero$, $\hzero$, and $\Dzero$ (with
$X = \Kpi$, $\Ktwopi$, or $\Kthreepi$). The $\Br(\Dstarzero \ra
\Dzero \pizero) \times \Br(\pizero \ra \gamgam)$ factor is only
present for the $\Bzerobar \ra \Dstarzero \hzero$ final states.
Note that the overall efficiency ${\cal E}$ for the $\Dstze
\eta^{\prime}$ decays is reduced with respect to the other
$\Bzerobar$ modes by the relatively small values of ${\Bsec}$.

In Table~\ref{tab:etaggeff} we display, as an example, the
contributions of the three $\Dzero$ final states in the decay mode
$\Bzerobar \ra \Dzero \eta(\gamgam)$. There are variations between
the acceptance and branching fraction for the three $\Dzero$ decay
modes leading to similar values of ${\cal E}$ for the three modes.
A similar conclusion holds for other $\Bzerobar \ra \Dstze \hzero$
final states.

To obtain branching fractions, the number of background subtracted
signal events, $S$, is divided by the number of $B\Bbar$ events in
the data sample, $N(\BBbar)$, and the overall efficiency, ${\cal
E}$:
\begin{equation}
{\Br} (\Bzerobar \ra \Dstze \hzero) = \frac {S} {N(\BBbar)\times
{\cal E}} {\rm .}
\end{equation}
These branching fraction calculations assume equal production of
$\Bzero \Bzerobar$ and $B^+B^-$ pairs at the \FourS\ resonance.

\begin{table*}
\caption{\label{tab:systerr} Systematic uncertainties of the
measured branching fractions in percent. The symbol ``-''
indicates that the systematic uncertainty is negligible.}
\begin{center}
\begin{tabular}{lcccccccccccccccccc}
\hline \hline
\\
Category && $\Dzero\pizero$ && $\Dstarzero\pizero$ &&
$\Dzero\eta(\gamgam)$ && $\Dzero\eta(\threepi)$ &&
$\Dstarzero\eta$ && $\Dzero\omega$ && $\Dstarzero\omega$ && $\Dzero\eta^{\prime}$ && $\Dstarzero\eta^{\prime}$\\
\hline
\\
Tracking                                    && 2.1  && 2.1  && 2.0  && 3.6  && 2.0  && 3.6  && 3.6  && 3.6  && 3.6   \ \\
Vertex-fit                                  && 1.4  && 1.4  && 1.4  && 2.5  && 1.4  && 2.5  && 2.5  && 1.4  && 1.4  \ \\
Kaon identification                         && 2.5  && 2.5  && 2.5  && 2.5  && 2.5  && 2.5  && 2.5  && 2.5  && 2.5  \ \\
$\gamma$, $\pizero$, and $\eta$ detection   && 5.2  && 8.1  && 3.7  && 6.0  && 6.8  && 5.9  && 9.1  && 3.5  && 6.5  \ \\
Cross-feed                                  && 1.0  && 0.7  && 1.4  && 0.5  && 2.4  && 0.6  && 1.7  &&  -   &&  -   \ \\
$\De$ resolution                            && 1.7  && 1.9  && 3.0  && 4.4  && 3.5  && 5.7  && 3.3  &&  -   && -    \ \\
\mes\ fit                                   && 0.3  && 3.2  && 4.5  && 4.8  && 8.4  && 3.0  && 10.3 && 2.3  && 2.3  \ \\
Peaking background                          && 3.3  && 6.3  && 3.2  && 2.0  && 0.5  && 3.4  && 4.0  &&  -   && -    \ \\
Event selection                             && 6.8  && 9.4  && 6.1  && 8.9  && 7.6  && 6.8  && 11.9 && 7.9  && 7.9  \ \\
${\cal B}(\Dstze)$ and ${\cal B}($\hzero$)$ && 4.6  && 6.6  && 4.4  && 4.6  && 6.3  && 4.3  && 6.4  && 5.6  && 7.3  \ \\
Number of $B\Bbar$ pairs                    && 1.1  && 1.1  && 1.1  && 1.1  && 1.1  && 1.1  && 1.1  && 1.1  && 1.1  \ \\
\mc\ statistics                             && 0.7  && 2.2  && 1.3  && 1.6  && 2.0  && 1.6  && 2.9  && 1.6  && 1.6  \ \\
\hline
\\
Total (\%)                                  && 11.1 && 16.4 && 11.2 && 14.5 && 15.8 && 13.5 && 20.8 && 11.7 && 13.7 \ \\
\hline \hline
\end{tabular}
\end{center}
\end{table*}

\begin{figure}
\begin{tabular}{lr} \hspace{-0.25cm}
\epsfig{file=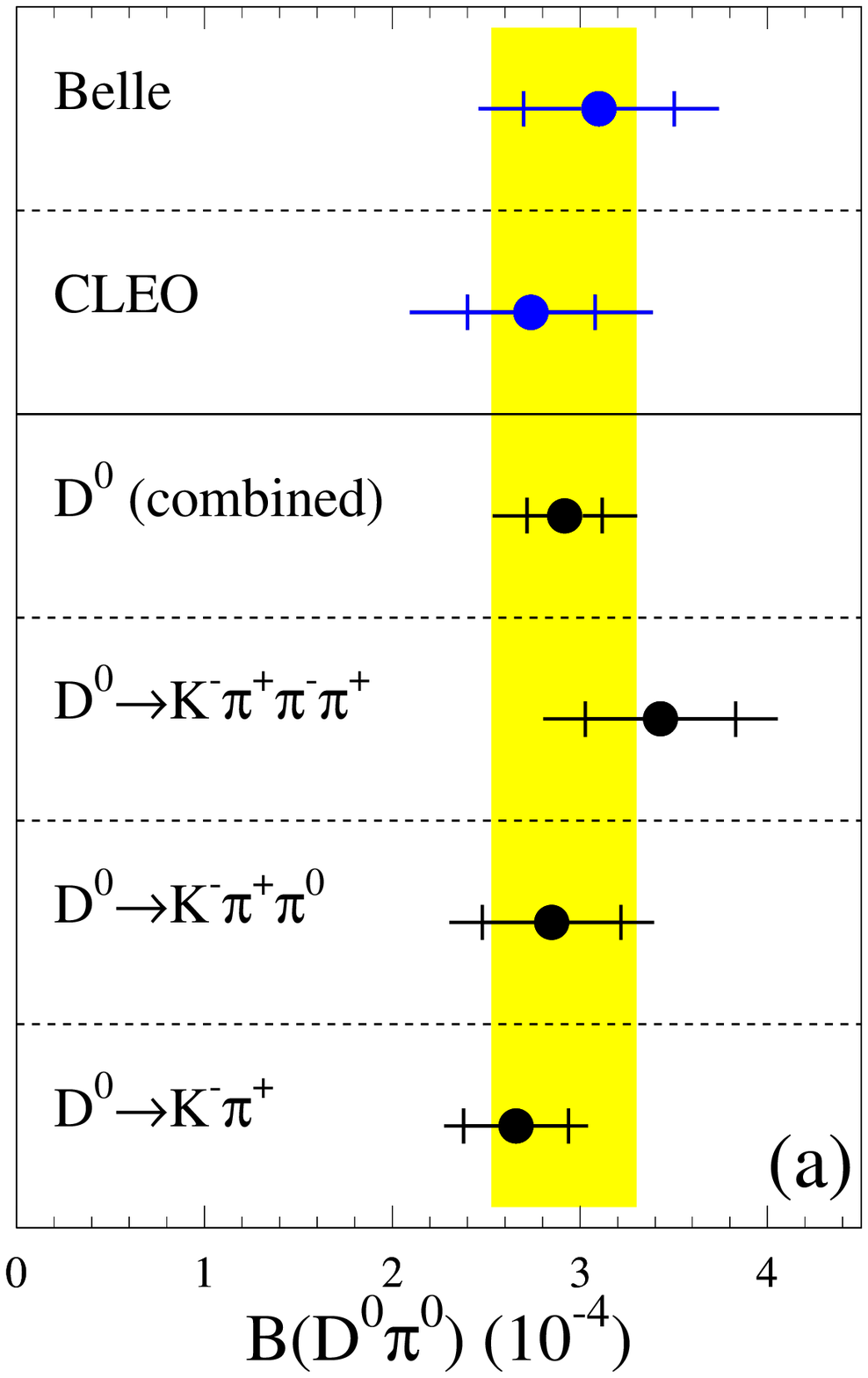,width=0.22\textwidth,height=0.275\textwidth}
&\hspace{-0.25cm}
\epsfig{file=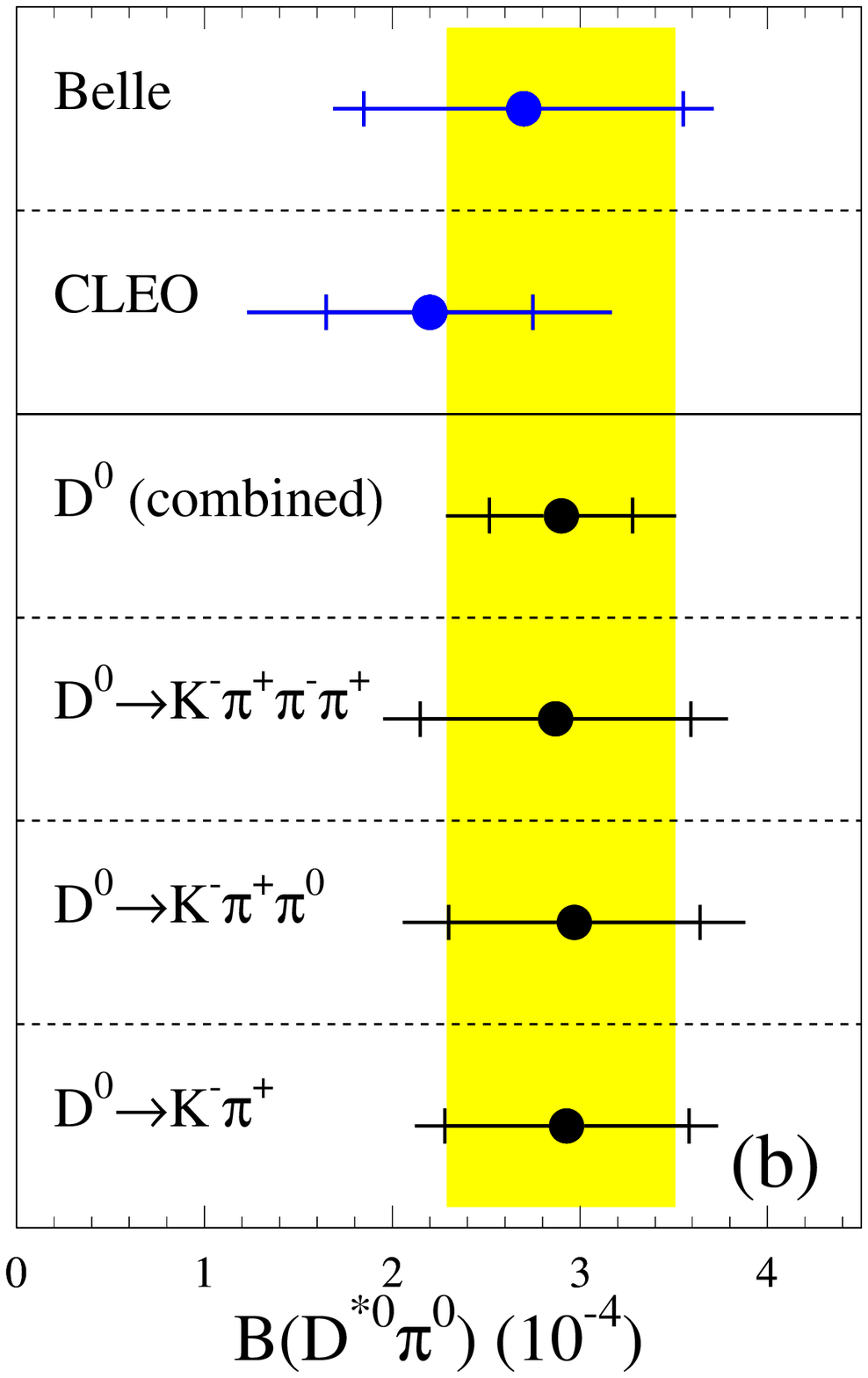,width=0.22\textwidth,height=0.275\textwidth}
\\
\end{tabular}
\begin{center}
\caption[D0stpi0br]{Measured branching fractions for each of the
three $\Dzero$ decay modes and for the combination of the three
for (a) $\Bzerobar \ra \Dzero \pizero$ and (b) $\Bzerobar \ra
\Dstarzero \pizero$. The shaded bands represent the results from
the present investigation. The length of the error bars is equal
to the sum in quadrature of the statistical and the systematic
uncertainty; the statistical contribution is superimposed on the
error bar. The CLEO~\cite{ref:CLEO} and Belle~\cite{ref:Belle}
results are also shown. \label{fig:d0_dstar0_pi0_br}}
\end{center}
\end{figure}

\section{Systematic Uncertainties}
\label{sec:SystUncert}

Systematic uncertainties are associated with the acceptance
corrections discussed in Sec.~\ref{sec:effcy}. The uncertainties
from the tracking-efficiency corrections are 0.8\% per charged
track. To take into account uncertainties caused by the vertex
reconstruction, we assign a systematic uncertainty equal to 1.1\%
per two-track vertex and 2.2\% per four-track vertex. For particle
identification the uncertainty is 2.5\% per $K^\pm$ track. The
uncertainties from the requirement that all the $\pi^\pm$
daughters must fail the tight kaon criterion are negligible.
Uncertainties in the acceptances for photon detection account for
imperfect simulation of photon-energy and position resolution,
thus affecting $\pizero$ and $\eta$ reconstruction efficiencies
and the $\De$ resolution. For the detection of isolated $\pizero$
and $\eta(\gamgam)$ mesons uncertainties of 5\% and 2.5\% are
used. These uncertainties are summed in quadrature, together with
other corrections that depend upon the energy of each $\gamma$
used to reconstruct the mesons.

We consider systematic uncertainties from other sources. For the
cross-feed fractions an uncertainty equal to 25\% of the estimated
fraction accounts for uncertainties in the branching fractions
reported in this study and used in the cross-feed determination.
This value is chosen conservatively; it corresponds to the
branching fraction measurement with the largest uncertainty
reported in this paper (see Table~\ref{tab:BF}).

The effect of the specific $\De$ range used to define the signal
region and based on the resolution measured from the \mc\
simulation has been estimated by varying the limits of the range
by $\pm \ 0.5\, \sigma$. The observed variations in the branching
fraction are used to determine the systematic uncertainty from
this source. In the case of the $\Dstze \pizero$ modes, we vary
the lower limit on the signal region definition ($-90 < \De <
100\, \MeV$) between $-110$ and $-60$ \MeV. Therefore, this
procedure also accounts for uncertainties in the
peaking-background estimates that are caused by the systematic
uncertainty of the energy resolution that originates from the EMC.

To evaluate the systematic uncertainty associated with using the
\mes\ resolution taken from \mc\ in the fit to data, we also let
it vary freely in that fit and half of the variation in the yields
is taken as the systematic error. We also investigate the
uncertainties in the combinatorial background due to setting the
value of the ARGUS shape parameter $\xi$ to the value obtained in
the fit to the data \mes\ distribution in the upper $\De$ sideband
$+ \ 6\,\sigma < \De < 350\,\MeV$. For the $\Dstze \eta^{\prime}$
analyses, the value of $\xi$ is obtained from the $\De$ sidebands
(see Sec.~\ref{sec:yieldextract}). We therefore vary the value of
$\xi$ by one standard deviation of the statistical error. In each
case we take half the variation observed as the systematic
uncertainty. Finally, the sum of the systematic errors from the
ARGUS shape parameter and the fixed Gaussian width is taken as the
systematic error for the \mes\ fitting procedure.

Systematic uncertainties in the peaking background arise from
imprecision in branching fractions and from statistical
uncertainties in the number of peaking-background events obtained
from the procedure described in Sec.~\ref{sec:pkBckgd}. For the
$\Dstze \pizero$ modes the systematic uncertainty associated with
the veto of the $\Bmi \ra \Dstze \rhomi$ background has been
studied and is part of the systematic uncertainty of the
background estimate. For these $\Dstze \pizero$ decay modes, we
remove the veto on the $\Bmi \ra \Dstze \rhomi$ background and we
include in the uncertainties half of the relative variation of the
branching fraction. Finally, we have explained in
Sec.~\ref{sec:pkBckgd} how the systematic uncertainty related to
the fitting method used in the calculation of the number of
peaking-background events is estimated. The variation of the
branching fraction due to the latter effect is small or negligible
(4\% at most) but is included in the systematic uncertainty from
peaking background.

\begin{figure}
\begin{tabular}{lr} \hspace{-0.25cm}
\epsfig{file=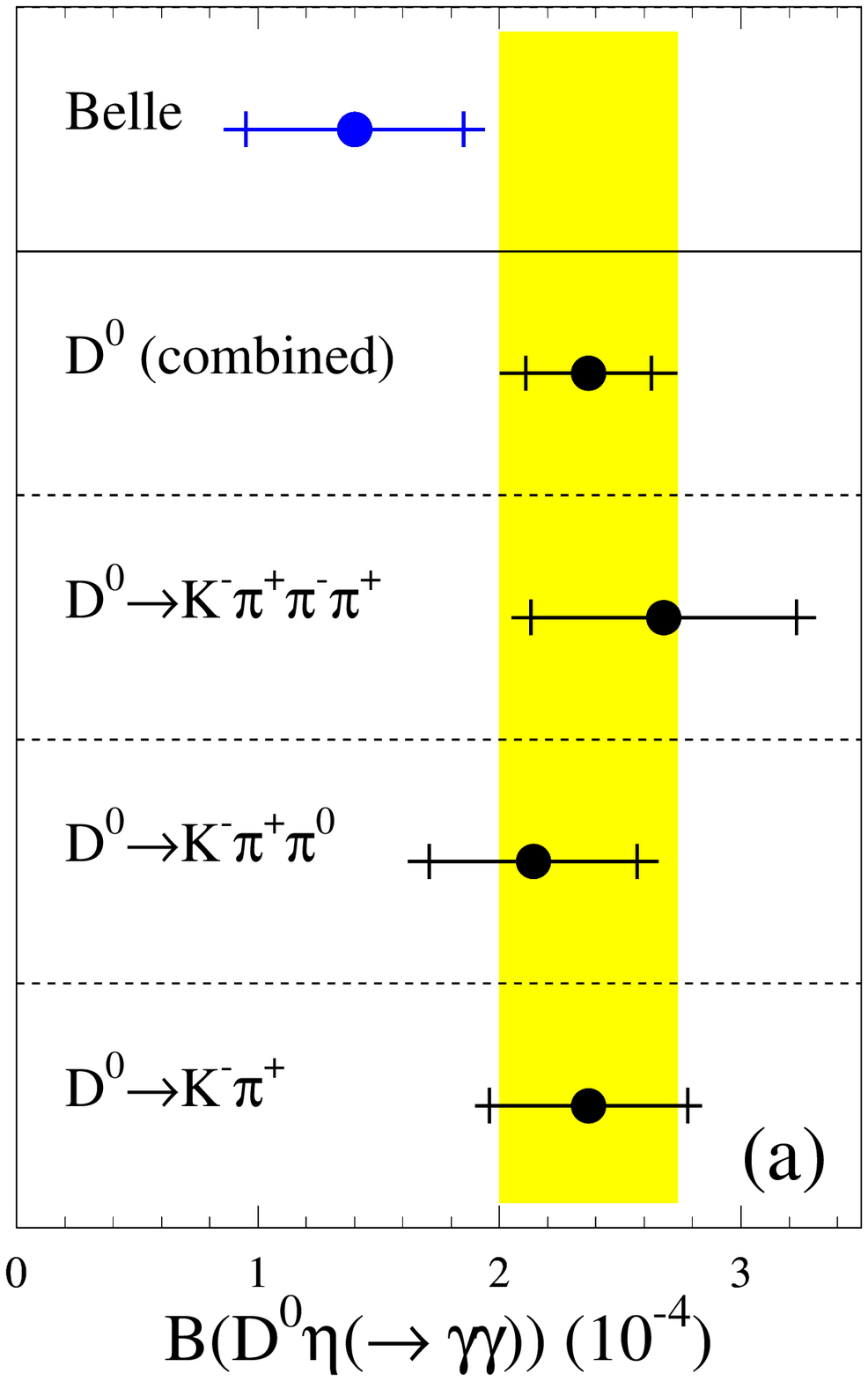,width=0.22\textwidth,height=0.275\textwidth}
&\hspace{-0.25cm}
\epsfig{file=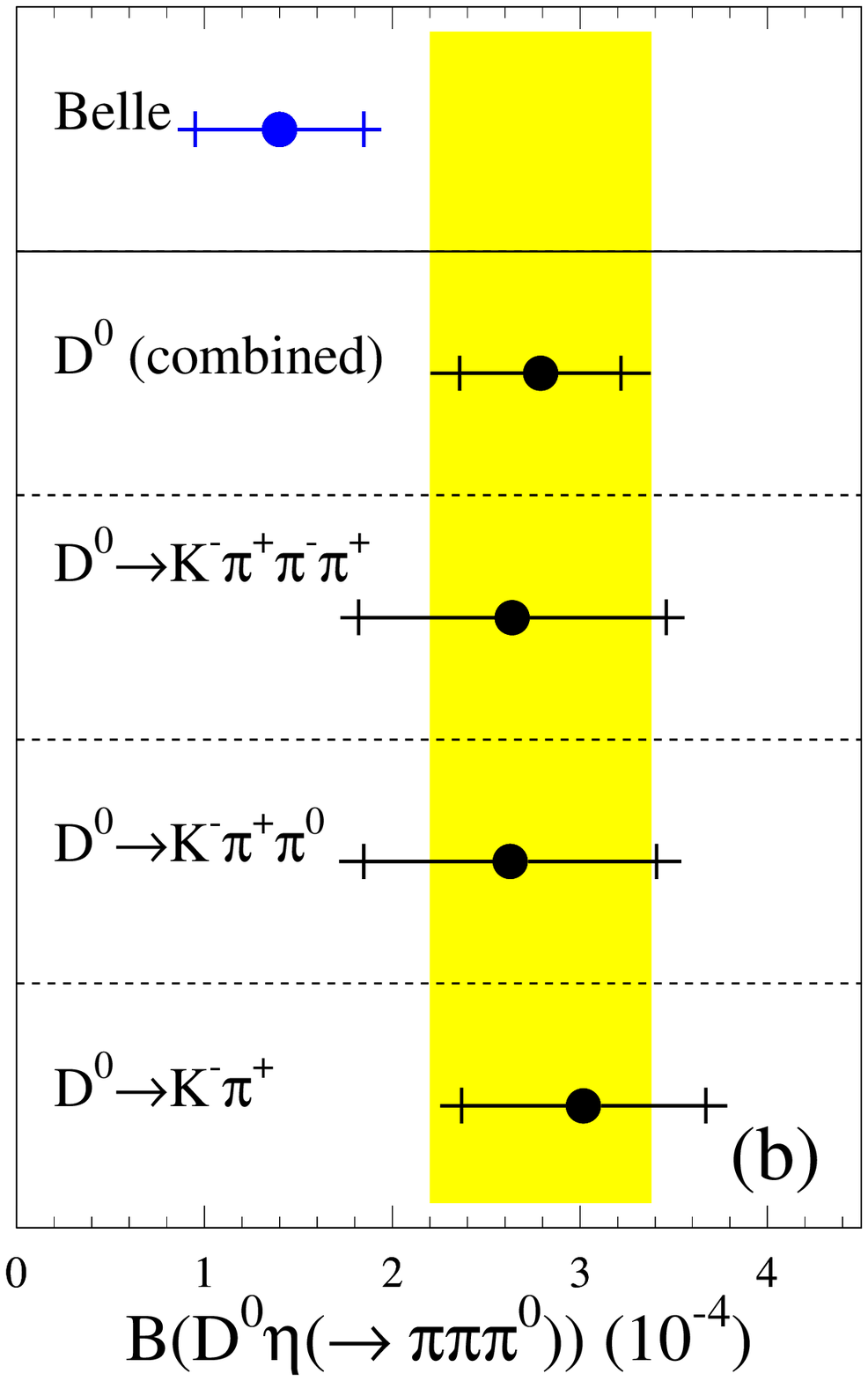,width=0.22\textwidth,height=0.275\textwidth}
\\
\hspace{-0.25cm}
\epsfig{file=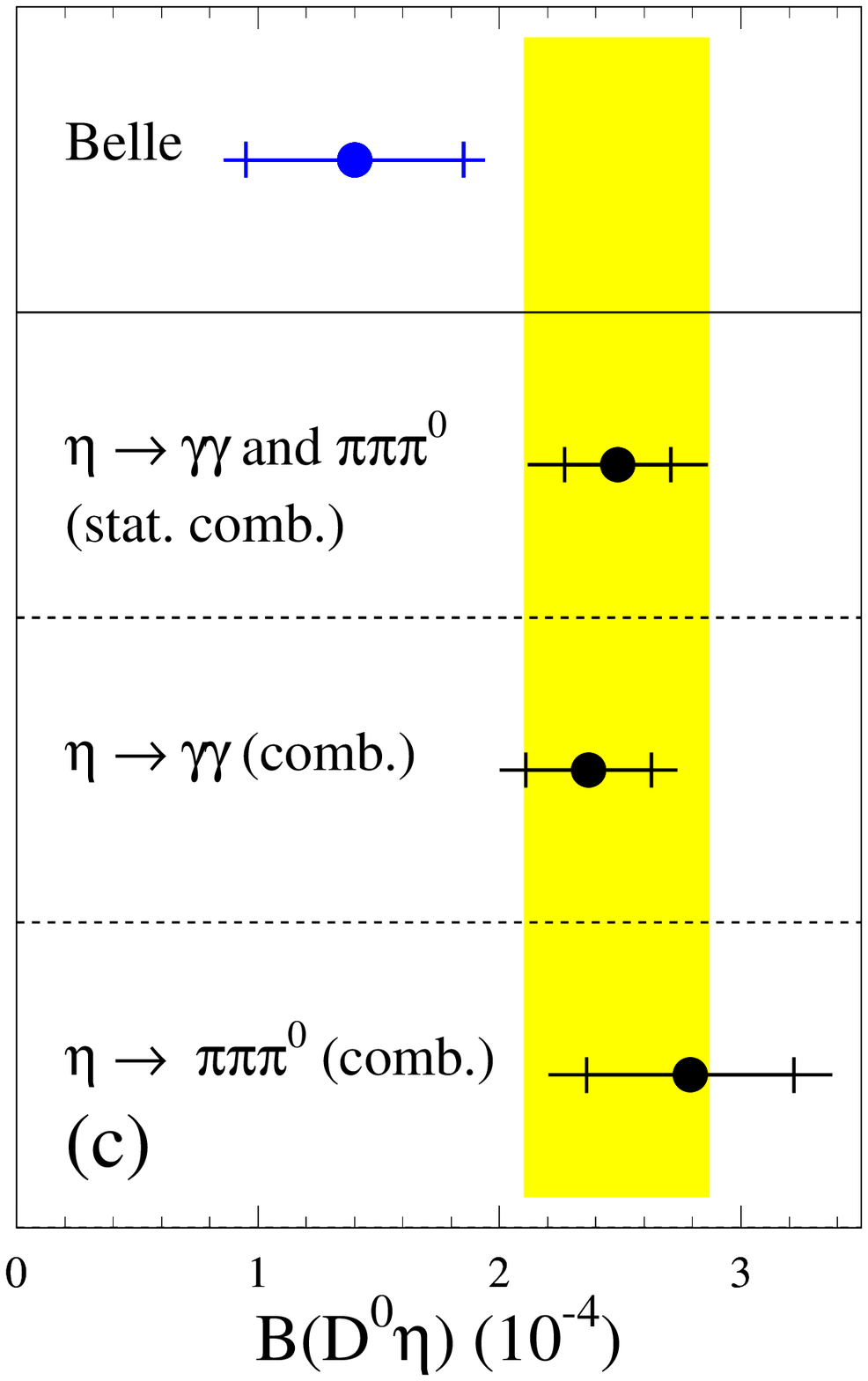,width=0.22\textwidth,height=0.275\textwidth}
&\hspace{-0.25cm}
\epsfig{file=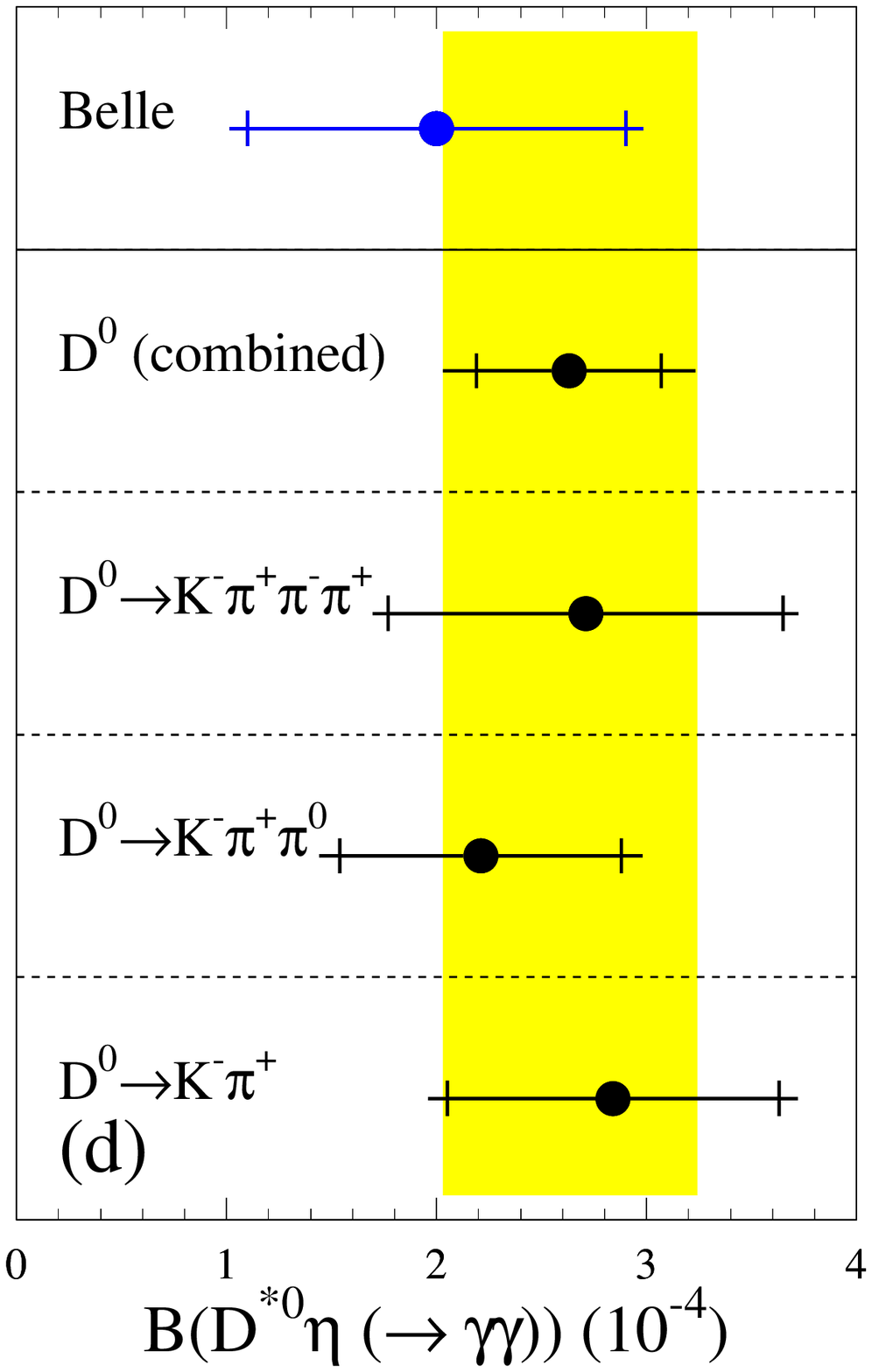,width=0.22\textwidth,height=0.275\textwidth}
\\
\end{tabular}
\begin{center}
\caption[D0stetabr]{Measured branching fractions for each of the
three $\Dzero$ decay modes and for the combination of the three
for (a) $\Bzerobar \ra \Dzero \eta(\ra \gamgam)$, (b) $\Bzerobar
\ra \Dzero \eta(\ra \threepi)$, (c) each of the $\Bzerobar \ra
\Dzero \eta$ modes and their combination, and (d) $\Bzerobar \ra
\Dstarzero \eta$. The branching fraction for the $\Dzero \eta$
channel is obtained as the average of the branching fraction of
each of the two $\eta$ decay modes, weighted by the statistical
uncertainties of these decays. The computation of the systematic
uncertainty includes both the correlated and uncorrelated
uncertainties of these two modes. The Belle~\cite{ref:Belle}
results are also shown. The error bars are as in
Fig.~\ref{fig:d0_dstar0_pi0_br}. \label{fig:d0_dstar0_eta_br}}
\end{center}
\end{figure}

\begin{figure}
\begin{tabular}{lr} \hspace{-0.25cm}
\epsfig{file=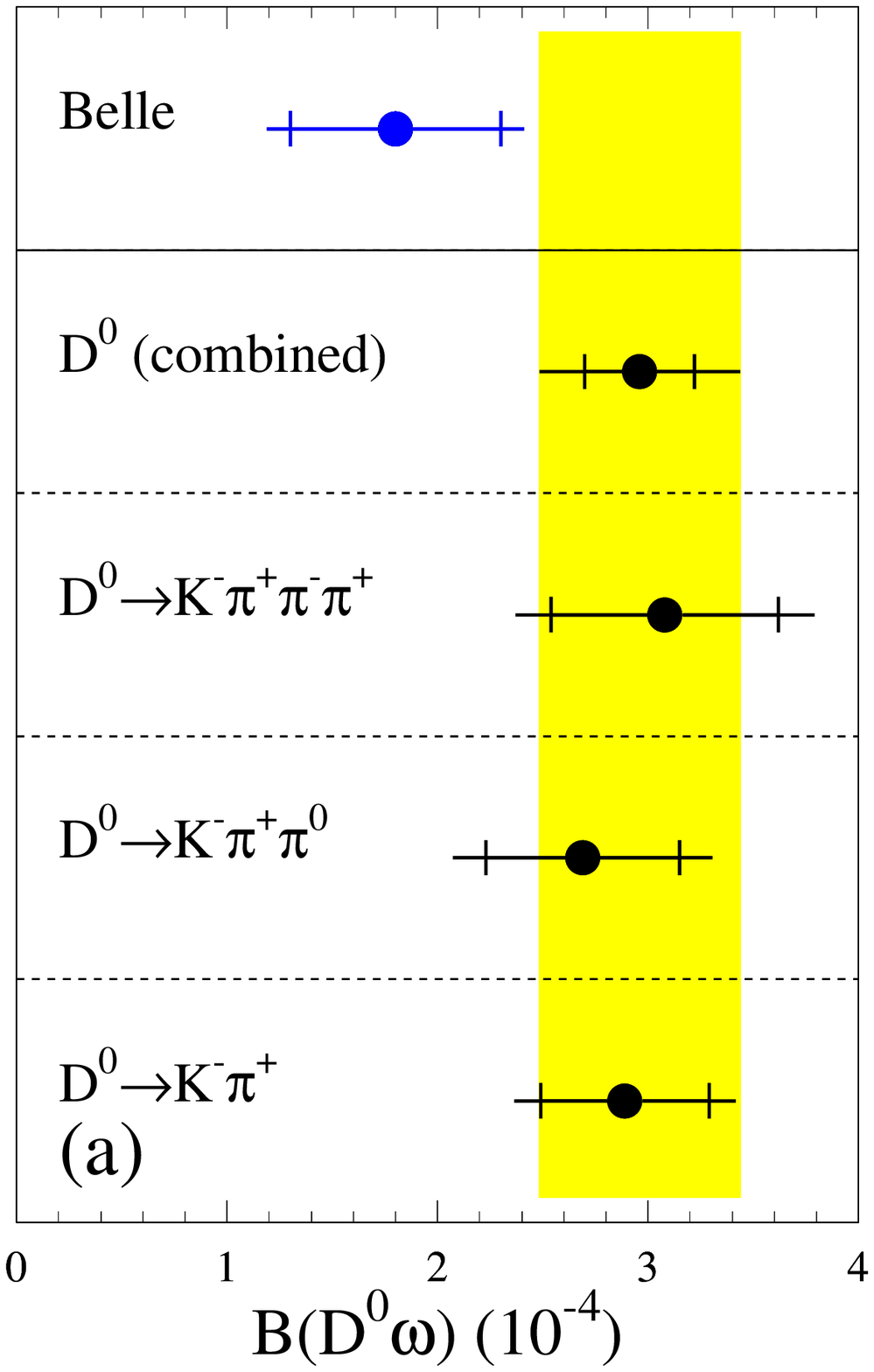,width=0.22\textwidth,height=0.275\textwidth}
&\hspace{-0.25cm}
\epsfig{file=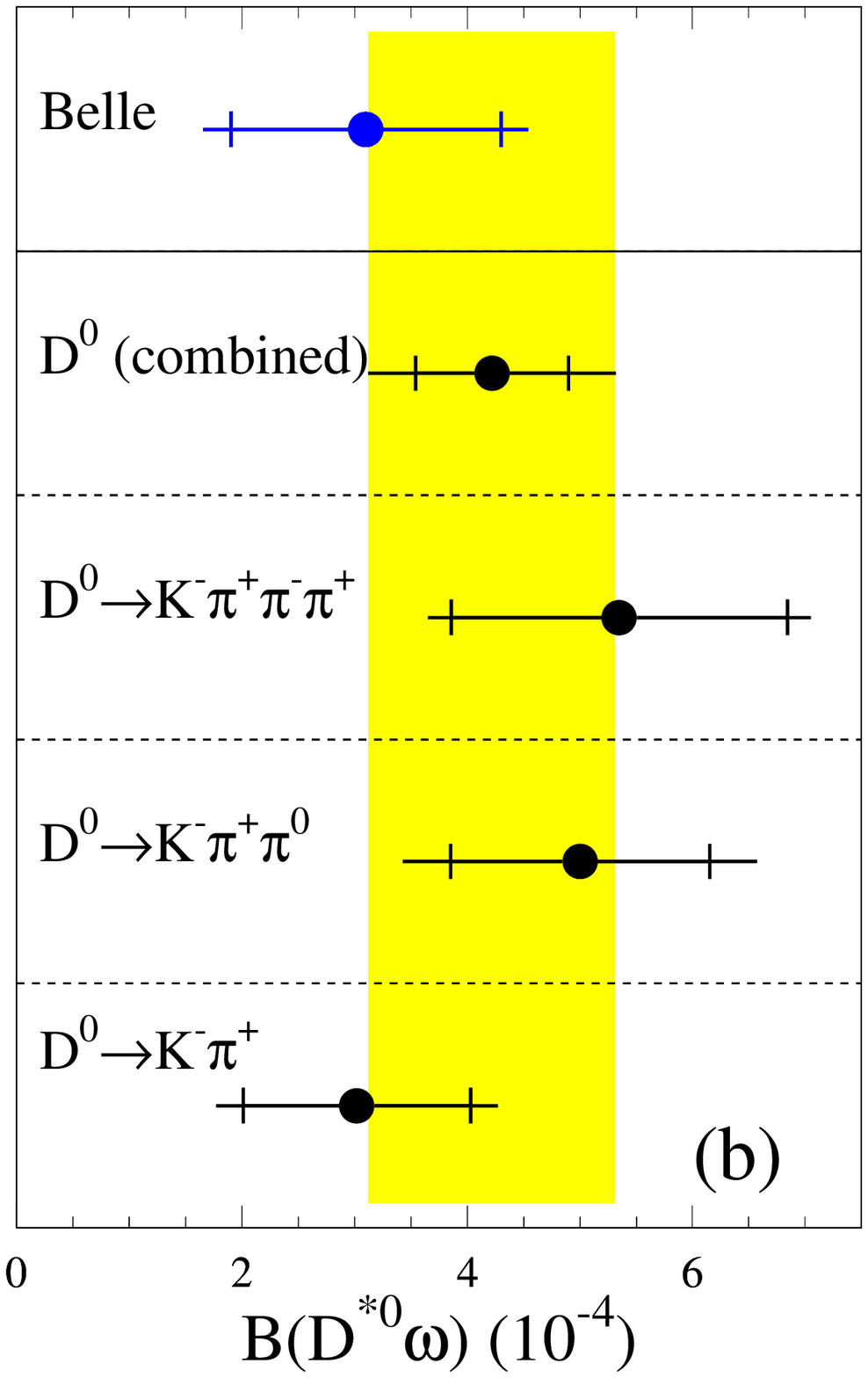,width=0.22\textwidth,height=0.275\textwidth}
\\
\multicolumn{2}{c}{\epsfig{file=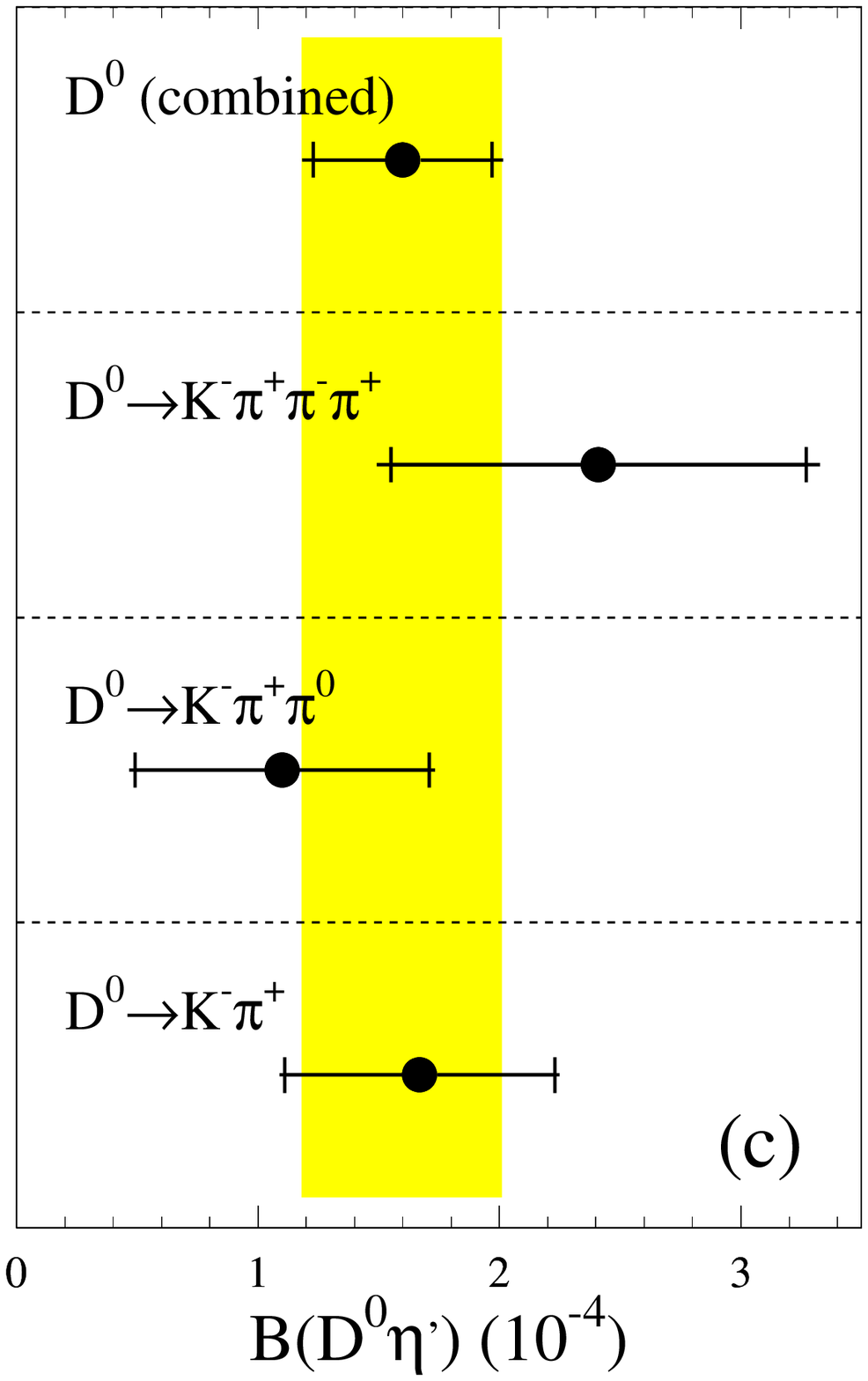,width=0.22\textwidth,height=0.275\textwidth}}
\\
\end{tabular}
\begin{center}
\caption[D0stomegabr]{Measured branching fractions for each of the
three $\Dzero$ decay modes and for the combination of the three
for (a) $\Bzerobar \ra \Dzero \omega$, (b) $\Bzerobar \ra
\Dstarzero \omega$, and (c) $\Bzerobar \ra \Dzero \eta^{\prime}$.
The Belle~\cite{ref:Belle} results, when existing, are also shown.
The errors bars are as in Fig.~\ref{fig:d0_dstar0_pi0_br}.
\label{fig:d0_dstar0_omega_br}}
\end{center}
\end{figure}

We vary the selection criteria applied to several other
uncorrelated variables such as invariant masses, event shape, and
helicity angles (see Secs.~\ref{sec:lightHad},
\ref{sec:CharmMesons}, and \ref{sec:ExclB}). We conservatively
assign a single systematic uncertainty due to the efficiencies
associated with these many selection criteria, equal to the
quadratic sum of the average of the absolute values of the
observed changes in branching fraction for each variable. None of
the various observed changes contribute in a dominant way to the
total systematic uncertainty due to event selection.

The uncertainties from the counting of $B\Bbar$ pairs, from the
branching fractions of $\Dstze$ and $\hzero$ secondary
decays~\cite{ref:PDG}, and from the statistics of the \mc\ samples
used to determine the signal acceptance, are also considered.

The systematic uncertainties described above are listed in
Table~\ref{tab:systerr} for all the modes reported in this paper.
It is seen that the dominant systematic uncertainties are due to
the event selection, from $\gamma$, $\pizero$, and $\eta$
detection, from the \mes\ fitting procedure, and from the $\Dstze$
and $\hzero$ branching fractions.

\begin{table*}
\caption{\label{tab:BFD0} Measured branching fractions for
$\Bzerobar \ra \Dstze \hzero$ $(\times 10^{-4})$. The measurements
are given for each of the three $\Dzero$ decay modes $\Kpi$,
$\Ktwopi$, and $\Kthreepi$. The first uncertainty is statistical
and the second systematic. }
\begin{center}
\begin{tabular}{lcrccclcrccclcrcccl}
\hline\hline \\
$\Bzerobar$ mode  &{ \  }& \multicolumn{5}{c}{$\Dzero \ra \Kpi$}
&{ \ }& \multicolumn{5}{c}{$\Dzero \ra \Ktwopi$} &{ \ }&
\multicolumn{5}{c}{$\Dzero \ra \Kthreepi$}  \\
{(decay channel)} &&&&&&&&&&&&&&&&&&  \\
\hline \\
$\Dzero \pizero$     &&  2.7&$\pm$&0.3&$\pm$&0.3 && 2.9&$\pm$&0.4&$\pm$&0.4 && 3.4&$\pm$&0.4&$\pm$&0.5     \ \\
$\Dstarzero \pizero$ &&  2.9&$\pm$&0.6&$\pm$&0.5 && 3.0&$\pm$&0.7&$\pm$&0.6 && 2.9&$\pm$&0.7&$\pm$&0.6     \ \\
$\Dzero \eta(\ra\gamgam)$  &&  2.4&$\pm$&0.4&$\pm$&0.2 && 2.1&$\pm$&0.4&$\pm$&0.3 && 2.7&$\pm$&0.5&$\pm$&0.3    \ \\
$\Dzero \eta(\ra\threepi)$ &&  3.0&$\pm$&0.6&$\pm$&0.4 && 2.6&$\pm$&0.8&$\pm$&0.5 && 2.6&$\pm$&0.8&$\pm$&0.4     \ \\
$\Dstarzero \eta(\ra\gamgam)$  &&  2.8&$\pm$&0.8&$\pm$&0.4 && 2.2&$\pm$&0.7&$\pm$&0.4 && 2.7&$\pm$&0.9&$\pm$&0.4    \ \\
$\Dzero\omega$     &&  2.9&$\pm$&0.4&$\pm$&0.3 && 2.7&$\pm$&0.5&$\pm$&0.4 && 3.1&$\pm$&0.5&$\pm$&0.5     \ \\
$\Dstarzero \omega$ &&  3.0&$\pm$&1.0&$\pm$&0.7 && 5.0&$\pm$&1.1&$\pm$&1.1 && 5.3&$\pm$&1.5&$\pm$&0.8    \ \\
$\Dzero \eta^{\prime}$  && 1.7&$\pm$&0.6&$\pm$&0.1 &&
1.1&$\pm$&0.6&$\pm$&0.2 &&
2.4&$\pm$&0.9&$\pm$&0.3 \ \\
 \hline\hline
\end{tabular}
\end{center}
\end{table*}

\begin{table}
\caption{\label{tab:BF} Measured branching fractions for
$\Bzerobar \ra \Dstze \hzero$ obtained by combining the three
$\Dzero$ decay modes. The first uncertainty is statistical and the
second systematic. The last column is the statistical
significance. The branching fraction for the $\Dzero \eta$ mode is
obtained as the average of the branching fractions of each of the
two $\eta$ decay modes, weighted by the statistical uncertainties
of these decays; the computation of the systematic uncertainty
includes both the correlated and uncorrelated errors of these two
modes. For the $\Dstze \eta^{\prime}$ modes, the number of
candidates is small, so Poisson statistics rather than Gaussian
statistics are used, and the value for the statistical
significance is defined as $\sqrt{2 \ln ({\cal L}_{max}/ {\cal
L}(0))}$, where ${\cal L}_{max}$ is the likelihood at the nominal
signal yield and ${\cal L}(0)$ is the likelihood with the signal
yield set to 0. For the $\Dstarzero \eta^{\prime}$ decay mode we
also quote a 90\% confidence level upper limit using Poisson
statistics.}
\begin{center}
\begin{tabular}{lcrccclcc}
\hline \hline \\
$\Bzerobar$ mode  & { \ } & \multicolumn{5}{c}{$\Br \ ( \times
10^{-4})$} & { \ } & statistical \
\\
(decay channel) & & \multicolumn{5}{c}{} & & significance \
\\
\hline      \\
$\Dzero \pizero$                               && 2.9&$ \pm$& 0.2&$ \pm$& 0.3 && $>6.5$        \ \\
$\Dstarzero \pizero$                           && 2.9&$ \pm$& 0.4&$ \pm$& 0.5 && $>6.5$         \ \\
$\Dzero\eta(\ra \gamgam)$ && 2.4&$ \pm$& 0.3&$ \pm$& 0.3 && $>6.5$         \ \\
$\Dzero \eta(\ra\threepi)$  && 2.8&$ \pm$& 0.4&$ \pm$& 0.4 && $6.2$         \ \\
$\Dzero\eta$ (combined)        && 2.5&$ \pm$& 0.2&$ \pm$& 0.3 && $>6.5$        \ \\
$\Dstarzero \eta(\ra \gamgam)$ && 2.6&$ \pm$& 0.4&$ \pm$& 0.4 && $5.5$         \ \\
$\Dzero \omega$                                && 3.0&$ \pm$& 0.3&$ \pm$& 0.4 && $>6.5$        \ \\
$\Dstarzero \omega$                            && 4.2&$ \pm$& 0.7&$ \pm$& 0.9 && $6.1$         \ \\
$\Dzero \eta^{\prime}$                         && 1.7&$ \pm$& 0.4&$ \pm$& 0.2 && $6.3$         \ \\
$\Dstarzero \eta^{\prime}$                     && 1.3&$ \pm$& 0.7&$ \pm$& 0.2 && $3.0$         \ \\
                                 &&  \multicolumn{5}{c}{$< 2.6$ (90\% CL)} &&    \ \\
\hline \hline
\end{tabular}
\end{center}
\end{table}
\begin{table}
\caption{\label{tab:BFRatios} Ratios of branching fractions for
$\Bzerobar \ra \Dstze \hzero$. The first uncertainty is
statistical, the second is systematic.}
\begin{center}
\begin{tabular}{lcrccclcc}
\hline \hline \\
${\cal B}$ ratio  & { \ } &              \multicolumn{5}{c}{this
experiment}    & { \ }  & theoretical prediction
\\
\hline \\
${{\Bzerobar \ra \Dzero \pizero} \over {\Bzerobar \ra \Dstarzero \pizero}}$  && $1.0$ & $\pm$ & $0.1$ & $\pm$ & $0.2$ && 1.20 \cite{ref:Chua} or $1.0\pm0.2$ \cite{ref:SCET}   \\
\\
${{\Bzerobar \ra \Dzero \etaM} \over {\Bzerobar \ra \Dstarzero \etaM}}$     && $0.9$ & $\pm$ & $0.2$ & $\pm$ & $0.1$ && 0.78 \cite{ref:Chua}     \\
\\
${{\Bzerobar \ra \Dzero \omegaM} \over {\Bzerobar \ra \Dstarzero \omegaM}}$  && $0.7$ & $\pm$ & $0.1$ & $\pm$ & $0.1$ && 0.41 \cite{ref:NeuSte} \\
\\
${{\Bzerobar \ra \Dzero \etaPM} \over {\Bzerobar \ra \Dstarzero \etaPM}}$ && $1.3$ & $\pm$ & $0.8$ & $\pm$ & $0.2$ && 0.64--0.78  \cite{ref:Deandrea}                    \\
\\
${{\Bzerobar \ra \Dzero \etaPM} \over {\Bzerobar \ra \Dzero \etaM}}$  && $0.7$ & $\pm$ & $0.2$ & $\pm$ & $0.1$ && 0.64--0.68  \cite{ref:Deandrea}                 \\
\\
${{\Bzerobar \ra \Dstarzero \etaPM} \over {\Bzerobar \ra \Dstarzero \etaM}}$      && $0.5$ & $\pm$ & $0.3$ & $\pm$ & $0.1$ && 0.67--0.68 \cite{ref:Deandrea} \\
\hline \hline
\end{tabular}
\end{center}
\end{table}

\begin{table}
\caption{\label{tab:BFTheory} Comparison of our measurements and
theoretical predictions of naive factorization for the branching
fractions of the color-suppressed $\Bzerobar$ decays reported in
this paper. For the experimental results the statistical and
systematic uncertainties are added in quadrature.}
\begin{center}
\begin{tabular}{lcccccccc}
\hline   \hline                      \\
$\Bzerobar$ decay mode          & { \ }  & ${\cal B}(\times 10^{-4})$ & { \ } & ${\cal B}(\times 10^{-4})$ \\
                           & { \ }  &  this experiment    & { \ }  & theory (factorization)   \\ \hline \\
$\Dzero \pizero$           && $2.9 \pm 0.4$  && 0.58 \cite{ref:Chua} (0.7 \cite{ref:NeuSte}) \\
$\Dstarzero \pizero$       && $2.9 \pm 0.6$  && 0.65 \cite{ref:Chua} (1.0 \cite{ref:NeuSte}) \\
$\Dzero \eta$              && $2.5 \pm 0.4$  && 0.34 \cite{ref:Chua} (0.5 \cite{ref:NeuSte}) \\
$\Dstarzero \eta$          && $2.6 \pm 0.6$  && 0.37 \cite{ref:Chua} (0.6 \cite{ref:NeuSte}) \\
$\Dzero \omega$            && $3.0 \pm 0.5$  && 0.66 \cite{ref:Chua}  (0.7 \cite{ref:NeuSte}) \\
$\Dstarzero \omega$        && $4.2 \pm 1.1$  && 1.7 \cite{ref:NeuSte}      \\
$\Dzero \eta^{\prime}$     && $1.7 \pm 0.4$  && 0.30--0.32 \cite{ref:Deandrea}  \\
$\Dstarzero \eta^{\prime}$ && $1.3 \pm 0.7$  && 0.41--0.47 \cite{ref:Deandrea} \\
\hline \hline
\end{tabular}
\end{center}
\end{table}


\section{Results}
\label{sec:Results}

\subsection{Branching fractions}

The branching fractions of the color-suppressed modes reported in
this paper and their statistical and systematic uncertainties are
listed in Table~\ref{tab:BFD0} for the three $\Dzero$ decay modes
$\Kpi$, $\Ktwopi$, and $\Kthreepi$. The measurements obtained by
combining the three $\Dzero$ decay modes are presented in
Table~\ref{tab:BF}. Except for the $\Dstarzero \eta^{\prime}$
decay channel all measurements have statistical significance in
excess of five-standard deviations. For the $\Dstarzero
\eta^{\prime}$ decay channel we quote a 90\% confidence level
upper limit using Poisson statistics. To aid in combining our
result with future results for $\Dstarzero \etap$ a central value
with statistical and systematic uncertainties is also given. For
the $\Dzero \eta$ decay mode the most precise result is obtained
by combining the $\eta \ra \gamgam$ and $\eta \ra \threepi$ decay
modes.

The results listed in Tables~\ref{tab:BFD0} and \ref{tab:BF} are
also presented in the summary Figs.~\ref{fig:d0_dstar0_pi0_br},
\ref{fig:d0_dstar0_eta_br}, and \ref{fig:d0_dstar0_omega_br} for
comparison. It is seen that, for a given $B$ decay, the three
measurements using the three $\Dzero$ decay modes are consistent
among themselves. Where available, previous results by the
CLEO~\cite{ref:CLEO} and Belle~\cite{ref:Belle} experiments are
also shown. The precision of the results on the branching
fractions presented in this paper can be compared to the precision
of existing measurements as listed in Tab.~\ref{tab:existingBFs}.

In some cases theoretical predictions are more precise for ratios
of branching fractions than for branching fractions
themselves~\cite{ref:Deandrea,ref:SCET}. An example is the ratio
of ${\cal B}(\Bzerobar \ra \Dstze \etap)$ to ${\cal B}(\Bzerobar
\ra \Dstze \eta)$~\cite{ref:Deandrea}. Systematic uncertainties
partly cancel in the measurement of ratios so they are also more
precisely determined experimentally. We compare measured ratios of
branching ratios to theoretical predictions in
Table~\ref{tab:BFRatios}.

\subsection{Isospin symmetry and decay amplitudes}
\label{sec:IsospinRel}

Isospin symmetry relates the amplitudes for the $B^- \ra \Dstze
\pimi$, $\Bzerobar \ra \Dstpl \pimi$, and $\Bzerobar \ra \Dstze
\pizero$ decay modes~\cite{ref:Rosner}. These amplitudes can be
expressed as~\cite{ref:NeuPet}
\begin{equation}
\begin{split}
{\cal A}(\Dstze \pimi) &= \sqrt{3}\,{\cal A}_{3/2,D^{(\ast)}}, \\
{\cal A}(\Dstpl \pimi) &=\sqrt{{1}/{3}}\,{\cal A}_{3/2,D^{(\ast)}}
+\sqrt{{2}/{3}}\,{\cal
A}_{1/2,D^{(\ast)}}, \ {\rm and} \\
\sqrt{2}{\cal A}(\Dstze \pizero) & = \sqrt{{4}/{3}}\,{\cal
A}_{3/2,D^{(\ast)}}-\sqrt{{2}/{3}}\,{\cal A}_{1/2,D^{(\ast)}},
\end{split}
\end{equation}
where the amplitudes ${\cal A}_{1/2,D^{(\ast)}}$ and ${\cal
A}_{3/2,D^{(\ast)}}$ correspond to pure $I=3/2$ and $I=1/2$
isospin eigenstates. This leads to the triangle relation:
\begin{equation}
{\cal A}(\Dstze \pimi) = {\cal A}(\Dstpl \pimi) + \sqrt{2}\, {\cal
A}(\Dstze \pizero).
\end{equation}
If the relative strong-interaction phase between the two-isospin
amplitudes ($\delta_{D^{(\ast)}}$) is equal to zero, the
interference between these isospin eigenstates is maximally
destructive for the color-suppressed $\Bzerobar \ra D^{(*)0} h^0$
decay, while it is respectively maximally constructive for the
color-allowed $\Bzerobar \ra D^{(*)+} h^-$ decay. It follows from
QCD factorization~\cite{ref:Beneke}, in the heavy-quark limit,
that
\begin{equation}
{\cal A}_{1/2,D^{(\ast)}}/ \sqrt{2} {\cal A}_{3/2,D^{(\ast)}}=
1+{\cal O}(\Lambda_{\rm QCD}/m_Q),
\end{equation}
where $m_Q$ represents $m_c$ or $m_b$ and where the correction to
``1'' is also suppressed by a power of $1/N_c$, the number of
colors~\cite{ref:NeuPet,ref:LLW}. The above relation also implies
that $\delta_{D^{(\ast)}}={\cal O}(\Lambda_{\rm QCD}/m_Q)$.
Final-state interactions (FSI) effects in the $I=3/2$ and $I=1/2$
channels might be expected to be independent, leading to a
non-zero phase difference $\delta_{D^{(\ast)}}$. If the value of
$\delta_{D^{(\ast)}}$ is large enough it will substantially undo
the destructive interference for the color-suppressed decay
$\Bzerobar \ra D^{(*)0} h^0$, increasing the associated branching
fraction.

Using the various equations listed above, the values from
Table~\ref{tab:BF} for ${\cal B}(\Bzerobar \ra \Dstze \pizero)$,
the Particle Data Group values~\cite{ref:PDG} for ${\cal B}(\Bmi
\ra \Dstarzero \pimi)$ and ${\cal B}(\Bzerobar \ra D^{*+} \pimi)$,
the recent measurements by the CLEO
collaboration~\cite{ref:cleoDpi} for ${\cal B}(\Bmi \ra \Dzero
\pimi)$ and ${\cal B}(\Bzerobar \ra D^{+} \pimi)$, and the $B$
meson lifetime ratio $\tau(\B^+)/\tau(\Bzero)= 1.083 \pm
0.017$~\cite{ref:PDG}, we calculate the value of the strong phase
difference $\vert \delta_{D} \vert = 30^\circ \pm 5^\circ$ for $D
\pi$ final states and $\vert \delta_{D^\ast} \vert = 33^\circ \pm
5^\circ$ for $D^{\ast}\pi$ final states. The ratio of isospin
amplitudes $\vert {\cal A}_{1/2,D^{(\ast)}}/ \sqrt{2} {\cal
A}_{3/2,D^{(\ast)}}\vert$ is found to be equal to $0.69\pm 0.09$
($0.76 \pm 0.08$).

\subsection{Discussion}
\label{sec:Discussion}

Significant non-zero strong interaction phases are evidence that
the naive factorization model is inadequate. Therefore, when
computing the decay amplitudes, instead of using the
parametrization with $a_1$ and $a_2$, the alternative
parametrization in terms of isospin amplitudes may be more
appropriate. Moreover, if we analyze the $B$ decays to $D^{(\ast)}
\pi$ final states without FSI~\cite{ref:NeuPet}, we compute a
value $\vert a_2 \vert=0.57 \pm 0.07$ ($0.56 \pm 0.08$). These
values are quite different from $\vert a_2 \vert=0.2$ to $0.3$
from charmonium final states and indicate as well the necessity of
including strong non factorizable and process-dependent FSI
effects in the description of $\Bzerobar \ra \Dstze \hzero$ modes.

Various theoretical approaches that relax the conditions of naive
factorization  are being pursued in an effort to understand the
emerging pattern of color-suppressed decay
rates~\cite{ref:NeuPet,ref:Chua,ref:SCET,ref:pQCD}.


\section{Summary}
\label{sec:Summary}

We present measurements of the branching fractions for the
color-suppressed decays $\Bzerobar \ra  {\Dzero \pizero}$,
${\Dstarzero \pizero}$, ${\Dzero \eta}$, ${\Dstarzero \eta}$,
${\Dzero \omega}$, ${\Dstarzero \omega}$, and $\Dzero
\eta^{\prime}$. Our results are in agreement with previous
measurements~\cite{ref:CLEO,ref:Belle} but are more precise.
Branching fractions  for ${\Bzerobar \ra \Dstarzero \eta}$,
${\Dstarzero \omega}$, and $\Dzero \eta^{\prime}$ are measured for
the first time with more than five-sigma statistical significance.
We also set an upper limit on the branching fraction for the
$\Dstarzero \eta^{\prime}$ decay.

All measured color-suppressed decays have similar branching
fractions with central values between $1.7 \times 10^{-4}$ and
$4.2 \times 10^{-4}$. They are all significantly larger than
theoretical expectations based on naive factorization and
therefore present a challenge for the theoretical interpretation.
These results strongly suggest the presence of final-state
re-scattering effects. \\ \\

\section{Acknowledgments}

We are grateful for the extraordinary contributions of our \pep2\
colleagues in achieving the excellent luminosity and machine
conditions that have made this work possible. The success of this
project also relies critically on the expertise and dedication of
the computing organizations that support \babar. The collaborating
institutions wish to thank SLAC for its support and the kind
hospitality extended to them. This work is supported by the US
Department of Energy and National Science Foundation, the Natural
Sciences and Engineering Research Council (Canada), Institute of
High Energy Physics (China), the Commissariat \`a l'Energie
Atomique and Institut National de Physique Nucl\'eaire et de
Physique des Particules (France), the Bundesministerium f\"ur
Bildung und Forschung and Deutsche Forschungsgemeinschaft
(Germany), the Istituto Nazionale di Fisica Nucleare (Italy), the
Foundation for Fundamental Research on Matter (The Netherlands),
the Research Council of Norway, the Ministry of Science and
Technology of the Russian Federation, and the Particle Physics and
Astronomy Research Council (United Kingdom). Individuals have
received support from the A. P. Sloan Foundation, the Research
Corporation, and the Alexander von Humboldt Foundation.

\end{document}

%% file: abstract.tex
Using a sample of $88.8 \times 10^6$ $B \kern
0.18em\overline{\kern -0.18em B}{}\xspace$ events collected with
the $\mbox{\slshape {B}\kern-0.1em{\smaller
A}\kern-0.1em{B}\kern-0.1em{\smaller A\kern-0.2em{R}}}$ detector
at the PEP-II storage rings at the Stanford Linear Accelerator
Center, we measure the branching fractions of seven
color-suppressed $B$-meson decays: ${\cal B} (\kern
0.18em\overline{\kern -0.18em B}{}\xspace^0 \ra D^0 \pi^0)$ $ = (
2.9 \pm 0.2 ({\rm stat.}) \pm 0.3 ({\rm syst.}) ) \times 10^{-4}$,
${\cal B} (\kern 0.18em\overline{\kern -0.18em B}{}\xspace^0 \ra
D^{*0} \pi^0)$ $= ( 2.9 \pm 0.4 ({\rm stat.}) \pm 0.5 ({\rm
syst.}) ) \times 10^{-4}$, ${\cal B} (\kern 0.18em\overline{\kern
-0.18em B}{}\xspace^0 \ra D^0 \eta) $ $= ( 2.5 \pm 0.2 ({\rm
stat.}) \pm 0.3 ({\rm syst.}) ) \times 10^{-4}$, ${\cal B} (\kern
0.18em\overline{\kern -0.18em B}{}\xspace^0 \ra D^{*0} \eta) $ $=
( 2.6 \pm 0.4 ({\rm stat.}) \pm 0.4 ({\rm syst.}) ) \times
10^{-4}$, ${\cal B} (\kern 0.18em\overline{\kern -0.18em
B}{}\xspace^0 \ra D^0 \omega) $ $= ( 3.0 \pm 0.3 ({\rm stat.}) \pm
0.4({\rm syst.}) ) \times 10^{-4}$, ${\cal B} (\kern
0.18em\overline{\kern -0.18em B}{}\xspace^0 \ra D^{*0} \omega)$ $=
( 4.2 \pm 0.7 ({\rm stat.}) \pm 0.9 ({\rm syst.}) ) \times
10^{-4}$, and ${\cal B} (\kern 0.18em\overline{\kern -0.18em
B}{}\xspace^0 \ra D^0 \eta^{\prime})$ $= ( 1.7 \pm 0.4 ({\rm
stat.}) \pm 0.2 ({\rm syst.}) ) \times 10^{-4}$. We set the 90\%
confidence-level upper limit: ${\cal B} (\kern
0.18em\overline{\kern -0.18em B}{}\xspace^0 \ra D^{*0}
\eta^{\prime})$ $< 2.6 \times 10^{-4}$. The channels $\kern
0.18em\overline{\kern -0.18em B}{}\xspace^0 \ra D^{*0} \eta$,
$D^{*0} \omega$, and $D^0 \eta^{\prime}$ are seen with more than
five-sigma statistical significance. All of these branching
fractions are significantly larger than theoretical expectations
based on the ``naive'' factorization model.

%% file: paper.bbl
\begin{thebibliography}{99}
\bibitem{ref:footnote1} Charge conjugate reactions are assumed throughout this paper
and branching fractions are averaged accordingly; also, the
superscript ``$^{(*)}$'' indicates that a symbol must be
considered both with and without the ``$^*$'' superscript.
\bibitem{ref:footnote2} Kinematic quantities are expressed in the laboratory frame unless
they are denoted with a superscript ``$^\ast$'', which indicates
quantities in the \epem\ center-of-mass frame.
\end{thebibliography}
